\documentclass[11pt]{article}
\usepackage{amsmath,amsfonts,amssymb,amsthm}
\usepackage{graphicx}
\usepackage{xcolor}
\usepackage{url}
\usepackage{geometry}
\usepackage{authblk}
\usepackage{float}
\usepackage{dcolumn}
\usepackage{multirow}
\usepackage{authblk}
\usepackage{adjustbox}
\usepackage{caption}

\graphicspath{{./images/}}
\usepackage[skip=0.5ex]{subcaption}
\usepackage{hyperref}

\hypersetup{
    colorlinks=true,
    linkcolor=blue,
    citecolor=red,
}
\usepackage{cleveref}
\crefname{figure}{figure}{figures}  
\Crefname{figure}{Figure}{Figures}  

\numberwithin{equation}{section}
\numberwithin{figure}{section}

\newcommand{\rom}[1] {\uppercase\expandafter{\romannumeral #1\relax}}
\begin{document}

\newcommand{\bea}{\begin{eqnarray}}
\newcommand{\eea}{\end{eqnarray}}
\newcommand{\be}{\begin{equation}}
\newcommand{\ee}{\end{equation}}

\definecolor{TW-color}{RGB}{100,0,100}
\definecolor{MR-color}{RGB}{0,0,255}
\definecolor{SK-color}{RGB}{0,255,0}
\definecolor{Error-color}{RGB}{250,0,0}
\newcommand{\TWedit}[1]{{\color{TW-color}#1}}
\newcommand{\Error}[1]{{\color{Error-color}#1}}
\newcommand{\SKedit}[1]{{\color{SK-color}#1}}
\newcommand{\MRedit}[1]{{\color{MR-color}#1}}
\title{\textbf{Type I/II Vortex Dynamics With Excited Normal Modes}}
\author[1]{Steffen Krusch\thanks{\tt S.Krusch@kent.ac.uk}}
\author[1]{Morgan Rees\thanks{\tt mjrr2@kent.ac.uk}}
\affil[1]{\it{\small{School of Mathematics, Statistics and Actuarial Sciences, University of Kent, Canterbury, United Kingdom}}}
\author[2,3]{Thomas Winyard\thanks{\tt twinyard001@dundee.ac.uk}}
\affil[2]{\it{\small{Division of Mathematics, University of Dundee, Dundee DD1 4HN, United Kingdom}}}
\affil[3]{\it{\small{Maxwell Institute of Mathematical Sciences and School of Mathematics, University of Edinburgh, Edinburgh, United Kingdom}}}
\date{}

\allowdisplaybreaks

\maketitle
\begin{abstract}
We investigate the effects of excited normal modes on vortex dynamics in the Abelian Higgs model in the type \rom{1} (static attraction) and type \rom{2} (static repulsion) regimes. We demonstrate that shape normal modes compete with the static inter-vortex forces. This can result in excited type \rom{1} (\rom{2}) vortices repelling (attracting). In addition we observe the existence of spectral walls that prevent the formation (break up) of bound states in type \rom{1} (\rom{2}) systems. We study the long lived quasi-bound states and orbits induced by their competing forces and the effective centrifugal forces. Finally, we observe that the effect is strong enough to induce vortex-antivortex pairs to follow long-lived orbits.
\end{abstract}
 
\section{\label{sec:intro} Introduction}

The Abelian-Higgs model \cite{higgs1966spontaneous} is a relativistic field theory whose static solutions in $(2+1)$-dimensions take the form of topologically stable solitons known as vortices. They are the simplest topological soliton that arise in gauge theories \cite{manton2004topological} and as such are an invaluable toy model for soliton dynamics for more general models of high energy physics.

The field theory consists of a complex scalar field $\phi$ coupled to a $U(1)$ gauge field $A$. Such a theory models phase transitions in cosmology where vortex excitations represent cross-sections of cosmic strings formed in the early universe \cite{vilenkin1985cosmic,vilenkin1994cosmic}. In addition, the static theory is equivalent to the effective Ginzburg-Landau theory \cite{ginzburg1950j}, which models magnetic field penetration in superconductors. Both these theories are governed by
a single parameter $\lambda$, which broadly splits solutions of $N$ vortices into three types: type I ($\lambda < 1$), where vortices exhibit long-range attraction and combine together to form large mega vortices of winding $N$, and type II ($\lambda > 1$), where vortices repel at long range such that there are no static solutions for more than a single vortex ($|N|>1$). Finally, in the critically coupled case ($\lambda = 1$) vortices exhibit no static interactions, and the space of static solutions is the $2N$ dimensional moduli space $\mathcal{M}_N = \mathbb{C}^N/S_N$ of unordered points in the plane (where $S_N$ is the permutation group)
\cite{taubes1,Manton:E_int}.

However, these two physical theories do diverge in their dynamics; the Abelian-Higgs model exhibits second-order dynamics with Lorentz invariance \cite{MORIARTY1988411,myers1992study,shellard1988vortex,Ruback,JD,samols,Stuart}, whereas the time-dependent Ginzburg-Landau model exhibits first-order dynamics \cite{dorsey1992vortex,MANTON1997114}. In this paper, we focus on second-order vortex dynamics which has been well studied in the low-energy limit for all values of $\lambda$ \cite{MORIARTY1988411,myers1992study,shellard1988vortex}. Most work has focused on the mathematically interesting critically coupled case, where low-energy dynamics can be approximated as free geodesic motion on the moduli space $\mathcal{M}_N$ of static solutions.

Recently it was shown numerically, for critical coupling, that the inclusion of normal modes can make the dynamic forces become attractive or repulsive \cite{SKMRTW,AMRW,alonso2026bps}. As a result, rather than the traditional $90^\circ$ scattering, two excited critically coupled vortices can scatter many times, with the number of scatterings shown to be chaotic \cite{SKMRTW,AMRW,alonso2026bps,alonsoizquierdo2026scatteringwobblingvortices}. These effects have also been studied concurrently in \cite{alonsoizquierdo2026scatteringwobblingvortices}, where spectral flow and mode excitation generate effective forces and resonant energy transfer, producing fractal structures in the scattering diagrams.

The moduli space approximation has been expanded to include these normal modes \cite{AMRW,alonso2026bps}. To understand the 2 vortex case one must match up the $N=2$ bound vortex normal modes to well separated $N=1$ vortices where the mode frequency changes continuously. It was shown that the frequency of some modes instead increase into the continuum, which produces a spectral wall that the vortices bounce off as energy is pumped into the continuous spectrum \cite{3Vortex,alonso2023spectral,AMRW,AORW,SW-1,SW-2}. This has now also been observed in the $N=3$ case \cite{3Vortex,alonso2026bps}.
Analogous mode-driven dynamics have since been identified in a range of related BPS solitons. Feshbach-type resonant modes have been shown to govern the energy transfer mechanism in BPS solitons more generally \cite{martin2025feshbach}, and a recent study of vortex-antivortex collisions in the deep type \rom{2} regime reveals a chaotic bounce window structure in the final state, driven by precisely such a quasi-normal mode \cite{bachmaier2026resonance}. The moduli space description of the single excited vortex has also been refined to incorporate the radially symmetric shape mode \cite{miguelez2025moduli}, alongside a collective coordinate model for the shape mode dynamics of two vortices at critical coupling \cite{AMMW} and a study of the radiation emitted by an excited vortex as the mode decays \cite{JJ}. The same resonant energy transfer mechanism underlies the well-known fractal pattern of bounce windows in kink-antikink scattering \cite{sug,CSW,MORW} and the chaotic decay of strongly excited oscillons \cite{blaschke2024amplitude}.

In this paper we will study the effects of normal modes on vortex interactions away from critical coupling, extending the analysis of \cite{kar112637}. Now there is a competition between the static inter-vortex forces and the dynamical forces from the normal modes. The structure of the paper is as follows. In \cref{sec:model} we review the Abelian-Higgs model and its static vortex solutions, and we set up the linearised eigenvalue problem for the normal modes for general $\lambda$ and topological charge $N$. In \cref{sec:InteractionEnergy} we describe how we approximate the inter-vortex interaction, by combining the static force with a mode-induced contribution that we extract from the angular frequency as the vortices move. The numerical scheme used for the dynamical simulations is set out in \cref{sec:InitialConfiguration}. We then briefly review scattering at critical coupling in \cref{sec:critcal_couling}, before considering the type \rom{1} regime in \cref{sec:TypeI} and the type \rom{2} regime in \cref{sec:TypeII}. In each case we observe spectral walls and quasi-stationary states, induced by the competition between the static force and the mode-induced force. Finally, in \cref{sec:VortexOrbits} we consider vortex orbits at critical coupling and away from it, and we show that the same mechanism supports long-lived vortex-antivortex orbits.

\section{The Model}
\label{sec:model}
The Abelian-Higgs model in $(2+1)$ dimensions is defined by the action
\begin{equation}
S = \int \int_{\mathbb{R}^2} \left[ -\frac{1}{4} f_{\mu\nu}f^{\mu \nu} + \frac{1}{2} \overline{D_\mu \phi}\, D^\mu \phi - \frac{\lambda}{8} (|\phi|^2 - 1)^2 \right] d^2x\, dt,
\label{eq:Action}
\end{equation}
where $\phi(t,x)$ is a complex scalar field (the Higgs field), with spatial coordinate $x \in \mathbb{R}^2$, and the 1-form $A_\mu(t,x) = (A_0, A_1, A_2)$ is a $U(1)$ gauge field. The covariant derivative is $D_\mu \phi = (\partial_\mu - i A_\mu)\phi$, and $f_{\mu\nu} = \partial_\mu A_\nu - \partial_\nu A_\mu$ is the field strength tensor. We choose the spacetime $\mathbb{R}^{2+1}$ metric signature $(+,-,-)$. As this is a gauge theory, the model is invariant under the following gauge transformations,
\begin{align}
\phi(x) &\mapsto e^{i\alpha(x)}\phi(x), &
A_\mu(x) &\mapsto A_\mu(x) + \partial_\mu \alpha(x).
\end{align}
Note that we have made use of rescalings such that the physical quantities of electric charge, Higgs vacuum expectation value and speed of light are all normalised to $1$. This means we are able to write the most general model in terms of a single real positive parameter $\lambda$ which directly determines the Higgs field mass.

Varying the action with respect to $\phi$ and $A_\mu$ yields the second-order Euler-Lagrange equations of motion
\begin{align}
D_\mu D^\mu \phi - \frac{\lambda}{2} (1 - |\phi|^2)\phi &= 0, \label{eq:eom1} \\
\partial_\mu f^{\mu \nu} + \frac{i}{2} \left( \bar{\phi} D^\nu \phi - \phi \overline{D^\nu \phi} \right) &= 0. \label{eq:eom2}
\end{align}

For finite energy, as $\rho = |x| \to \infty$, field configurations must satisfy $f_{12} \to 0$, $D_\mu \phi \to 0$, and $|\phi| \to 1$. This constrains the Higgs field at infinity to lie on the unit circle, $\phi_\infty = \lim_{\rho \to \infty} \phi(x) \in S^1$. Hence, on the boundary $\phi_\infty: S^1 \to S^1$ and therefore
has an associated winding number $N \in \mathbb{Z}$. As the covariant derivative vanishes, we can equate the angular derivative of the phase and the angular component of the gauge field $A_\theta$. Using Stokes' theorem this leads to the field strength being quantised,
\begin{equation}
N = \frac{1}{2\pi} \int_{\mathbb{R}^2} f_{12} \, d^2x.
\label{eq:TopCharge}
\end{equation}
Note that (in reference to the static Ginzburg-Landau model) we refer to the static part of the field strength tensor $f_{12} = \partial_1 A_2 - \partial_2 A_1$ as the magnetic field $B = f_{12}$ orthogonal to the plane. As the Higgs field is continuous the winding number $N$ counts the number of zeroes of $\phi$ with multiplicity. We will consider these zeroes to be the positions of individual vortices.

\subsection{Static vortex solutions}
To find static solutions of the equations of motion \cref{eq:eom1,eq:eom2}, we make use of the principle of symmetric criticality and consider the fields retain the radial symmetry of the static energy. Namely, we use the ansatz,
\begin{align}
\phi &= f(\rho) \, e^{iN\theta}, &
(A_0, A_\rho, A_\theta) &= (0, 0, a_\theta(\rho)),
\label{eq:polarAnsatz}
\end{align}
where $(\rho,\theta)$ are the polar coordinates in the plane, and we have chosen the radial gauge \(A_\rho = 0\). Substituting this ansatz into \cref{eq:Action} and assuming a static configuration, we obtain the reduced static energy
\begin{equation}
V = \pi \int_0^{\infty} \left( \frac{1}{\rho^2} \left( \frac{d a_\theta}{d \rho} \right)^2 + \left( \frac{d f}{d \rho} \right)^2 + \frac{1}{\rho^2} (N - a_\theta)^2 f^2 + \frac{\lambda}{4} (1 - f^2)^2 \right) \rho \, d\rho,
\label{eq:polarEnergy}
\end{equation}
which yields the Euler-Lagrange ODEs,
\begin{align}
f'' + \frac{1}{\rho} f' - \frac{1}{\rho^2} f (N - a_\theta)^2 - \frac{\lambda}{2} f (f^2 - 1) &= 0, \label{eq:eom_radial1} \\
a_\theta'' - \frac{1}{\rho} a_\theta' + (N - a_\theta) f^2 &= 0. \label{eq:eom_radial2}
\end{align}
The following boundary choices ensure finite energy and regularity at the origin,
\begin{align}
f(0) &= 0,&
\lim_{\rho \to \infty} f(\rho) &= 1,\\
a_\theta(0) &= 0,&
\lim_{\rho \to \infty} a_\theta(\rho) &= N.
\end{align}
It will be convenient later to expand the solution near the origin as,
\begin{align}
f(\rho) &= \rho^N F(\rho^2), & a_\theta(\rho) &= \rho^2 \, G(\rho^2),
\end{align}
where $F(\rho^2)$ and $G(\rho^2)$ are radial power series in $\rho^2$ with non-zero leading coefficients. Thus, we can write the original fields as the following radially symmetric solution of degree $N$,
\begin{equation}
\begin{aligned}
&\phi = (x_1 + i x_2)^N F(x_1^2 + x_2^2), \\
&A_\mu = (A_0, A_1, A_2) = \begin{pmatrix} 0 \\ -x_2 \, G(x_1^2 + x_2^2) \\ x_1 \, G(x_1^2 + x_2^2) \end{pmatrix},
\end{aligned}
\label{eq:Ansatz}
\end{equation}
where \(F(\rho^2)\) and \(G(\rho^2)\) are nonlinear functions, expandable as a power series near \(\rho = 0\). This yields the Euler-Lagrange equations of motion
\begin{align}
8 \rho^2 F'' + 8 F' - \lambda \rho^{2N} F^3 + \lambda F - 2 \rho^2 F G^2 + 4 N (F G + 2 F') &= 0, \label{eq:eom_power1} \\
4 \rho^2 G'' + 8 G' + \rho^{2(N-1)} F^2 (N - \rho^2 G) &= 0. \label{eq:eom_power2}
\end{align}
The coupled nonlinear system \cref{eq:eom_power1,eq:eom_power2} or \cref{eq:eom_radial1,eq:eom_radial2} can now be solved numerically using a one-dimensional arrested Newton flow algorithm with \(4^{\mathrm{th}}\)-order finite differences \cite{TomANF1}.

\subsection{Vortex normal modes}
\label{sec:Linearisation}
We will take a similar approach to the original work by Goodband and Hindmarsh in \cite{Hindmarsch}, which studied vortex normal modes for several values of $\lambda.$ Our discussion is closely related to \cite{alonso2026spectral} where the complete spectrum is analysed in greater detail, and we include our calculations here to present a consistent and self-contained approach to the numerical simulations of excited vortices. 
These modes have also been studied using different methods for $\lambda = 1$ \cite{Alonso_Izquierdo_2016,alonso2023spectral} and all $\lambda$ \cite{TomMartinShortrange}. An analogous geometric analysis of the Jacobi operator has recently been carried out for vortices in the gauged $\mathbb{C}P^1$ sigma model \cite{gavrea2026shape}, where the shape mode eigenvalue lies close to the continuum threshold and is therefore weakly bound.

To proceed we consider perturbations of the fields $(\phi, A)$ around the background of a static vortex solution $(\phi_s, a)$.
It is convenient to rewrite the vector gauge potential in terms of total angular momentum states,
\begin{align}
    a_+ &= a_1 + i a_2, & a_- &= a_1 - i a_2,
    \label{eq:gauge}
\end{align}
where $a_- = \overline{a}_+$ and $(a_1, a_2) = \left( -\frac{\sin\theta}{\rho} a_\theta(\rho), \frac{\cos\theta}{\rho} a_\theta(\rho) \right)$, with $a_\theta(\rho)$ a radial profile function from \cref{eq:polarAnsatz}. We consider the quantities
\begin{align}
    \psi(x) &= \phi(x) - \phi_s(x), & \chi_\pm(x) &= A_\pm(x) - a_\pm(x),
    \label{eq:pert}
\end{align}
where $(\phi_s(x), a_\pm(x))$ is a static solution of \cref{eq:eom1,eq:eom2}. A given configuration is therefore close to the static vortex when the perturbation $\psi$ and $\chi_\pm$ are small. This gives a correction to the action \cref{eq:Action} of the form
\begin{equation}
    S = S(\phi_s, a_\pm) + \epsilon^2 S_2 + \mathcal{O}(\epsilon^3),
    \label{eq:action}
\end{equation}
where $\epsilon \ll 1$ is the magnitude of the perturbation, and
\begin{align}
    S_2 &= \frac{1}{2} \int \xi^\dagger \mathcal{D} \xi \, d^2 x, & \xi^\dagger(\mathbf{x}) &= (\chi_- e^{i \omega t}, \chi_+ e^{-i \omega t}, \overline{\psi} e^{i \omega t}, \psi e^{-i \omega t}),
    \label{eq:S2}
\end{align}
where $\xi$ is the collected perturbations, $\omega$ is the angular frequency of the linear mode, and $t$ denotes time. Note that the linear action term vanishes as $(\phi_s, a)$ is a solution of the field equations. Since $\epsilon$ is small we neglect all terms higher than quadratic, leaving only linear corrections to the resulting equations of motion. The perturbed fields can now be written,
\begin{align}
    \phi(x) &= \phi_s(x) + \epsilon \psi(x) e^{-i \omega t}, & \overline{\phi(x)} &= \overline{\phi_s(x)} + \epsilon \overline{\psi(x)} e^{i \omega t}, \notag \\
    A_+(x) &= a_+(x) + \epsilon \chi_+(x) e^{-i \omega t}, & A_-(x) &= a_-(x) + \epsilon \chi_-(x) e^{i \omega t},
    \label{eq:pertAns1}
\end{align}
where we define $(A_0, A_\rho, A_\theta) = (0, 0, a_\theta(\rho))$ in the radial gauge $A_\rho = 0$, as in \cref{eq:polarAnsatz}.

We remove any remaining gauge freedom by choosing a gauge condition such that any linear derivative terms in the resulting equations of motion vanish \cite{Cheng:1984vwu},
\begin{equation}
    \partial_\mu \chi^\mu - (\overline{\psi} \phi_s - \overline{\phi_s} \psi) = 0.
    \label{eq:gaugeCond}
\end{equation}
Note that the Lorenz gauge ($\partial_\mu A^\mu = 0$) is automatically satisfied by this gauge choice. We make this choice such that it matches our gauge choice for the numerical dynamic field simulations later.

Substituting the perturbation ansatz \cref{eq:pertAns1} into the equations of motion and linearising about the static vortex configuration, we obtain an eigenvalue problem for the normal mode frequencies. Separating the time dependence yields the eigenvalue equation
\begin{equation}
    \mathcal{D}_{\mathrm{LG}} \begin{pmatrix} \chi_+ \\ \chi_- \\ \psi \\ \overline{\psi} \end{pmatrix} = \omega^2 \begin{pmatrix} \chi_+ \\ \chi_- \\ \psi \\ \overline{\psi} \end{pmatrix},
    \label{eq:eig}
\end{equation}
where
\[
    \mathcal{D}_{\mathrm{LG}} = \begin{pmatrix}
        D_1 & 0 & A & B \\
        0 & D_1 & C & E \\
        E & B & D_2 & V_1 \\
        C & A & V_2 & D_3
    \end{pmatrix}.
\]
Here
\begin{align}
        D_1 &= -\Delta + |\phi_s|^2, \notag \\
        D_2 &= -\Delta - i (a_+ + a_-) \partial_1 + (a_+ - a_-) \partial_2 + \frac{\lambda}{2} (2 |\phi_s|^2 - 1) + a_+ a_- + |\phi_s|^2,\notag \\
        D_3 &= -\Delta + i (a_+ + a_-) \partial_1 - (a_+ - a_-) \partial_2 + \frac{\lambda}{2} (2 |\phi_s|^2 - 1) + a_+ a_- + |\phi_s|^2, \notag\\
        A &= i \overline{\partial_1 \phi_s} + \overline{\partial_2 \phi_s} + \overline{\phi_s} a_+, \quad
        B = -i \partial_1 \phi_s - \partial_2 \phi_s + \phi_s a_+, \quad
        C = i \overline{\partial_1 \phi_s} - \overline{\partial_2 \phi_s} + \overline{\phi_s} a_-, \notag\\
        E &= -i \partial_1 \phi_s + \partial_2 \phi_s + \phi_s a_-, \quad
        V_1 = \frac{\lambda}{2} \phi_s^2 - \phi_s^2 \quad {\rm and} \quad V_2 = \frac{\lambda}{2} \overline{\phi_s}^2 - \overline{\phi_s}^2, \notag
\end{align}
with the Laplacian defined as $\Delta = \partial_{xx} + \partial_{yy}$.

We reduce the spectral problem by exploiting the rotational symmetry of the vortex background \cite{AD1}. Decomposing the perturbations into angular Fourier modes labelled by an integer wave number $k$, we obtain a family of radial eigenvalue problems. The wave number $k$ corresponds to the angular symmetry of the perturbation: $k=0$ represents radially symmetric (shape) modes, whilst $k \neq 0$ represents modes with 
$k$-fold rotational structure. For an $N$-vortex, the observed physically relevant modes satisfy $k\leq N,$ see \cite{alonso2026spectral} for a more detailed discussion.

For $k=0$ shape modes, we use the ansatz
\begin{align}
    \chi_+ &= -i e^{i \theta} v(\rho), & \chi_- &= i e^{-i \theta} v(\rho), \label{eq:k0_chi} \\
    \psi &= e^{i N \theta} u(\rho), & \overline{\psi} &= e^{-i N \theta} u(\rho), \label{eq:k0_psi}
\end{align}
where $v(\rho)$, $u(\rho)$ are radial eigenfunctions for the perturbations, found numerically. 

Substituting the ansatz \cref{eq:k0_chi,eq:k0_psi} into the eigenvalue problem \cref{eq:eig} and using the polar form of the background fields, we obtain the coupled radial eigenvalue problem
\begin{align}
    -\frac{d^2 v}{d\rho^2} - \frac{1}{\rho} \frac{d v}{d\rho} + \left( \frac{1}{\rho^2} + f^2(\rho) \right) v + \frac{2 N}{\rho} \left( 1 - \frac{a_\theta}{N} \right) f(\rho) u &= \omega^2 v, \label{eq:k0_v} \\
    -\frac{d^2 u}{d\rho^2} - \frac{1}{\rho} \frac{d u}{d\rho} + \left( \frac{(N - a_\theta)^2}{\rho^2} + \frac{3\lambda}{2} f^2(\rho) - \frac{\lambda}{2} \right) u + \frac{2 N}{\rho} \left( 1 - \frac{a_\theta}{N} \right) f(\rho) v &= \omega^2 u, \label{eq:k0_u}
\end{align}
where $\phi_s = f(\rho) e^{i N \theta}$ and $a_\theta(\rho)$ are from \cref{eq:polarAnsatz}, and $a_\pm = \mp \frac{i a_\theta(\rho)}{\rho} e^{\mp i \theta}$.

The case $k=0$ is treated separately because the radially symmetric perturbations decouple into a simpler two-component system involving only $v(\rho)$ and $u(\rho)$. For $k \neq 0$, the angular structure introduces an additional radial function $w(\rho)$, leading to a three-component system.
\begin{align}
    \chi_+ &= -\frac{i}{2} \left( e^{i (k+1) \theta} \left( s(\rho) - \frac{k}{\rho} v(\rho) \right) + e^{-i (k-1) \theta} \left( s(\rho) + \frac{k}{\rho} v(\rho) \right) \right), \label{eq:kn0_chi+} \\
    \chi_- &= \frac{i}{2} \left( e^{-i (k+1) \theta} \left( s(\rho) - \frac{k}{\rho} v(\rho) \right) + e^{i (k-1) \theta} \left( s(\rho) + \frac{k}{\rho} v(\rho) \right) \right), \label{eq:kn0_chi-} \\
    \psi &= e^{i N \theta} \left( \cos(k \theta) u(\rho) - i k \sin(k \theta) w(\rho) \right), \label{eq:kn0_psi} \\
    \overline{\psi} &= e^{-i N \theta} \left( \cos(k \theta) u(\rho) + i k \sin(k \theta) w(\rho) \right), \label{eq:kn0_psibar}
\end{align}
where
\begin{equation}
s(\rho) = v'(\rho) - \rho f(\rho) w(\rho), \label{eq:s}
\end{equation}
and $v(\rho)$, $u(\rho)$, $w(\rho)$ are radial eigenfunctions for the perturbations. This gives the second-order ODEs
{\small
\begin{align}
     \omega^2 v &=-\frac{d^2 v}{d\rho^2} - \frac{1}{\rho} \frac{d v}{d\rho} + \left( \frac{k^2}{\rho^2} + f^2(\rho) \right) v + 2 \left( f(\rho) + \rho f'(\rho) \right) w, \label{eq:kn0_v} \\
    \omega^2 u &=-\frac{d^2 u}{d\rho^2} - \frac{1}{\rho} \frac{d u}{d\rho} + \left( \frac{k^2 + (N - a_\theta)^2}{\rho^2} + \frac{3\lambda}{2} f^2(\rho) - \frac{\lambda}{2} \right) u -\frac{2 (N - a_\theta) (k^2 + \rho^2 f^2(\rho))}{\rho^2} w\notag\\&\quad+\frac{2 (N - a_\theta) f(\rho)}{\rho} \frac{d v}{d\rho}, \label{eq:kn0_u} \\
    \omega^2 w &=-\frac{d^2 w}{d\rho^2} - \frac{1}{\rho} \frac{d w}{d\rho} + \left( \frac{k^2 + (N - a_\theta)^2}{\rho^2} + \left( 1 + \frac{\lambda}{2} \right) f^2(\rho) - \frac{\lambda}{2} \right) w - \frac{2 (N - a_\theta)}{\rho^2} u + \frac{2 f'(\rho)}{\rho} v. \label{eq:kn0_w}
\end{align}
}
We employ a central second-order finite-difference scheme to discretise the ODEs \cref{eq:k0_v}--\cref{eq:k0_u} for $k=0$, writing the eigenvalue problem as a $2 \times 2$ block matrix with entries of size $M \times M$. For $k \neq 0$, we discretise \cref{eq:kn0_v}--\cref{eq:kn0_w} as a $3 \times 3$ block matrix. Details of the discretisation and boundary conditions are given in \cref{appendix:dis,appendix:bc}.

\begin{figure}
    \centering
    \begin{minipage}{0.99\textwidth}
        \includegraphics[width=\linewidth]{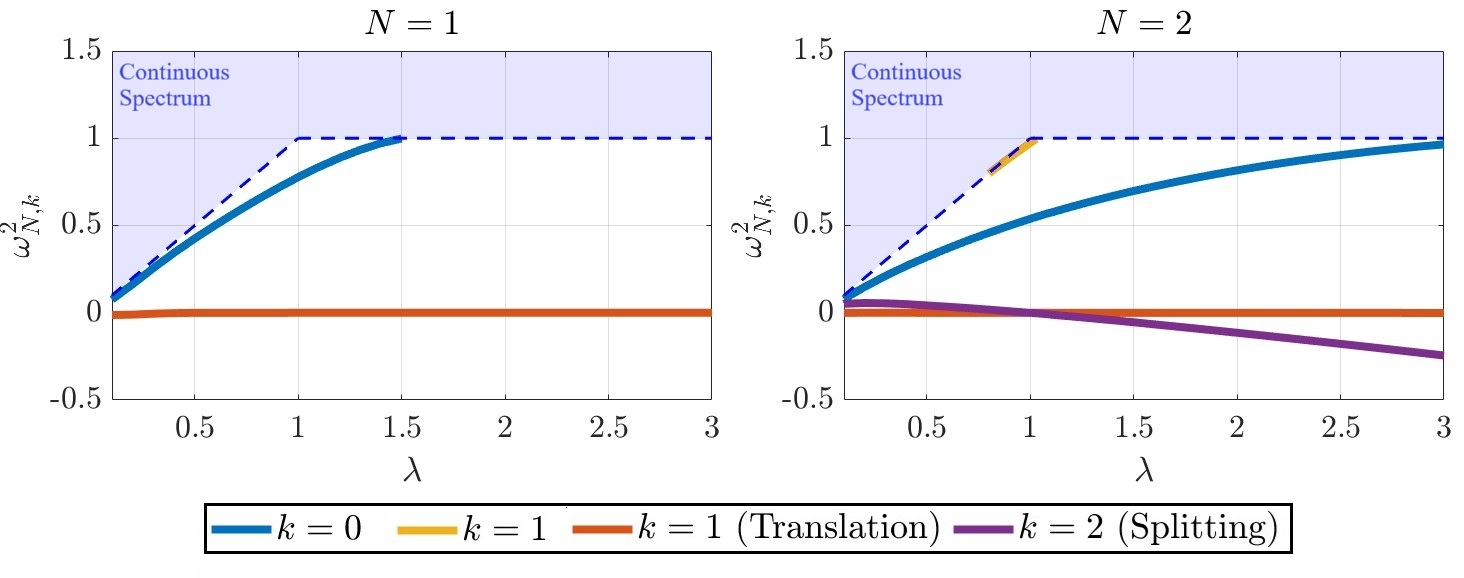}
        \caption{The frequency of the vortex normal modes as a function of $\lambda \in [0.1,3]$ for $N = 1,2$ and wave number $k \leq N$. The shaded region indicates the continuous spectrum.}
        \label{fig:N1Spectra}
    \end{minipage}
\end{figure}

We plot the eigenvalues for $\lambda \in [0.1,3]$ for $N=1$ and $N=2$ vortices in \cref{fig:N1Spectra}. Note that we normalise the eigenfunctions using the L$_2$ norm,
\begin{equation}
    \pi \int_0^\infty \left( v(\rho)^2 + u(\rho)^2 + w(\rho)^2 \right) \rho \, d\rho = 1, \label{eq:L2_norm}
\end{equation}
where $w(\rho) = 0$ for $k=0$.

The $k = 0$ mode is a radially symmetric shape mode, and exists in the discrete spectrum at $N=1$ for $\lambda \in [0.1,1.5]$. 
This is the only shape mode for $N=1,$ also see \cite{alonso2026spectral}.
For $N = 2$, the $k = 0$ mode exists in the discrete spectrum for $\lambda \in [0.1,3]$. Also at $N=2$, we have the $k=1$ shape mode, which exists only for $\lambda \in [0.8,1.03]$. For both $N=1$ and $N=2$, the translation mode is the $k=1$ zero mode, satisfying $k \leq N$. At $N=2$, there is additionally a $k=2$ splitting mode, which is a zero mode at critical coupling $(\lambda = 1)$, satisfying $k \leq N$.

\section{Approximating Vortex Interactions}
\label{sec:InteractionEnergy}
We now consider how the inter-vortex forces are modified by exciting normal modes. The long and short range static forces of vortices have been well studied \cite{speight1997static,Manton:E_int,TomMartinShortrange}. However, in general, the static forces need to be found numerically due to the non-linear nature of the equations of motion \cite{TomMartinShortrange}.

To understand the long-range behaviour, we write the fields as a perturbation about the vacuum $(f, a_\theta) = (1, N)$. We then substitute this into \cref{eq:eom_radial1,eq:eom_radial2} and retain only up to linear order terms in perturbations. The resulting system of ODEs decouple and can be solved to find,
\begin{align}
    f(\rho) &\sim 1 - \frac{q}{2\pi} K_0(\sqrt{\lambda} \rho), &
    a_\theta(\rho) &\sim N - \frac{m}{2\pi} \rho K_1(\rho),
    \label{eq:linear}
\end{align}
where $q$ and $m$ are $\lambda$-dependent constants, and $K_0$, $K_1$ are modified Bessel functions of the second kind. We can then model a vortex in the linear system of ODEs as a point source leading to the fields in \cref{eq:linear} \cite{speight1997static}.  Each point source has an associated scalar charge $q$ and magnetic dipole moment $m$ and two point sources have the interaction energy,
\begin{equation}
    E_{\text{int}}(s) = -\frac{q}{2\pi} K_0(\sqrt{\lambda} s) + \frac{m}{2\pi} K_0(s),
    \label{eq:AsyInt}
\end{equation}
where $s$ is their separation, and $s \gg 1,1/\sqrt{\lambda}$. The first term, mediated by the Higgs field with length scale $1/\sqrt{\lambda}$ is negative, contributing an attractive interaction. The second term, mediated by the magnetic field with length scale $1$ is positive contributing a repulsive interaction. Hence, for $\lambda < 1$ ($\lambda >1$) the leading length scale corresponds to the Higgs field (magnetic field) leading to attraction (repulsion) \cite{speight1997static,Manton:E_int}. At critical coupling ($\lambda = 1$), where $q = m$, the contributions from the Higgs field and magnetic field cancel, leading to no long-range interaction between static vortices. Note that the interaction energy between a vortex anti-vortex pair has the opposite sign for the magnetic term, hence both terms mediate an attractive force for all $\lambda$.

Short range vortex interactions have also been studied for all $\lambda$ in \cite{TomMartinShortrange}, where a formula for the interaction for $N$ vortices was derived from the spectral data of the Jacobi operator for the
cocentred $N$-vortex. A key consequence of this approach is that the interaction energy of a vortex pair with separation $s$ varies as $s^4,$ 
and the following formula holds
\begin{equation}
    E_{\text{int}}(s) = E_{\text{int}}(0) + \frac{1}{2} \omega_{\lambda;2,2}^2(0) \frac{s^4}{32 \beta(\lambda)},
    \label{eq:SRInt}
\end{equation}
where $E_{\text{int}}(0)$ is the static energy of the radially symmetric 2-vortex, $\omega_{\lambda;2,2}^2(0)$ is the angular frequency for the 2-vortex splitting mode (see \cref{sec:Linearisation}) and $\beta(\lambda)$ is a $\lambda$-dependent, numerically calculated constant.

For general separation $s,$ we can calculate the static inter-vortex forces by solving the field equations in 2-dimensions at fixed separations, using an arrested Newton flow algorithm to find the minimal energy solutions. We pin the vortices at the desired distance $s \in [0, 10]$, with step $h_s = 0.1$. We hence have the static interaction per vortex as
\begin{equation}
    E_{\text{Static}}(s) = \frac{1}{2} \left( V_2^{\lambda}(s) - 2 V_1^{\lambda}(0) \right),
    \label{eq:Estatic}
\end{equation}
where $V_1^{\lambda}(0)$ is the static energy of the $N=1$ vortex, and $V_2^{\lambda}(s)$ is the static energy of the minimised $N=2$ solution, where vortices are positioned 
at $\pm \frac{s}{2}.$

To understand how normal mode excitations modify the inter-vortex interaction, we perform dynamical simulations of a vortex pair initialised with a small velocity and a small perturbation to the shape mode. As the vortices move, we track both their separation and the instantaneous angular frequency of the mode oscillation. We calculate the angular frequency
\begin{equation}
    \omega(t) = \frac{\pi}{\Delta t},
    \label{eq:omega}
\end{equation}
where $\Delta t$ is the time difference between two consecutive peaks or troughs in the static energy as a function of time $t$. By simultaneously tracking the vortex positions, we obtain $\omega^2$ as a function of the separation $s.$
The results are displayed in \cref{fig:l0.9_sepFreq,fig:l0.5_sepFreq,fig:l11_sepFreq,fig:l2_sepFreq}.Assuming the $k=0$ shape mode is the only excited shape mode, the mode interaction energy is calculated from the numerically determined frequency, such that
\begin{equation}
    E_{\text{Mode}}(s) = \frac{1}{2} \epsilon^2 \left( \omega^2(s) - \max_{a \in \mathbb{R}} \omega^2(a) \right),
    \label{eq:Emode}
\end{equation}
where $\epsilon = \sqrt{\frac{I(0)}{\omega_{1,0}^2}}$, with $I(0)$ being the intensity of the excitation, and the energy is normalised such that it is asymptotically zero. We therefore have the total interaction energy per vortex, 
$E_{\text{Int}}(s) = E_{\text{Static}}(s) + E_{\text{Mode}}(s)$.

\section{Numerical Methods}
\label{sec:InitialConfiguration}

We now consider numerical solutions to the time dependent equations of motion \cref{eq:eom1,eq:eom2}. Our initial conditions will take the form of well separated vortices of various degree with excited normal modes. We first construct a single excited vortex which is a solution of the linearised equation of motion in \cref{eq:eig}. Throughout, we assume that the $k=0$ shape mode is the only shape mode excited per vortex. Hence, we write the initial configuration as,
\begin{align}
    \phi_1(t, x) &= \mathcal{R}((x_1 + i x_2)^N) F(x_1^2 + x_2^2) + \epsilon \psi_1(x) \cos{(\omega t - \sigma(0))}, \notag \\
    \phi_2(t, x) &= \mathcal{I}((x_1 + i x_2)^N) F(x_1^2 + x_2^2) + \epsilon \psi_2(x) \cos{(\omega t - \sigma(0))}, \notag \\
    A_{\mu}(t, x) &= \begin{pmatrix}
        0 \\
        -x_2 G(x_1^2 + x_2^2) + \epsilon \chi_1(x) \cos{(\omega t - \sigma(0))} \\
        x_1 G(x_1^2 + x_2^2) + \epsilon \chi_2(x) \cos{(\omega t - \sigma(0))}
    \end{pmatrix},
    \label{eq:StationaryConfig}
\end{align}
where $\sigma(0)$ is the initial phase of the mode, $\psi_i$, $\chi_i$ are perturbations, $F$ and $G$ are solutions of \cref{eq:eom_power1,eq:eom_power2}, $\omega$ is the angular frequency, and $\epsilon$ is the magnitude of the perturbation.

We set the velocity of the excited vortex by performing a Lorenz boost of this normal excitation. This leads to the configuration,
\begin{align}
    \tilde{\phi_1}(t, x) &= \mathcal{R}((\gamma (x_1 + v t) + i x_2)^N) F(\gamma^2 (x_1 + v t)^2 + x_2^2) + \epsilon \psi_1(\tilde{x}) \cos{(\omega \gamma (t + v x_1) - \sigma(0))}, \notag \\
    \tilde{\phi_2}(t, x) &= \mathcal{I}((\gamma (x_1 + v t) + i x_2)^N) F(\gamma^2 (x_1 + v t)^2 + x_2^2) + \epsilon \psi_2(\tilde{x}) \cos{(\omega \gamma (t + v x_1) - \sigma(0))}, \notag \\
    \tilde{A_{\mu}}(t, x) &= \begin{pmatrix}
        -\gamma v x_2 G(\gamma^2 (x_1 + v t)^2 + x_2^2) + \gamma v \epsilon \chi_1(\tilde{x}) \cos{(\omega \gamma (t + v x_1) - \sigma(0))} \\
        -\gamma x_2 G(\gamma^2 (x_1 + v t)^2 + x_2^2) + \gamma \epsilon \chi_1(\tilde{x}) \cos{(\omega \gamma (t + v x_1) - \sigma(0))} \\
        \gamma (x_1 + v t) G(\gamma^2 (x_1 + v t)^2 + x_2^2) + \epsilon \chi_2(\tilde{x}) \cos{(\omega \gamma (t + v x_1) - \sigma(0))}
    \end{pmatrix},
    \label{Ansatz2}
\end{align}
where $\tilde{x} = (\gamma (x_1 + v t), x_2)$ is the Lorentz-boosted spatial coordinate, and $\gamma = (1 - v^2)^{-1/2}$ is the Lorentz factor. For large $\epsilon$, non-linear terms in \cref{eq:action} become significant, and the energy varies up to $\mathcal{O}(\epsilon^3)$ for a $\pi$-shift in phase.

To study vortex scattering we must consider excited well separated vortices, such that we can use the Abikrosov ansatz \cite{abrikosov1957magnetic} to construct field configurations for well-separated, Lorentz-boosted vortices with excited shape modes,
\begin{align}
    \hat{\phi} &= \prod_i \tilde{\phi}(x - d_i), &
    \hat{A_{\mu}} &= \sum_i \tilde{A_{\mu}}(x - d_i),
    \label{AbikEq}
\end{align}
where $d_i$ are the vortex zero positions, and the approximation holds when all separations are much larger than the vortex core size $\max(1/\sqrt{\lambda},1)$.

For our dynamical simulations, we use a lattice of $601 \times 601$ points, with spacing $h = 0.1$. We use a $6^{\rm th}$-order symplectic time evolution, and $4^{\rm th}$-order finite difference scheme for spatial derivatives. We also employ natural boundary conditions \cite{TomANF2} so that the lattice can be updated on the boundary. Damping boundary conditions are imposed to absorb radiation, such that it is not reflected back into the system. For a detailed discussion of the numerics, see \cite{SKMRTW}.

\section{Critical Coupling\label{sec:critcal_couling}}
Recent work has examined vortex dynamics with excited normal modes at critical coupling $(\lambda = 1)$ \cite{AMRW,SKMRTW}. As shown in \cref{fig:N1Spectra}, there is a single normal mode for $N=1$ ($k=0$) and two for $N=2$ ($k=0,1$) at $\lambda = 1$.

\begin{figure}
    \centering
    \includegraphics[width=0.6\linewidth]{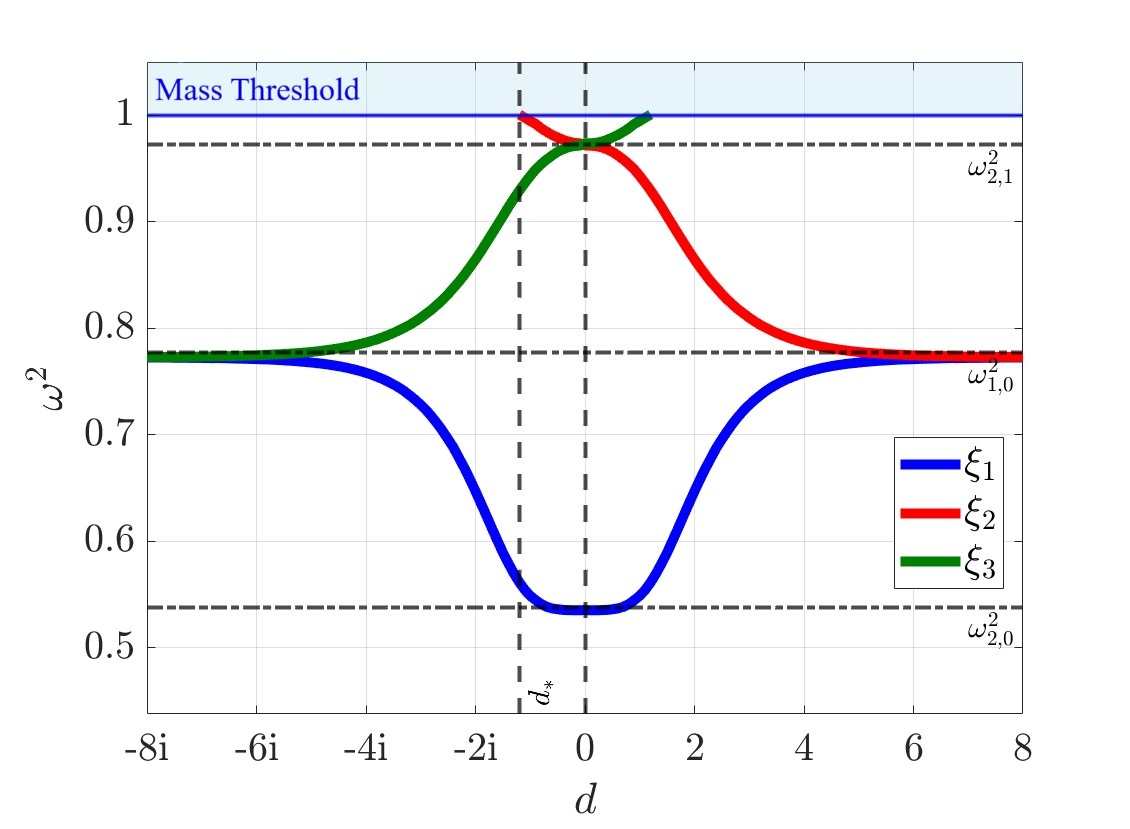}
    \caption{The angular frequency of the normal modes for a 2-vortex configuration at critical coupling $\lambda = 1$, as a function of the half-separation $d$. The three branches correspond to the in-phase excitation $\xi_1$, out-of-phase excitation $\xi_2$, and a third branch $\xi_3$ which emerges from the continuous spectrum. As $d \to 0$, the $\xi_1$ and $\xi_2$ branches tend towards the $k=0$ and $k=1$ modes of the coincident $N=2$ vortex respectively (see \cref{fig:N1Spectra}). The shaded region indicates the continuous spectrum, and the point where $\xi_2$ and $\xi_3$ meet the continuous spectrum defines the spectral wall at $d = d_*$.}
\label{fig:SpectralFlowN1}
\end{figure}

For a head-on two-vortex collision, the initial excitation decomposes into in-phase ($\xi_1$) and out-of-phase ($\xi_2$) modes. As the vortices approach, these frequencies split,
tending towards the $k=0$ and $k=1$ modes of the coincident $N=2$ vortex. Beyond coincidence, $\xi_1$ repeats with right-angle separation, whilst $\xi_2$ pairs with a third branch $\xi_3$ tending towards the mass threshold, introducing a spectral wall \cite{AMRW,alonso2026bps}. When the vortices are well separated, there are two degenerate bound modes: the in-phase superposition $\xi_1$ and the out-of-phase superposition $\xi_2$. Denoting the $k=0$ shape mode of the $i^{\rm th}$ single $N=1$ vortex as $|i\rangle$, these are defined as
\begin{equation}
    \xi_1 = \frac{I(0)}{2} \left(|1\rangle + |2\rangle \right),
    \label{eq:N2config1}
\end{equation}
\begin{equation}
    \xi_2 = \frac{I(0)}{2} \left(|1\rangle - |2\rangle \right),
    \label{eq:N2config2}
\end{equation}
where 
$I(0) = \frac{1}{2}(\epsilon\omega)^2$ 
is the initial intensity of the excitation.
 The resulting spectral structure is plotted in \cref{fig:SpectralFlowN1}, where notably the number of bound modes varies with separation. This admits a low-energy scattering description on $\mathbb{C}^3$, consistent with other soliton models \cite{halcrow2021consistent}.

\begin{figure}
    \centering
    \includegraphics[width=0.6\linewidth]{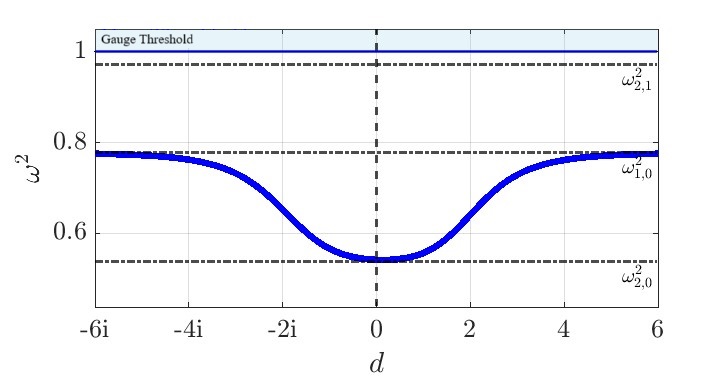}
    \caption{Flow of the angular frequency $\omega^2$ as a function of vortex distance from the origin $d$, where $s = 2d$. Initial velocity $v_{\rm in} = 0.01$, initial intensity $I(0) = 0.01$, where $I(0) = \frac{1}{2}(\epsilon\omega)^2$.}
    \label{fig:l1_sepFreq}
\end{figure}

In \cite{SKMRTW}, we studied scattering along the $\xi_1$ branch \cref{eq:N2config1}, finding that the decreasing frequency signals an attractive intervortex force. This is confirmed in \cref{fig:l1_sepFreq}: as the vortices scatter, energy transfers from the mode to the kinetic degrees of freedom, with $\omega^2$ dropping with separation. The scattering behaviour depends sensitively on initial velocity, intensity, and phase of the shape mode, including the formation of quasi-bound states.
Reference \cite{AMRW} followed instead
the upper two modes \cref{eq:N2config2}. The increasing frequency along $\xi_2$ induces a repulsive force; however, at a specific excitation intensity, $\xi_2$ transitions into $\xi_3$ and the vortices reach $d_*$, where the mode frequency enters the continuous spectrum.

The remainder of this paper examines how this behaviour changes for $\lambda \neq 1$, as the splitting mode ($k=2$) for $N=2$ becomes energetic rather than a zero mode and static intervortex forces emerge.

\section{Type \texorpdfstring{\rom{1}}{I} Dynamics}
\label{sec:TypeI}
For $\lambda < 1$ (type \rom{1}) vortices exhibit attractive static forces.

We expect the same phenomena to occur here, except that because of the static force, it is unlikely that the vortices will become trapped in the spectral wall. Instead, we might observe the existence of a spectral wall by a change in velocity of vortices as the frequency enters the continuum.
We have many regions of interest to consider. For the case of type \rom{1} vortices, the excited scattering is not symmetric from $x_1$ to $x_2$, noting that our initial configuration places the vortices on the $x_1$-axis.

\begin{figure}
\centering
\includegraphics[width=0.65\linewidth]{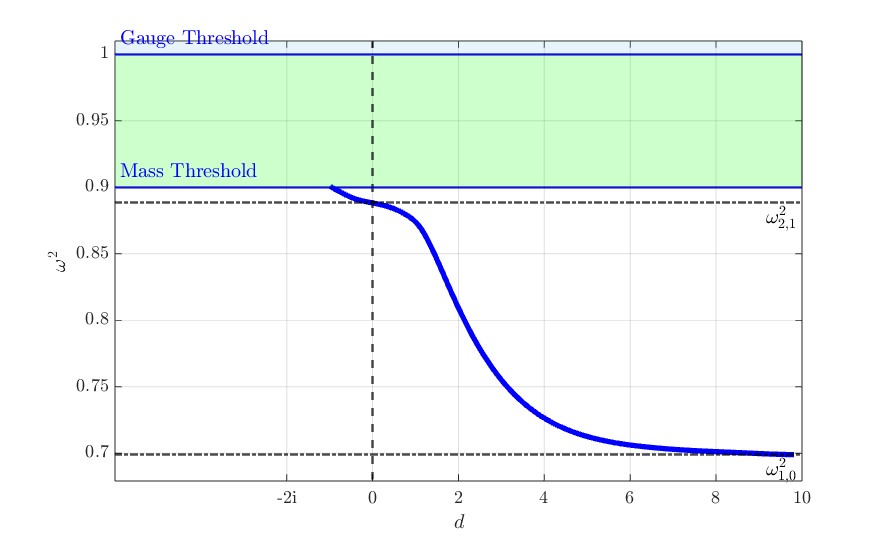}
\caption{Angular frequency flow for a two-vortex system ($\lambda = 0.9$) as a function of vortex separation $d$ from the origin (blue). The green area indicates the gauge threshold, and the blue area the mass threshold. A $d^4$ approximation is shown in dashed red.}
\label{fig:l0.9_sepFreq}
\end{figure}

\begin{figure}
\centering
\includegraphics[width=0.7\linewidth]{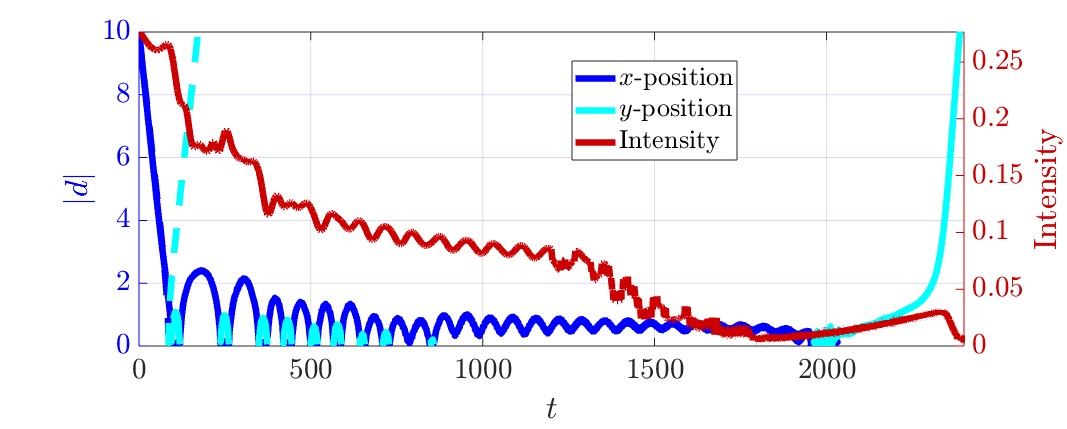}
\caption{Trajectories of a two-vortex system ($\lambda = 0.9$, $v_{\rm in} = 0.1$, $I(0) = 0.3$) as a function of time, where $I(0) = \frac{1}{2}(\epsilon \omega)^2$. The blue line shows the $x$-direction distance from the origin, cyan the $y$-direction, and red the excitation intensity per vortex. Dashed blue and cyan lines show unexcited scattering with the same parameters.}
\label{fig:l0.9_trajectories1}
\end{figure}

\begin{figure}
\centering
\includegraphics[width=0.7\linewidth]{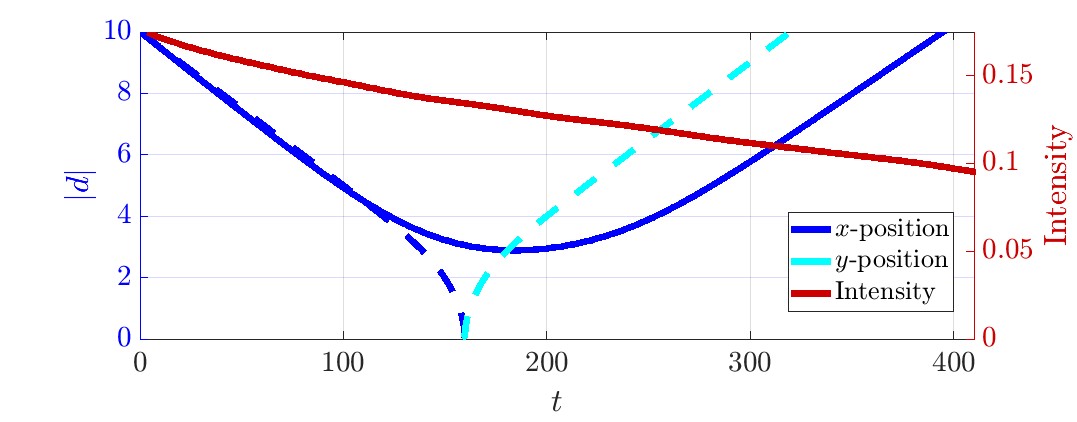}
\caption{Trajectories of a two-vortex system ($\lambda = 0.9$, $v_{\rm in} = 0.05$, $I(0) = 0.3$) as a function of time, where $I(0) = \frac{1}{2}(\epsilon \omega)^2$. The blue line shows the $x$-direction distance from the origin, red the excitation intensity per vortex, and dashed blue and cyan lines show unexcited scattering in the $x_1$ and $x_2$-directions, respectively.}
\label{fig:l0.9_trajectories2}
\end{figure}

It can be seen in \cref{fig:l0.9_sepFreq} that as the vortices move closer to the origin, the frequency increases. For positive $d$, the frequency interpolates between the asymptotic values $\omega_{1,0}^2 = 0.7136541$, which is the $N=1$, $k = 0$ shape mode, and the $N=2$, $k=1$ mode at the coincident configuration, $\omega_{2,1}^2 = 0.8883169$. We see that after the vortices pass through the coincident configuration, the frequency increases further, hitting the continuous spectrum at $d\approx -1i$. This would suggest the presence of a spectral wall. To confirm this, we can observe the trajectories of the vortices in a scattering solution, see \cref{fig:l0.9_trajectories1,fig:l0.9_trajectories2}.

We observe two interesting phenomena when observing the trajectories in \cref{fig:l0.9_trajectories1}. Note that blue is the position in $x_1$, and cyan is the position in $x_2$. Firstly, from the blue line we can see that the vortices initially travel at a near-constant velocity. After the vortices scatter, we can see from the cyan line that the vortices hit the spectral wall and bounce back. This can be seen in the difference in the bounce size between the blue and cyan bounces. The distance in which the vortices bounce back in $x_2$ is the same as the distance where the frequency hits the continuous spectrum, which confirms the presence of a spectral wall.

Additionally, we can see that after the vortices have bounced a number of times, the vortices settle in the $x_1$-plane at a fixed distance. This is not a result of the spectral wall, instead it is due to the net force being zero. We notice that the intensity of the excitation, displayed as the red line in \cref{fig:l0.9_trajectories1}, decreases overall. This is due to a decay in the excitation as it radiates energy. We observe temporary dips in the intensity as the vortices change direction, as seen at the peaks of the bounces in the $x_1$ direction. This is due to the energy transfer mechanism, where the energy from the excitation is transferred to the kinetic energy. In addition, we see drops in intensity at the peaks of the bounces in the $x_2$-direction. This is because energy is transferred from the excitation to the spectral wall, and the resulting effect is that the vortices bounce off the spectral wall.

\begin{figure}
\centering
\begin{minipage}{0.24\textwidth}
\centering
\includegraphics[width=\linewidth]{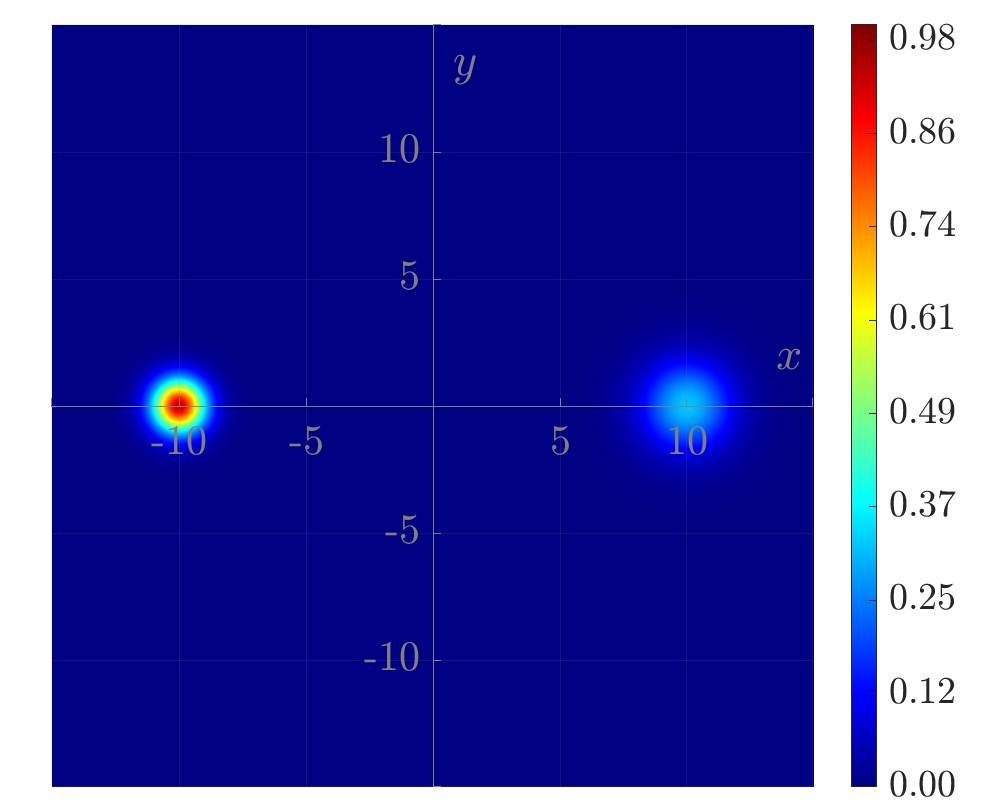}
\caption*{$t = 0$}
\end{minipage}
\begin{minipage}{0.24\textwidth}
\centering
\includegraphics[width=\linewidth]{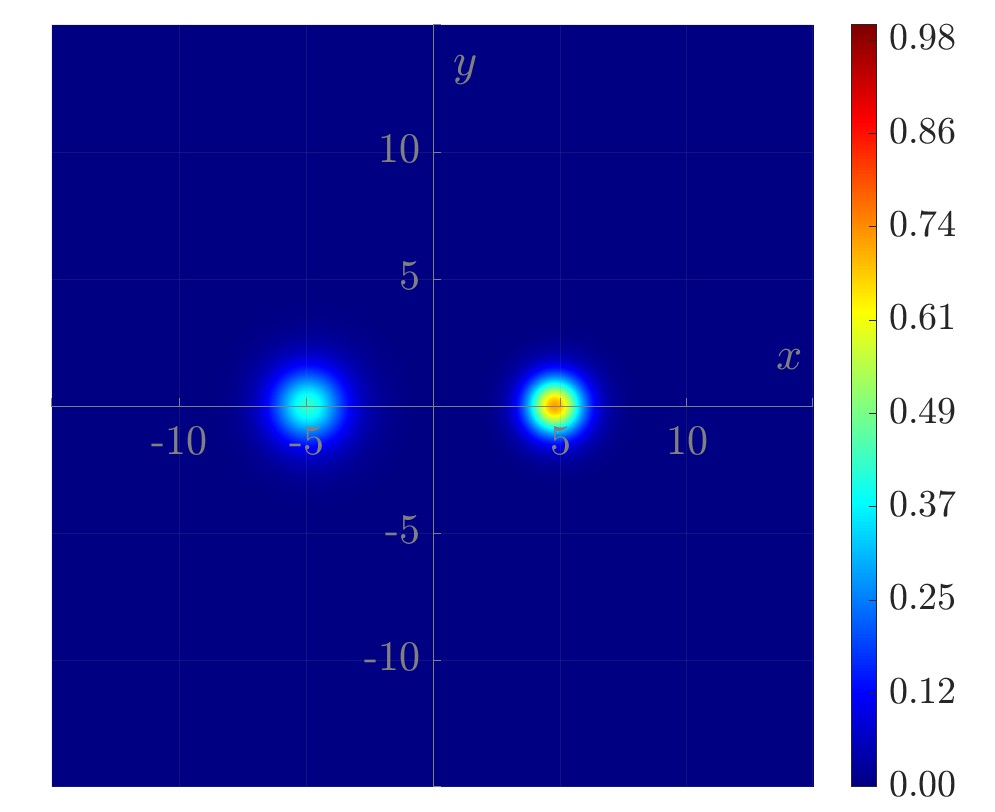}
\caption*{$t = 50$}
\end{minipage}
\begin{minipage}{0.24\textwidth}
\centering
\includegraphics[width=\linewidth]{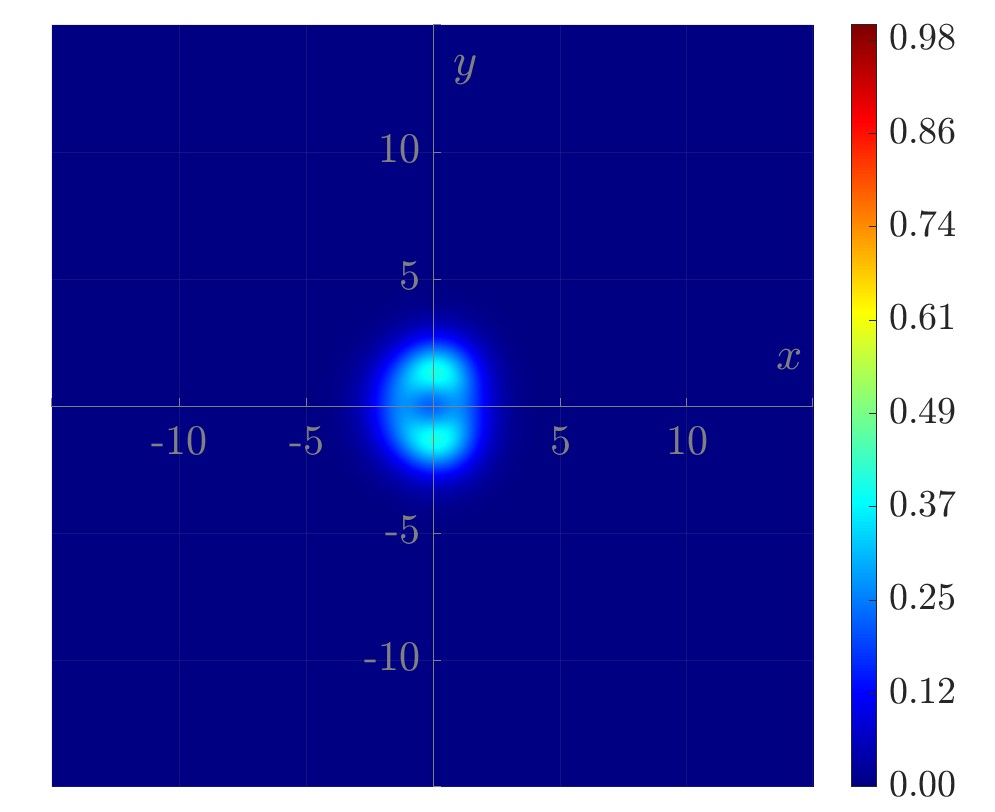}
\caption*{$t = 100$}
\end{minipage}
\begin{minipage}{0.24\textwidth}
\centering
\includegraphics[width=\linewidth]{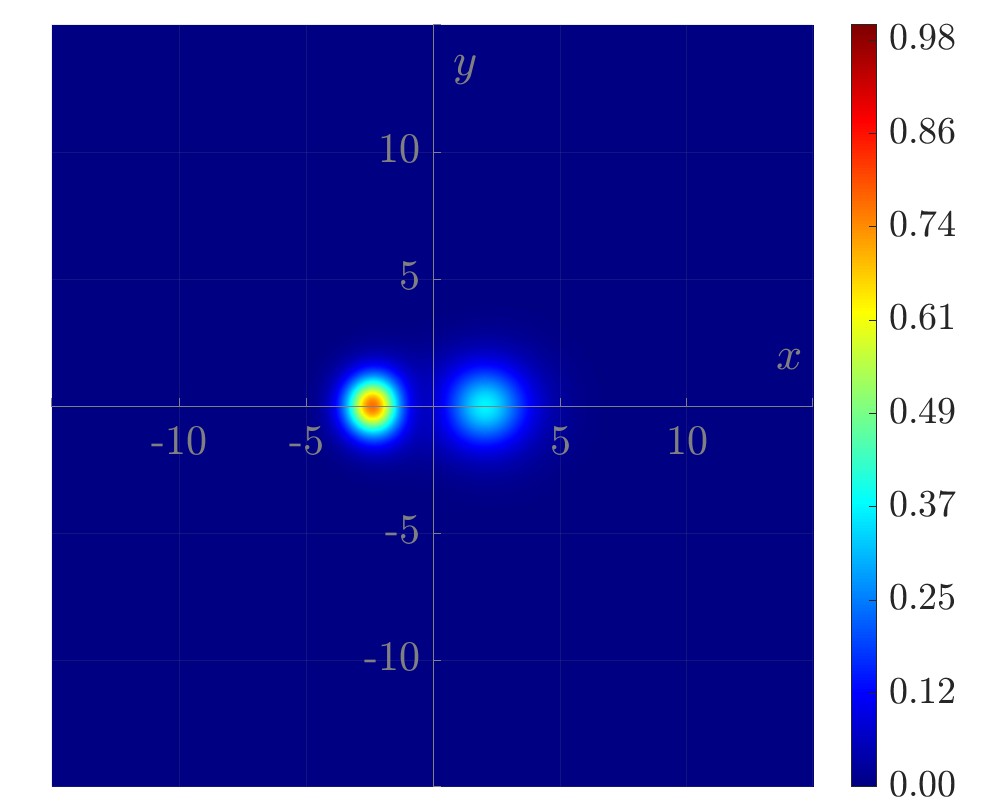}
\caption*{$t = 150$}
\end{minipage}
\begin{minipage}{0.24\textwidth}
\centering
\includegraphics[width=\linewidth]{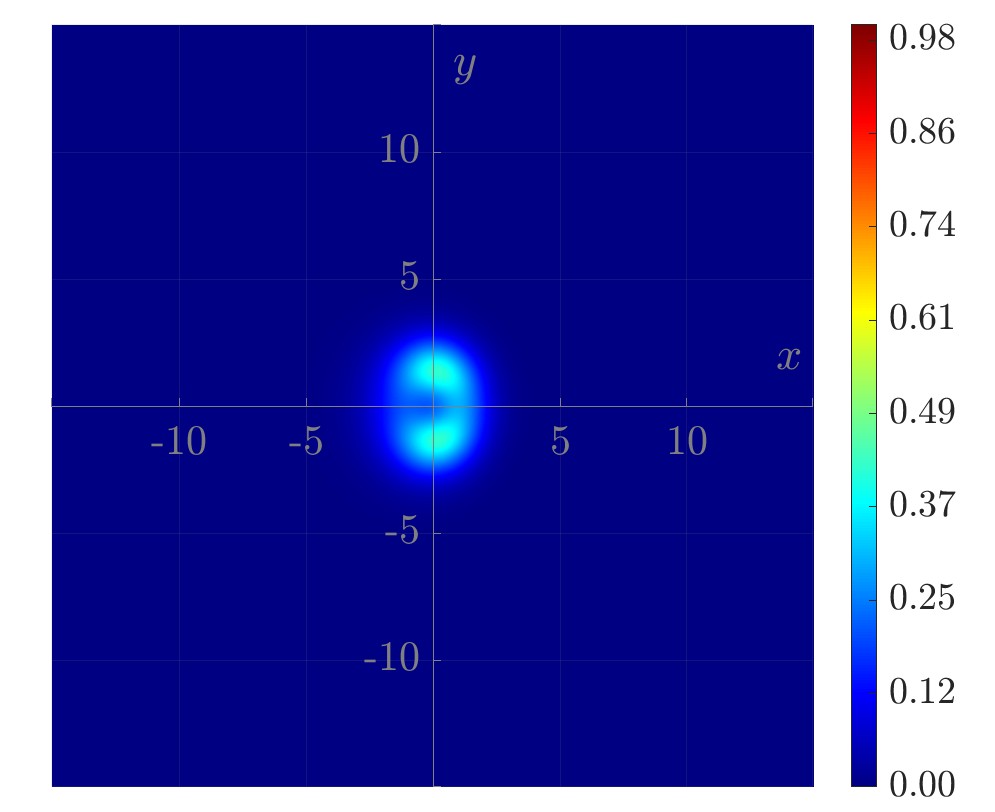}
\caption*{$t = 250$}
\end{minipage}
\begin{minipage}{0.24\textwidth}
\centering
\includegraphics[width=\linewidth]{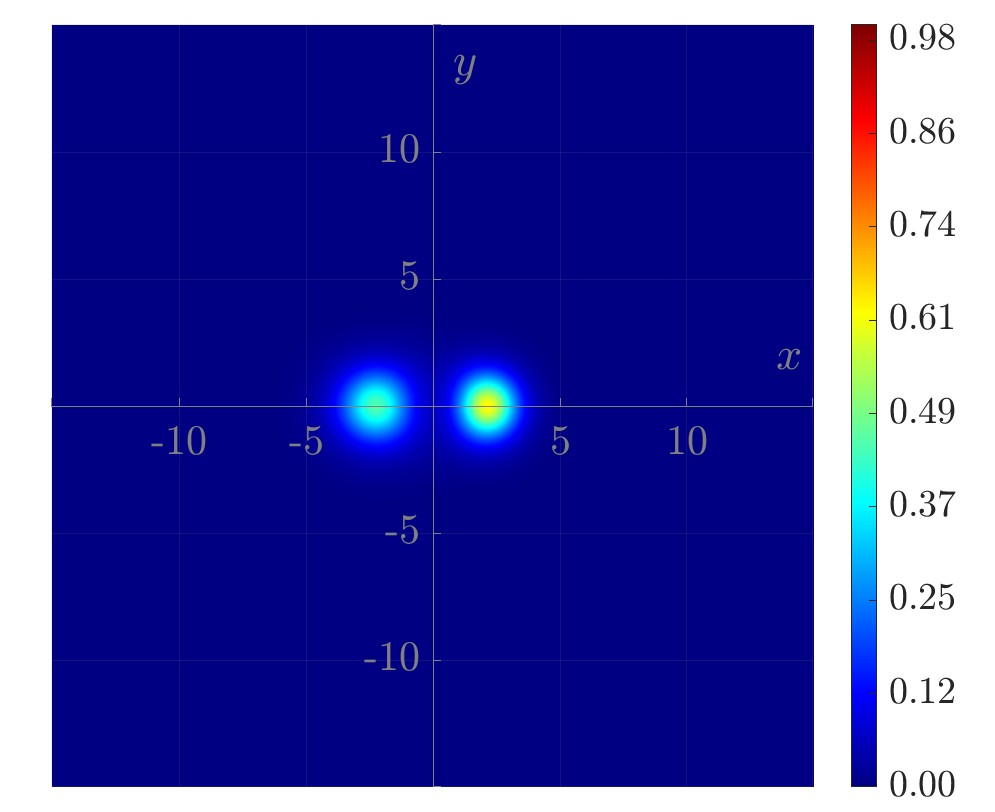}
\caption*{$t = 300$}
\end{minipage}
\begin{minipage}{0.24\textwidth}
\centering
\includegraphics[width=\linewidth]{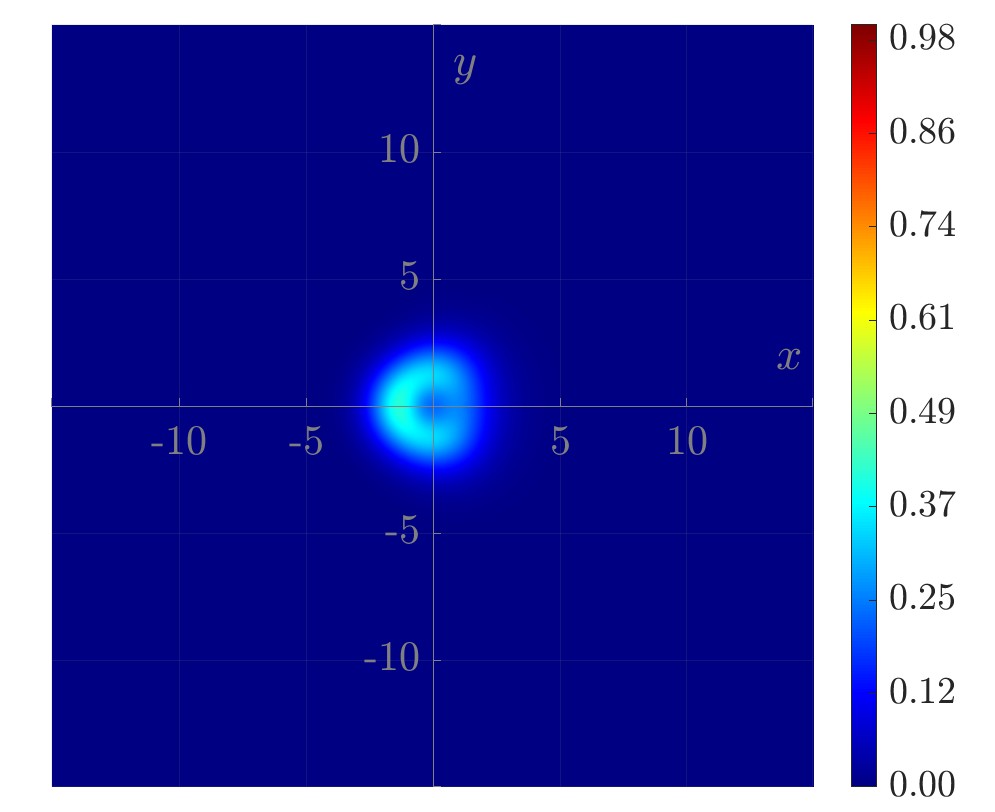}
\caption*{$t = 350$}
\end{minipage}
\begin{minipage}{0.24\textwidth}
\centering
\includegraphics[width=\linewidth]{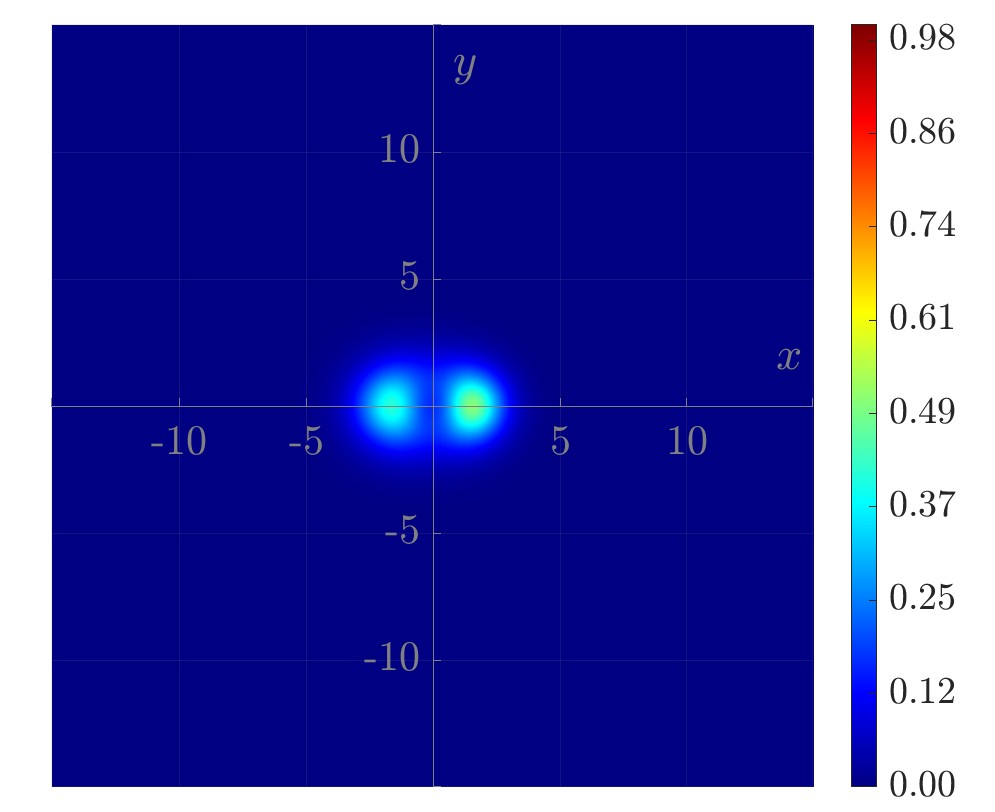}
\caption*{$t = 400$}
\end{minipage}
\begin{minipage}{0.24\textwidth}
\centering
\includegraphics[width=\linewidth]{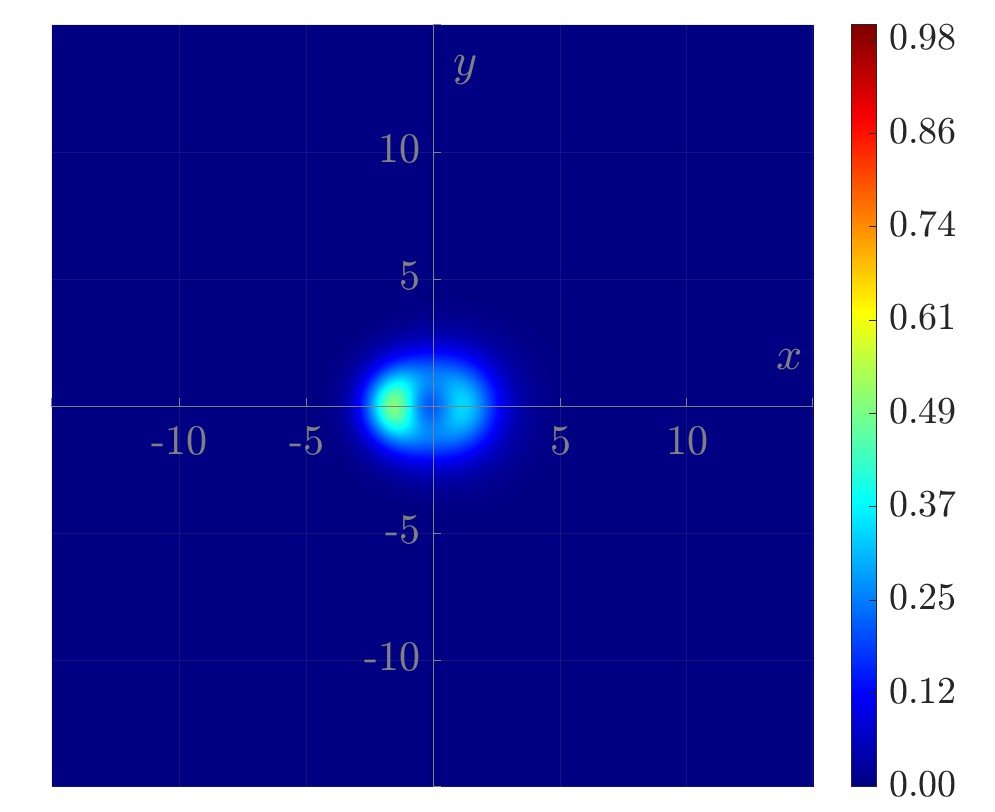}
\caption*{$t = 600$}
\end{minipage}
\begin{minipage}{0.24\textwidth}
\centering
\includegraphics[width=\linewidth]{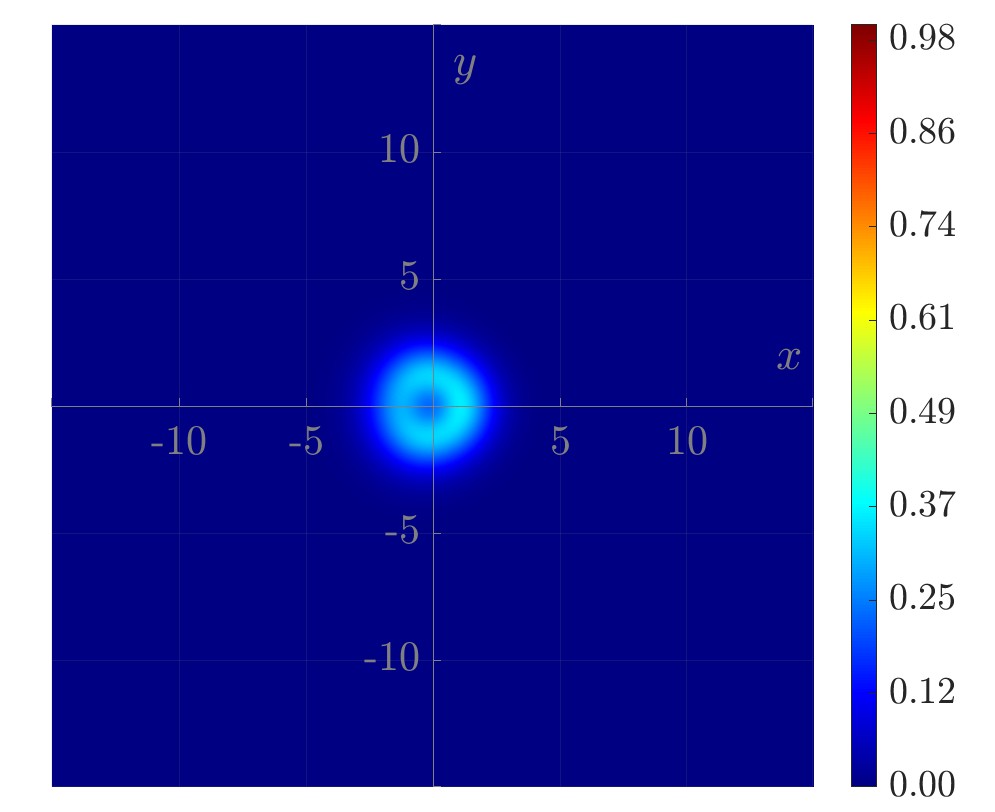}
\caption*{$t = 850$}
\end{minipage}
\begin{minipage}{0.24\textwidth}
\centering
\includegraphics[width=\linewidth]{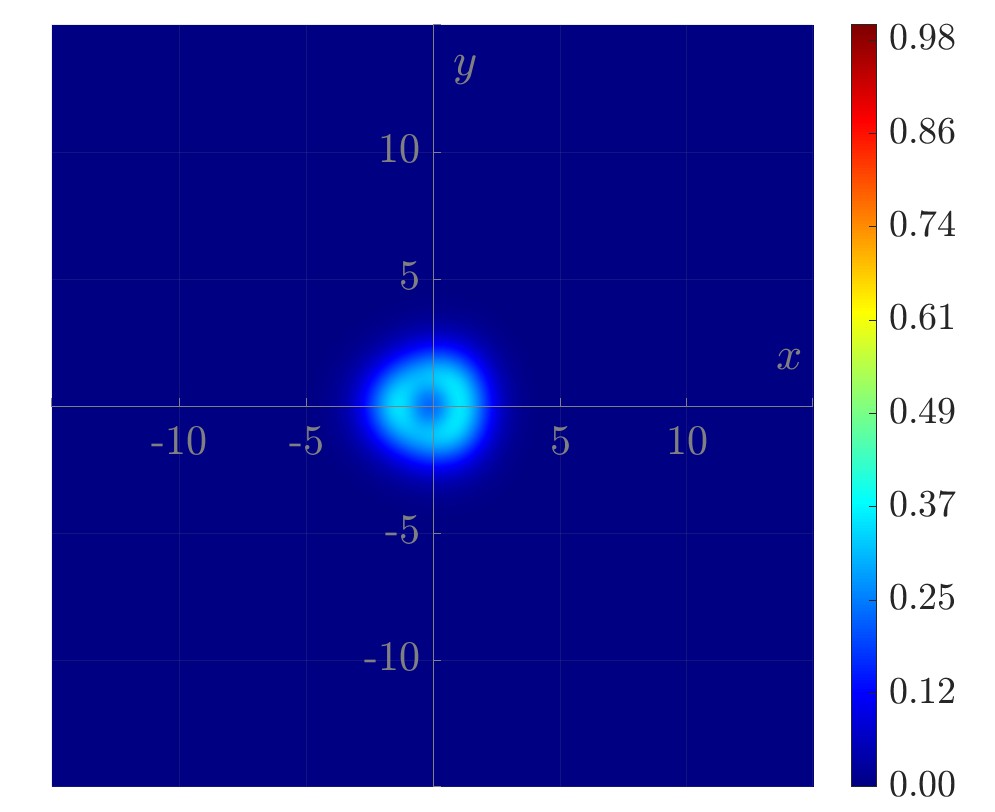}
\caption*{$t = 1000$}
\end{minipage}
\begin{minipage}{0.24\textwidth}
\centering
\includegraphics[width=\linewidth]{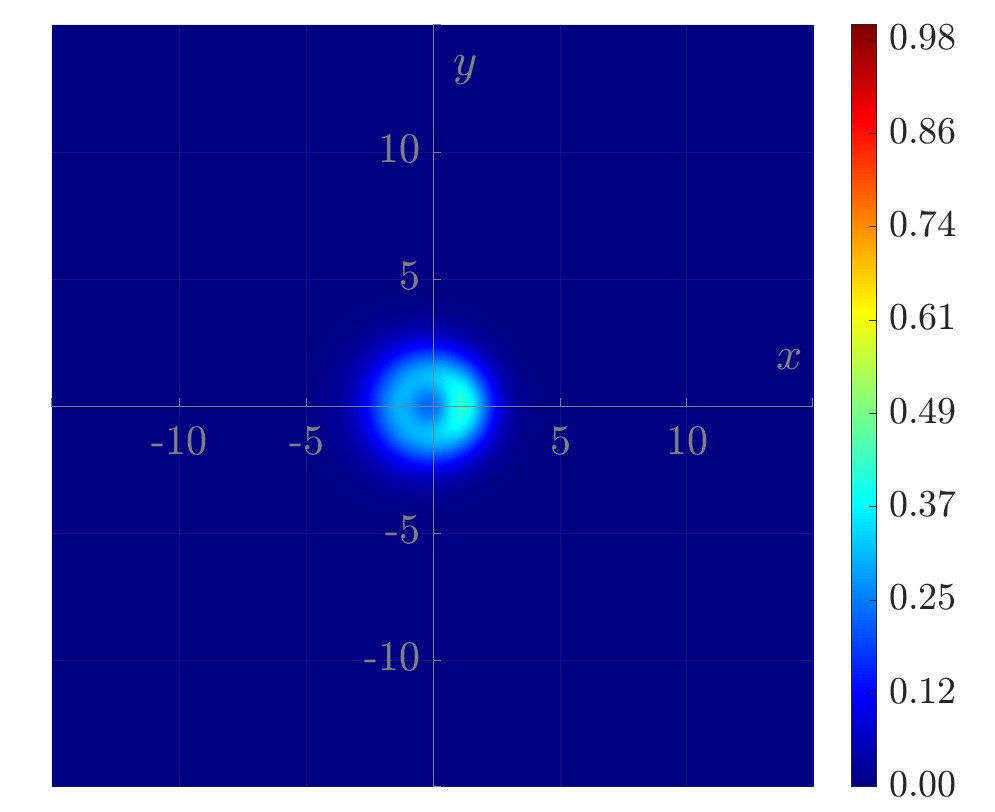}
\caption*{$t = 1400$}
\end{minipage}
\begin{minipage}{0.24\textwidth}
\centering
\includegraphics[width=\linewidth]{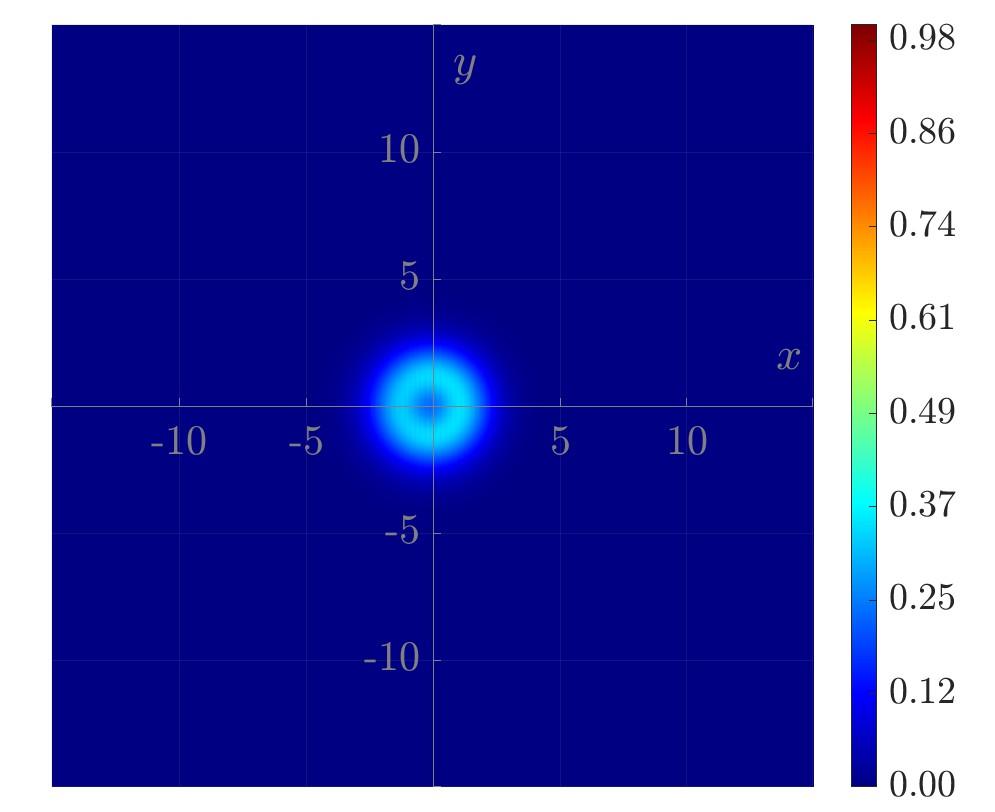}
\caption*{$t = 2000$}
\end{minipage}
\begin{minipage}{0.24\textwidth}
\centering
\includegraphics[width=\linewidth]{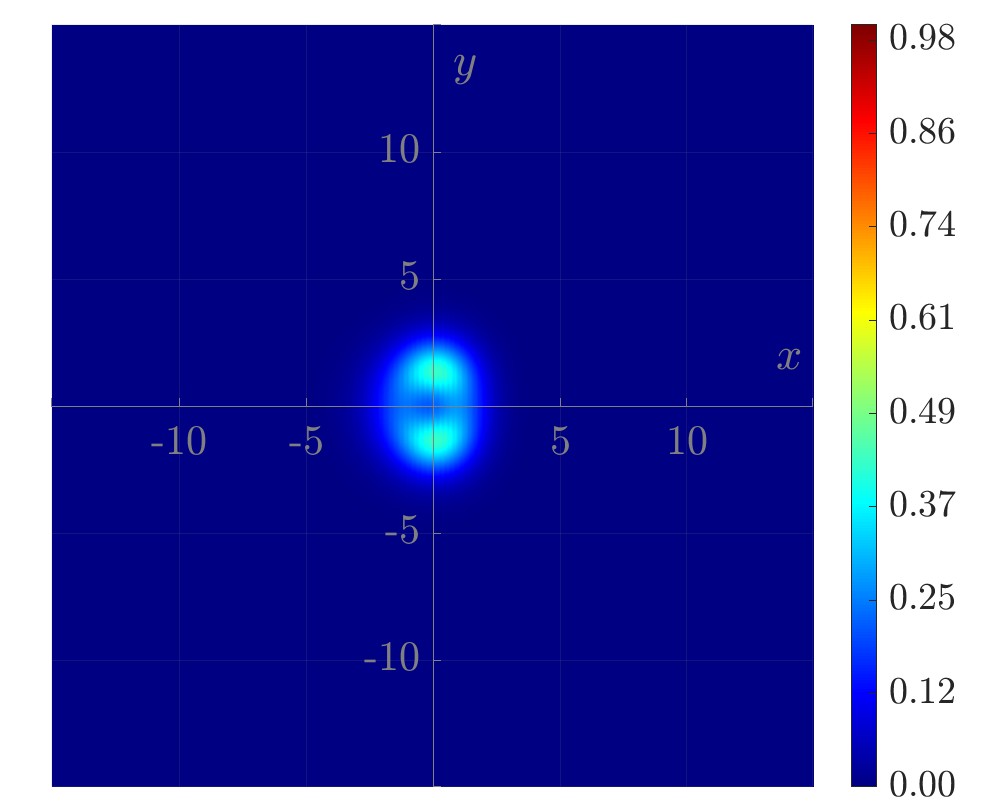}
\caption*{$t = 2200$}
\end{minipage}
\begin{minipage}{0.24\textwidth}
\centering
\includegraphics[width=\linewidth]{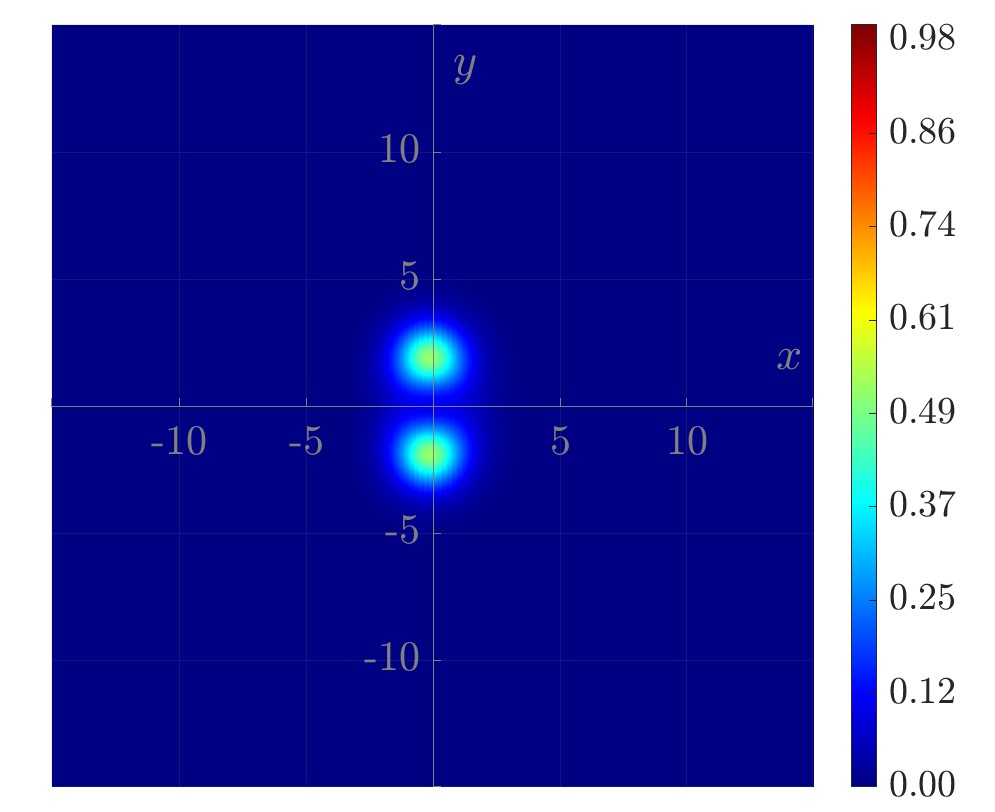}
\caption*{$t = 2300$}
\end{minipage}
\begin{minipage}{0.24\textwidth}
\centering
\includegraphics[width=\linewidth]{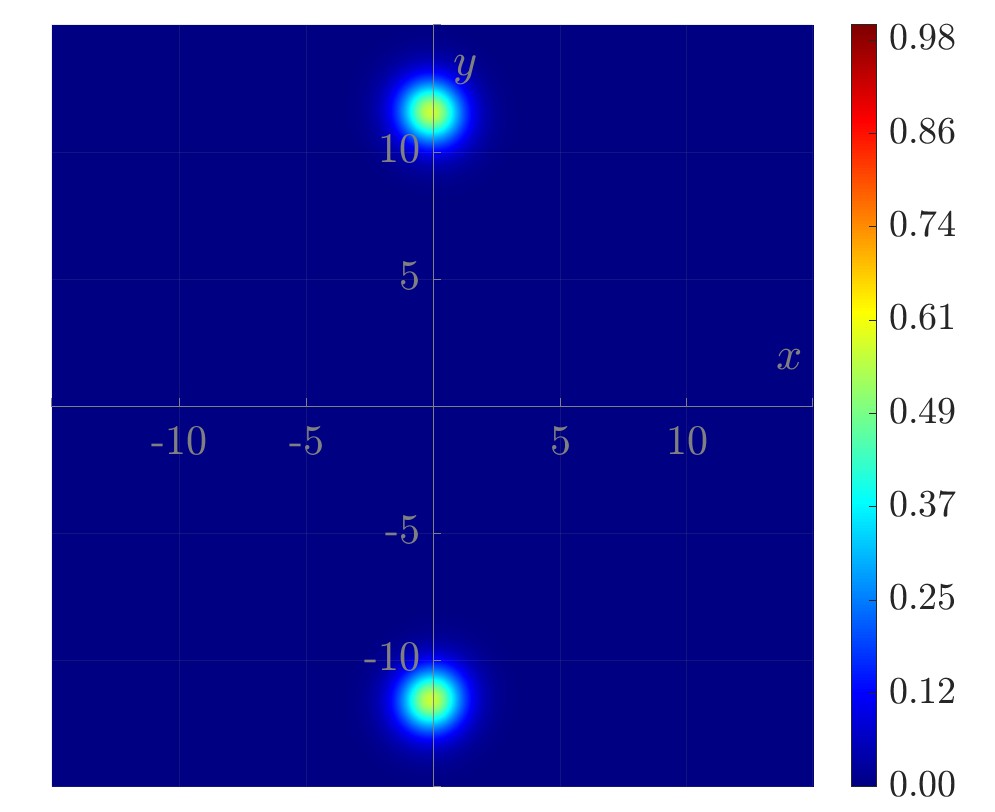}
\caption*{$t = 2400$}
\end{minipage}
\caption{Snapshots of energy density for a two-vortex scattering ($\lambda = 0.9$, $v_{\rm in} = 0.1$, $I(0) = 0.3$), with out-of-phase $k=0$ shape modes, as seen in \cref{fig:l0.9_trajectories1}.}
\label{fig:l0.9_snapshot}
\end{figure}

We can explore some dynamical snapshots of \cref{fig:l0.9_trajectories1} displaying a heat plot of the energy density. The simulation shown in \cref{fig:l0.9_snapshot} displays the scattering of two 1-vortices with $\lambda = 0.9$. We have excited the $k = 0$ shape mode on each vortex, with a relative phase of $\pi$ to induce a repulsive force. We have chosen an initial intensity of $I(0) = 0.3$, where $I(0) = \frac{1}{2}(\epsilon \omega)^2$, and initial velocity $v_{\rm in} = 0.1$.

\Cref{fig:l0.9_snapshot} shows how the vortices scatter multiple times, which is expected in the type \rom{1} regime. It can also be seen that the vortices initially start out of phase, as seen by the difference in the peaks of the energy density. After the vortices scatter, the peaks are equivalent, highlighting the asymmetry in the scattering process.

\Cref{fig:l0.9_trajectories2} shows the scattering of a two-vortex system at $\lambda = 0.9$, whereby the intensity of the excitation is large enough such that the repulsion from the mode dominates the scattering process. As such, the vortices never meet and are repelled towards infinity.

\begin{figure}
\centering
\begin{minipage}{0.75\textwidth}
\includegraphics[width=\linewidth]{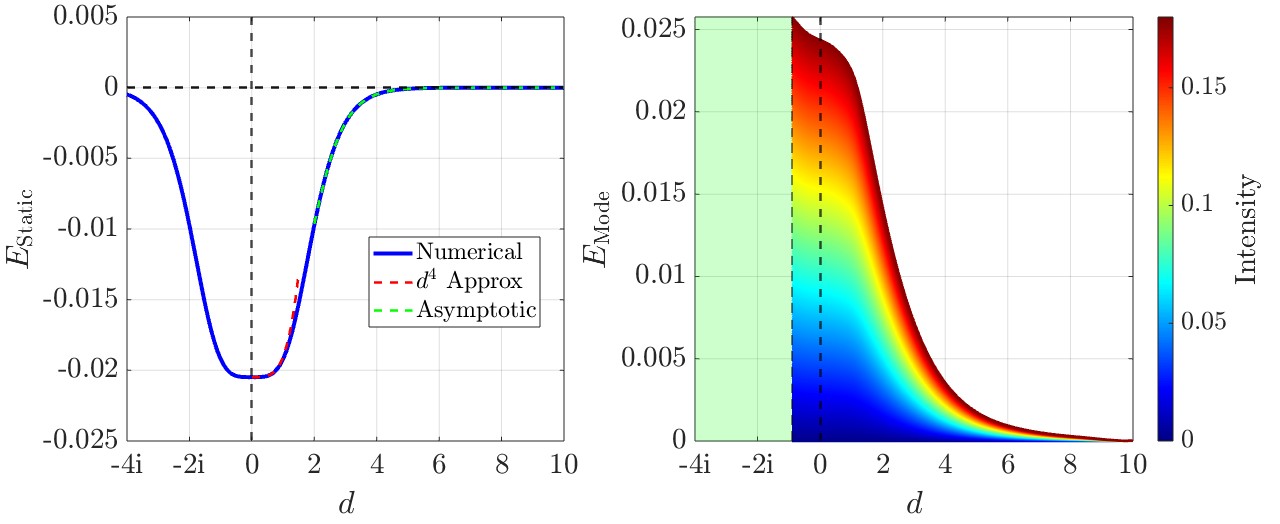}
\end{minipage}
\begin{minipage}{0.75\textwidth}
\includegraphics[width=\linewidth]{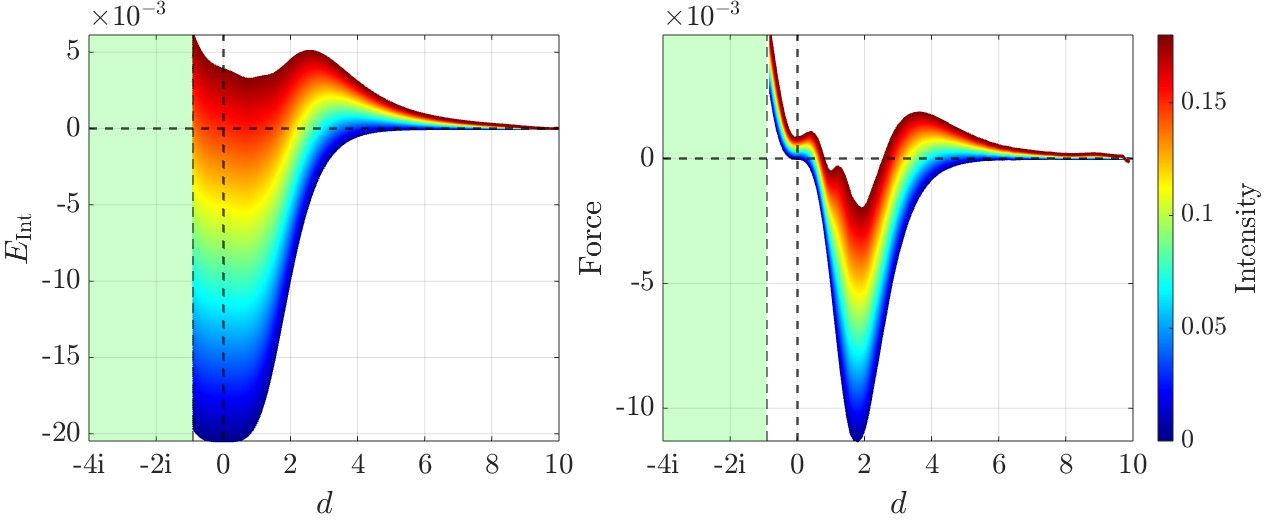}
\end{minipage}
\caption{Interaction energies and forces ($\lambda = 0.9$) as a function of vortex separation $d$. Top left: static force (blue) via \eqref{eq:Estatic}, with $d^4$ (dashed red, \eqref{eq:SRInt}) and asymptotic (dashed green, \eqref{eq:AsyInt}) approximations. Top right: mode interaction energy via \eqref{eq:Emode}. Bottom left: total interaction energy (static plus mode). Bottom right: total force, $F = -\frac{\partial E_{\rm int}}{\partial d}$.}
\label{fig:Int_09}
\end{figure}

We display the static force, seen on the top left of \cref{fig:Int_09}, calculated using \cref{eq:Estatic}, and the force induced by excitation of the shape mode, see the top right of \cref{fig:Int_09}, calculated using \cref{eq:Emode}. We are interested in the total interaction energy, the sum of the static force and the mode interaction, as seen on the bottom left of \cref{fig:Int_09}. 
We also display the total net force of the interaction, calculated as $F = -\frac{\partial E_{\rm int}}{\partial d}$, seen in the bottom right of \cref{fig:Int_09}.

We interpret the interaction energy, seen from \cref{fig:Int_09} as follows. For $|d| \ll 1$, the frequency and static force can both be approximated by $d^4$, therefore, if the coefficient of the interaction of the mode is smaller than that of the static force, then the two vortices will stay coincident at the origin; otherwise, the mode induced force will force them to separate. For $d > 1$, there exists a local extremum in the interaction energy at precisely the distance observed in \cref{fig:l0.9_trajectories1}. This explains the existence of this quasi-stationary state as the net force $F = -\frac{\partial}{\partial d}E_{\rm Int} = 0$. 

For $d < 0$, we see that the interaction energy is highly positive due to the existence of the spectral wall. This makes it extremely difficult for vortices to move past this region. Additionally, for large $d$, we see that if the intensity of the perturbation is too large, then the vortices will never meet and will repel.

\begin{figure}
\centering
\includegraphics[width=0.7\linewidth]{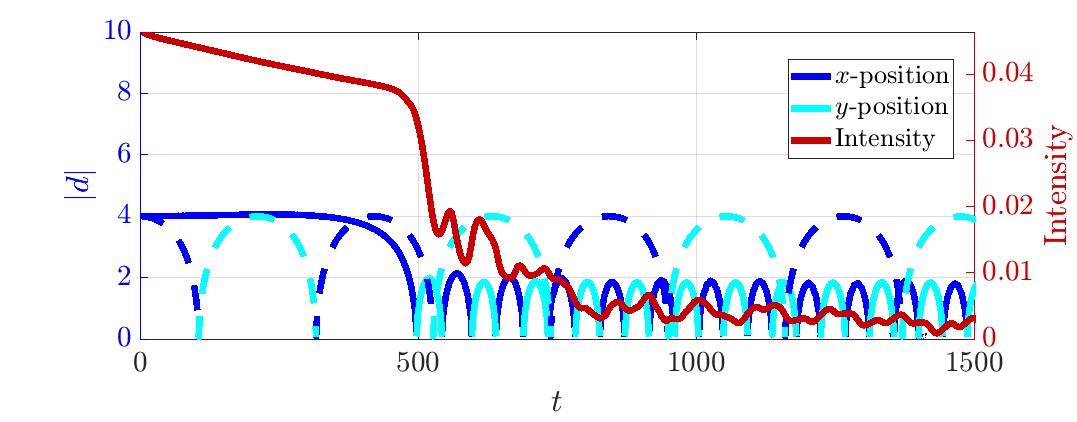}
\caption{Trajectories of a two-vortex system ($\lambda = 0.9$, $v_{\rm in} = 0$, $I(0) = 0.05$) near the local extremum, where $I(0) = \frac{1}{2}(\epsilon \omega)^2$. The blue line shows the $x_1$-direction distance from the origin, cyan the $x_2$-direction, and red the excitation intensity per vortex. Dashed blue and cyan lines show unexcited scattering in the $x_1$ and $x_2$-directions, respectively.}
\label{fig:trajectories_localMax_l09}
\end{figure}

We can further explore the local extremum, by choosing an initial configuration where the vortices are situated near the local extremum; see \cref{fig:trajectories_localMax_l09}.

It can be seen in \cref{fig:trajectories_localMax_l09} that initially the vortices form a quasi-stationary state, where they stay at a fixed position for a period of time. This is because the net force is zero.
If the intensity of the excitation is large, the vortices will repel, as seen in \cref{fig:l0.9_trajectories2}, while if the intensity is small, then the static force will dominate and the vortices will be attracted toward the origin, as seen by the blue line in \cref{fig:trajectories_localMax_l09}. We observe that this indeed happens in \cref{fig:trajectories_localMax_l09} as the intensity indicated by the red line decays, the quasi-stationary state decays and the vortices accelerate towards the origin.

It is important to note that the $k=1$ mode exists only in the discrete spectrum for $\lambda > 0.8$, see \cref{sec:Linearisation}. As such, we must also explore what happens when two vortices scatter along the $x_1$-axis for $\lambda < 0.8$. In this regime, the angular frequency of the vortices will hit the continuous spectrum before the vortices collide. This could suggest the presence of a spectral wall, in which case we should see the vortices slow down in the region where the angular frequency hits the continuum.

\begin{figure}
\centering
\includegraphics[width=0.7\linewidth]{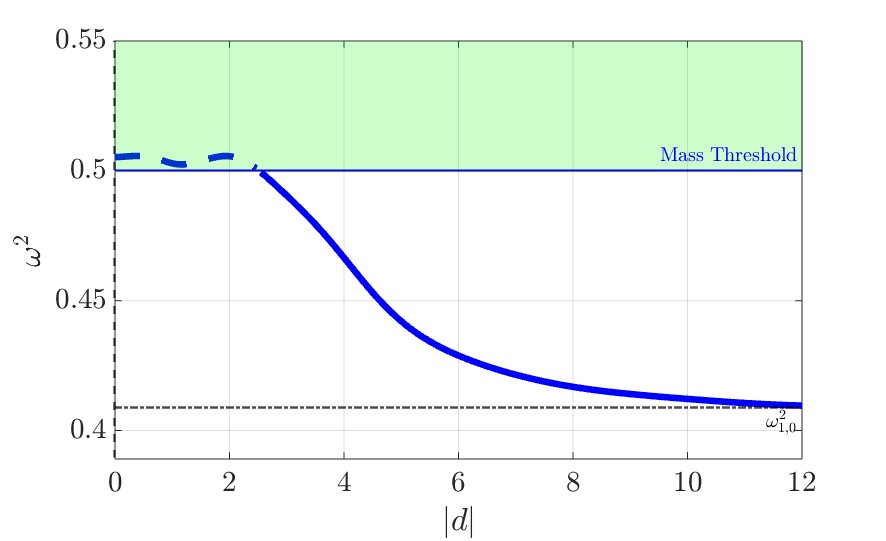}
\caption{Angular frequency flow for a two-vortex system ($\lambda = 0.5$) as a function of vortex separation $d$ from the origin (blue). The green area indicates the gauge threshold.}
\label{fig:l0.5_sepFreq}
\end{figure}

We can plot the angular frequency as a function of the distance from the origin for a 2-vortex system with $\lambda = 0.5$, see \cref{fig:l0.5_sepFreq}.

\begin{figure}
\centering
\begin{minipage}{0.7\textwidth}
\includegraphics[width=\linewidth]{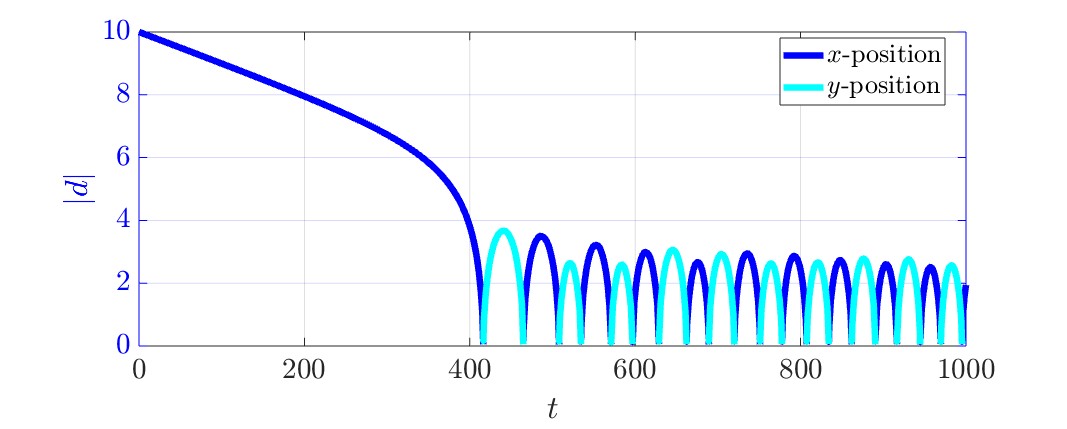}
\end{minipage}
\begin{minipage}{0.7\textwidth}
\includegraphics[width=\linewidth]{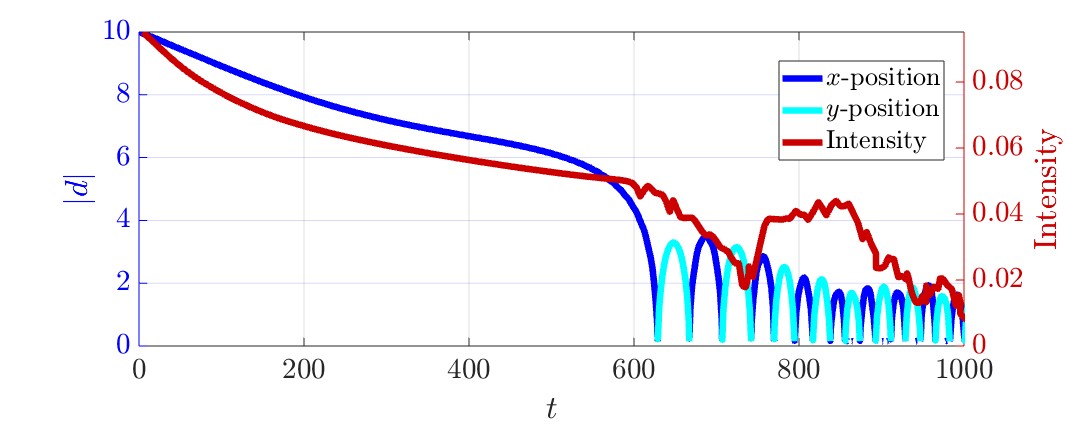}
\end{minipage}
\caption{Trajectories of a two-vortex system ($\lambda = 0.5$, $v_{\rm in} = 0.01$, $I(0) = 0.1$) as a function of time, where $I(0) = \frac{1}{2}(\epsilon \omega)^2$. Top: unexcited scattering. Bottom: excited scattering, with blue ($x_1$-direction), cyan ($x_2$-direction), and red (intensity) lines.}
\label{fig:l0.5_trajectories}
\end{figure}

It can be seen in \cref{fig:l0.5_sepFreq} that as the vortices approach the origin, the frequency increases from the asymptotic value $\omega_{1,0}^2 = 0.4254454$, hitting the continuous spectrum at $d\approx2.5$. To confirm the existence of a spectral wall, we will consider the trajectories of the vortices, see \cref{fig:l0.5_trajectories}.

We can see in \cref{fig:l0.5_trajectories} that the bounces in $x_2$, indicated by the cyan lines, are smooth, which is expected as the frequency of the shape mode is in the continuous spectrum. More interestingly, the bounces in $x_1$, indicated by the blue lines, show some irregularities near the peak of the bounces.

\begin{figure}
\centering
\begin{minipage}{0.75\textwidth}
\includegraphics[width=\linewidth]{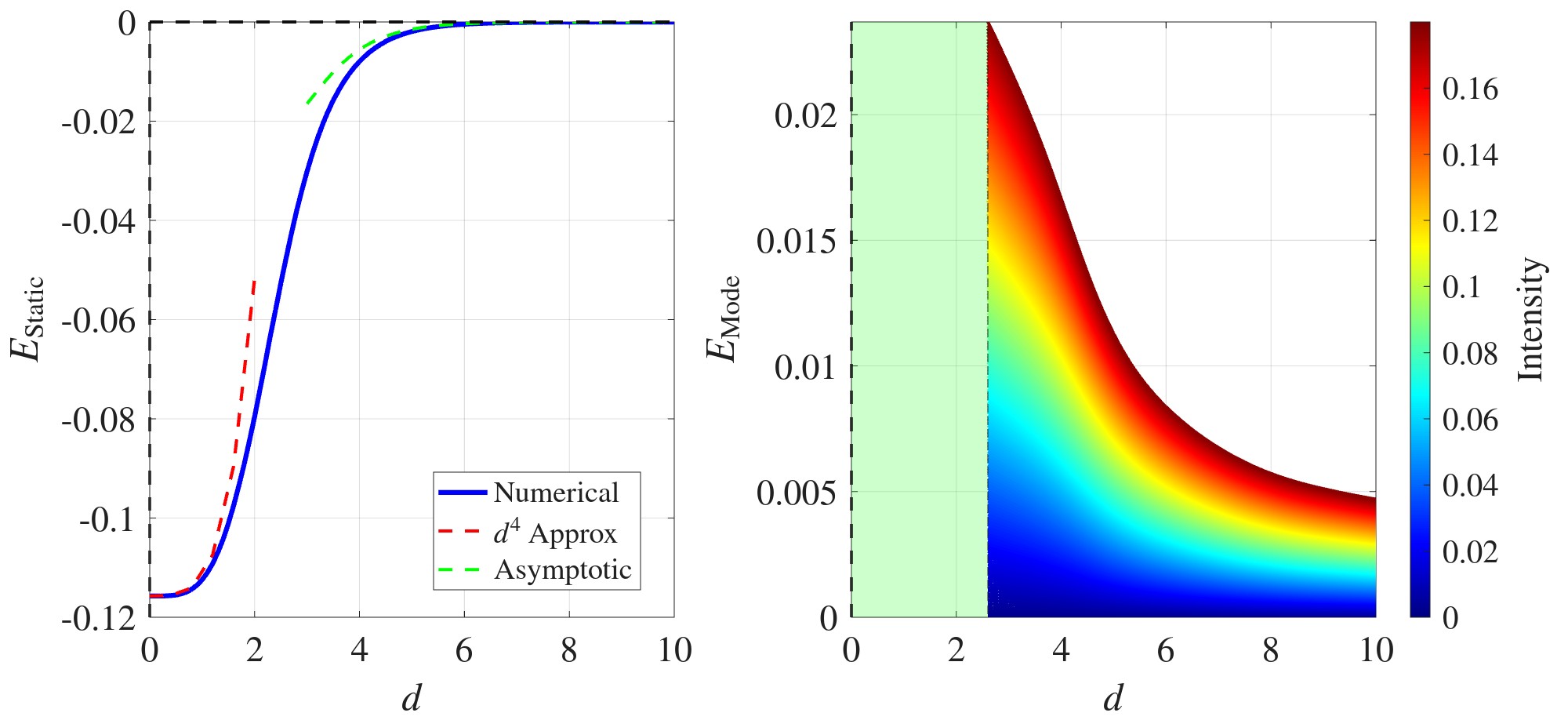}
\end{minipage}
\begin{minipage}{0.75\textwidth}
\includegraphics[width=\linewidth]{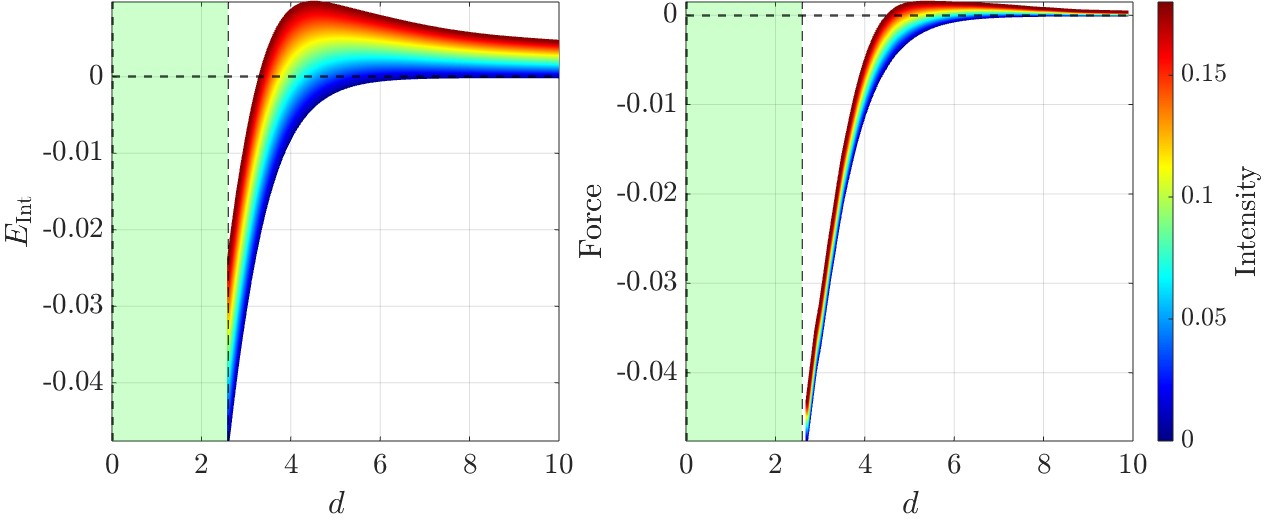}
\end{minipage}
\caption{Interaction energies and forces ($\lambda = 0.5$) as a function of vortex separation $d$. Top left: static force (blue) via \eqref{eq:Estatic}, with $d^4$ (dashed red, \eqref{eq:SRInt}) and asymptotic (dashed green, \eqref{eq:AsyInt}) approximations. Top right: mode interaction energy via \eqref{eq:Emode}. Bottom left: total interaction energy (static plus mode). Bottom right: total force, $F = -\frac{\partial E_{\rm int}}{\partial d}$.}
\label{fig:Int_05}
\end{figure}

Little effect is noticed when the vortices first approach each other, due to the high acceleration as a result of the static force, so it is difficult to confirm the existence of a spectral wall.
Again, to gain insight into the full behaviour of the 2-vortex system with $\lambda = 0.5$, we can calculate the interaction energy; see \cref{fig:Int_05}.

We see from \cref{fig:Int_05} that even though the frequency hits the continuous spectrum before the vortices coincide, there is also a local extremum at $d \in (3,4.5)$, meaning that the vortices could form a quasi-stationary state at this fixed distance, as the net force is zero. The irregularities we observe in the blue line in \cref{fig:l0.5_trajectories} occur approximately the distance of the local extremum, suggesting that the vortices cannot move past this potential barrier, while the intensity of the excitation is large. For $|d| < 3$, that static force is large and hence the vortices want to attract, as seen by the negative interaction energy. This explains the bounces seen in \cref{fig:l0.5_trajectories}, where the vortices bounce multiple times.

\section{Type \texorpdfstring{\rom{2}}{II} Dynamics}
\label{sec:TypeII}
The static force for type \rom{2} vortices is repulsive, hence it is natural to consider mode excitations in the attractive channel. As such, we consider the in-phase superposition of vortices with excited internal shape modes. Since we only consider the attractive channel, the spectral flow of the mode structure is symmetric from $x_1$ to $x_2$.

\begin{figure}
\centering
\includegraphics[width=0.7\linewidth]{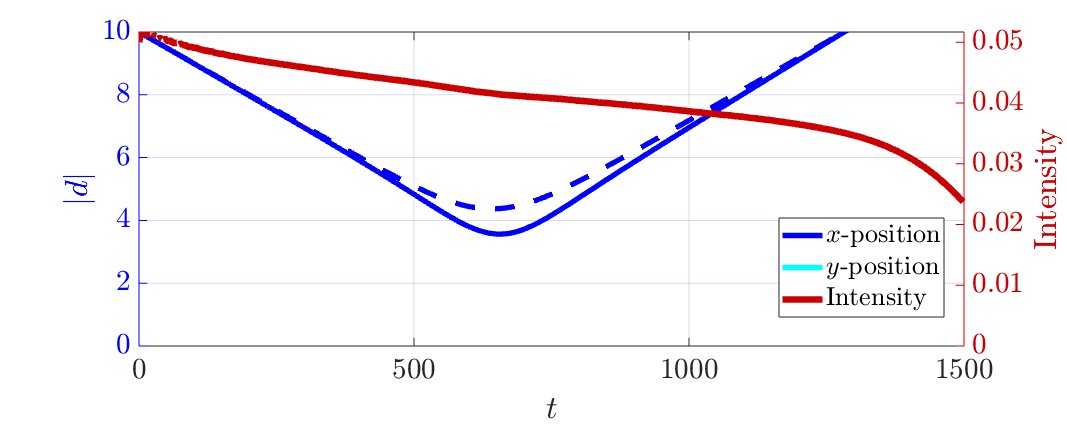}
\caption{Trajectories of a two-vortex system ($\lambda = 1.1$, $v_{\rm in} = 0.01$, $I(0) = 0.05$) as a function of time, where $I(0) = \frac{1}{2}(\epsilon \omega)^2$. The blue line shows the $x_1$-direction distance from the origin, red the excitation intensity per vortex, and dashed blue the unexcited scattering in the $x_1$-direction.}
\label{fig:l1.1_trajectories2}
\end{figure}

\begin{figure}
\centering
\includegraphics[width=0.7\linewidth]{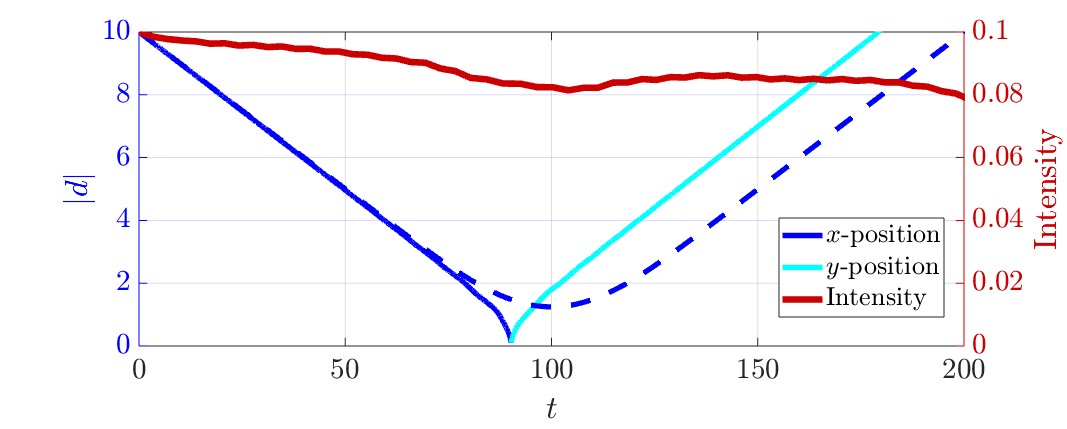}
\caption{Trajectories of a two-vortex system ($\lambda = 1.1$, $v_{\rm in} = 0.1$, $I(0) = 0.1$) as a function of time, where $I(0) = \frac{1}{2}(\epsilon \omega)^2$. The blue line shows the $x_1$-direction distance from the origin, cyan the $x_2$-direction, red the excitation intensity per vortex, and dashed blue the unexcited scattering in the $x_1$-direction.}
\label{fig:l1.1_trajectories3}
\end{figure}

\begin{figure}
\centering
\includegraphics[width=0.7\linewidth]{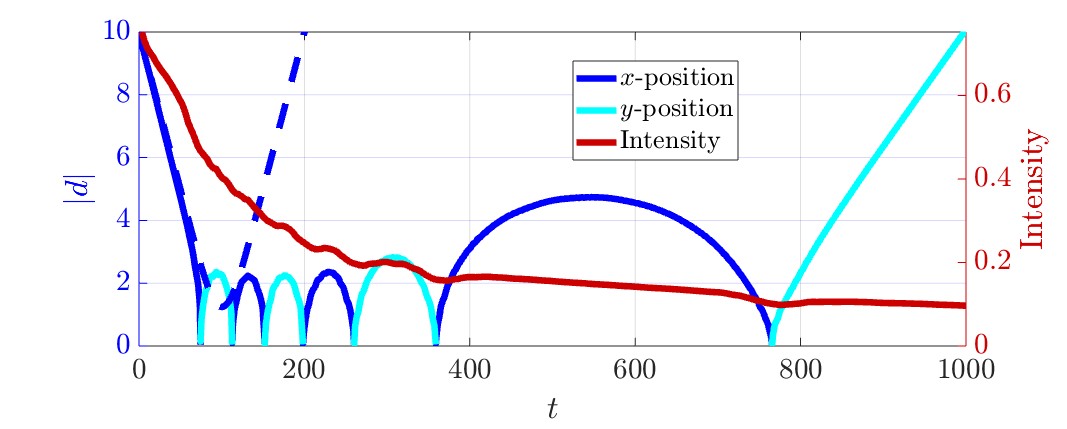}
\caption{Trajectories of a two-vortex system ($\lambda = 1.1$, $v_{\rm in} = 0.1$, $I(0) = 0.7$) as a function of time, where $I(0) = \frac{1}{2}(\epsilon \omega)^2$. The blue line shows the $x_1$-direction distance from the origin, cyan the $x_2$-direction, red the excitation intensity per vortex, and dashed blue the unexcited scattering in the $x_1$-direction.}
\label{fig:l1.1_trajectories1}
\end{figure}

We observe in \cref{fig:l1.1_trajectories1,fig:l1.1_trajectories2,fig:l1.1_trajectories3} the distance of the vortices from the origin as a function of time. If the vortices are on the $x_1$-axis, we plot their distance in blue, and if they are on the $x_2$-axis, we plot the distance in cyan. We also overlay the intensity of the excitation as a function of time, displayed in red. Additionally, we also plot the distance of the vortices from the origin as a function of time for a dynamical simulation with the same parameters, but in the absence of a mode excitation. This data is displayed in the same colours as before, but with dashed lines.

\Cref{fig:l1.1_trajectories2} shows a simulation with an initial intensity of the excitation $I(0) = 0.05$, initial velocity $v_{\rm in} = 0.01$, and the vortices are initially positioned at $x_1 = \pm10$. We see from the blue line indicating the position in the $x_1$-plane that the vortices do not meet the coincident configuration. This is because the excitation and kinetic motion of the vortices are too small to overcome the static force. We observe the red line in \cref{fig:l1.1_trajectories2} that describes the intensity of the excitation as the vortices evolve. We note that the intensity drops as the vortices change direction and increases once they are moving back towards their initial position. This is because energy is transferred from the mode to the kinetic energy.

Let us now consider a dynamical simulation, with initial velocity $v_{\rm in} = 0.1$, initial intensity of the excitation, $I(0) = 0.1$, where $I(0) = \frac{1}{2}(\epsilon \omega)^2$, and the vortices are initially positioned at $x_1 = \pm10$.

The dashed blue line in \cref{fig:l1.1_trajectories3}, which indicates the distance from the origin of the vortices in the $x_1$-direction, that the static force dominates the scattering, and the vortices are repelled to infinity before meeting at the origin. Alternatively, if we observe the solid blue line, which corresponds to a similar configuration, except with an initial intensity of the excitation per vortex of $I(0) = 0.1$, that we can overcome the static repulsion, and we observe that the vortices scatter at $90^{\deg}$. The red line indicating the intensity of the excitation over time gradually decreases due to the decay of the mode, however, as observed in \cite{SKMRTW}, there is an energy transfer mechanism whereby energy is transferred from the mode to the kinetic energy, which we see at $t \approx 100$, when the vortices meet at the origin. Moreover, we form the conjecture that there is not enough energy in the mode to result in a quasi-bound state.

In \cref{fig:l1.1_trajectories1}, we again show the position of the vortices from the origin, for an initial configuration with $v_{\rm in} = 0.1$, initial intensity of the excitation $I(0) = 0.7$, where $I(0) = \frac{1}{2}(\epsilon \omega)^2$, and the vortices initially positioned at $x_1 = \pm10$.

We can see that initially, the vortices form a quasi-bound state, where they scatter multiple times. After some time, we see from the cyan line at $t \approx 500$ that the vortices separate significantly more, then move back towards each other. Interestingly, they do not move all the way to the origin, but instead slow, until they change direction and escape to infinity.

\begin{figure}
\centering
\begin{minipage}{0.24\textwidth}
\centering
\includegraphics[width=\linewidth]{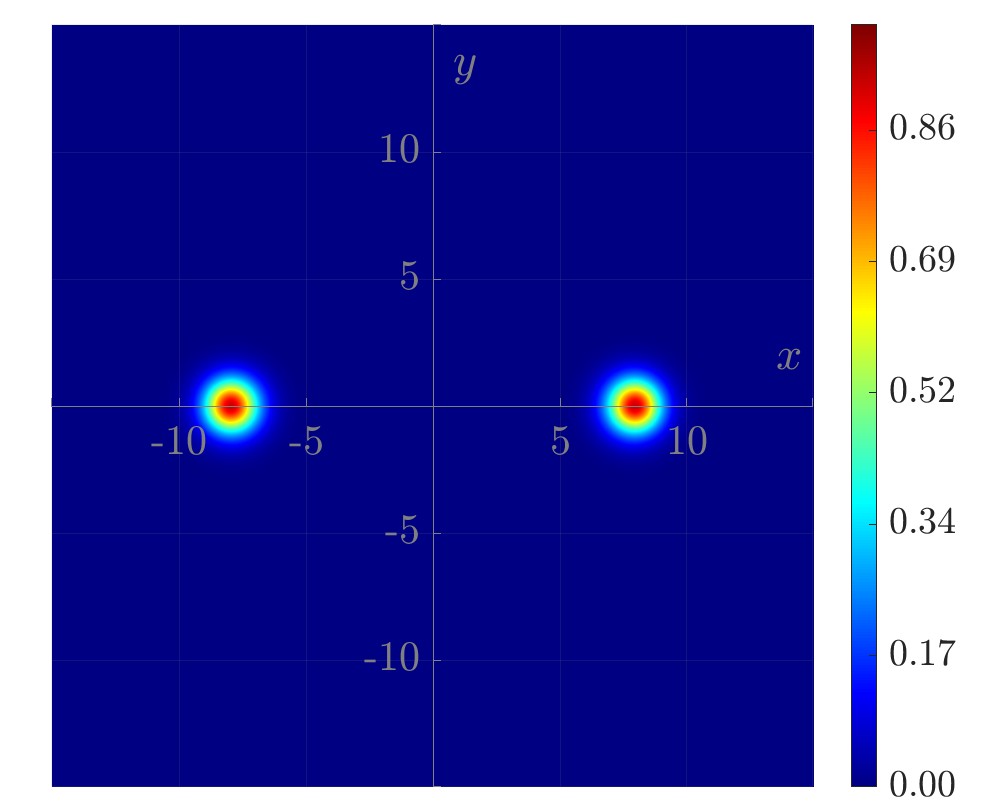}
\caption*{$t = 20$}
\end{minipage}
\begin{minipage}{0.24\textwidth}
\centering
\includegraphics[width=\linewidth]{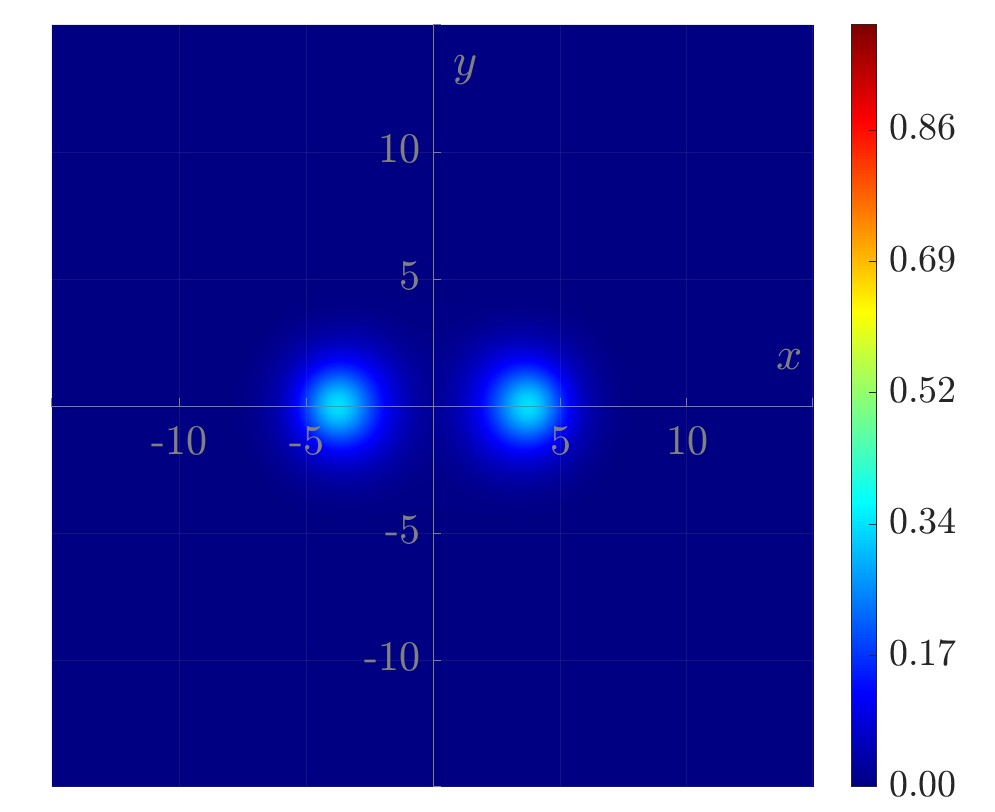}
\caption*{$t = 60$}
\end{minipage}
\begin{minipage}{0.24\textwidth}
\centering
\includegraphics[width=\linewidth]{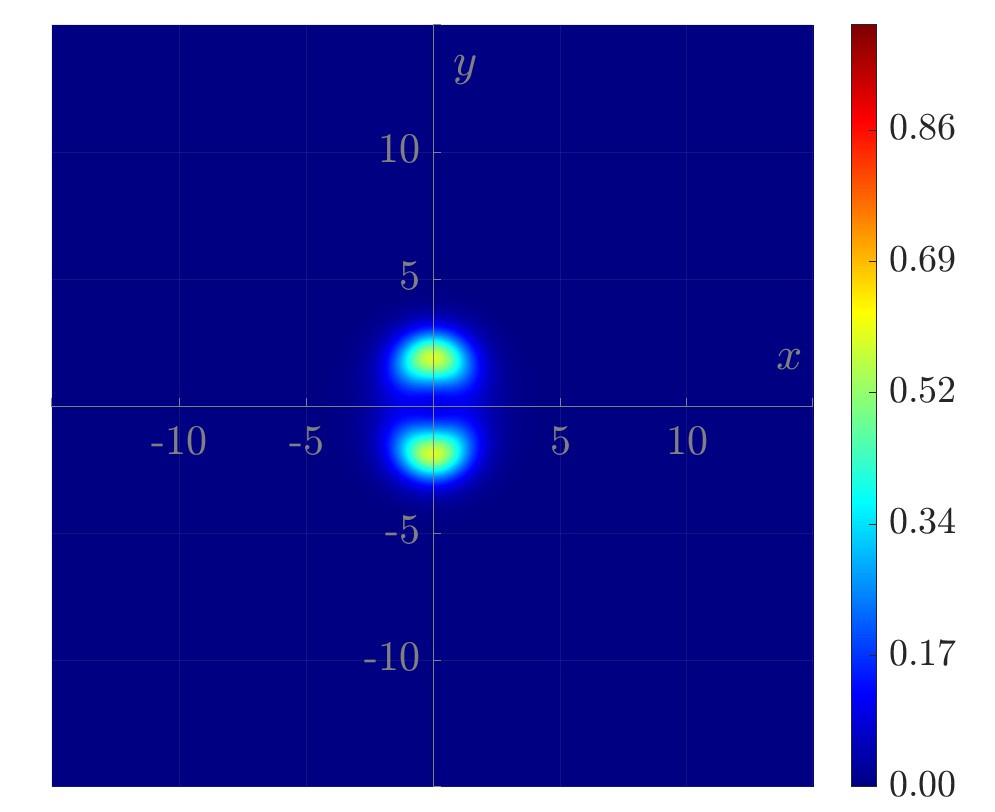}
\caption*{$t = 80$}
\end{minipage}
\begin{minipage}{0.24\textwidth}
\centering
\includegraphics[width=\linewidth]{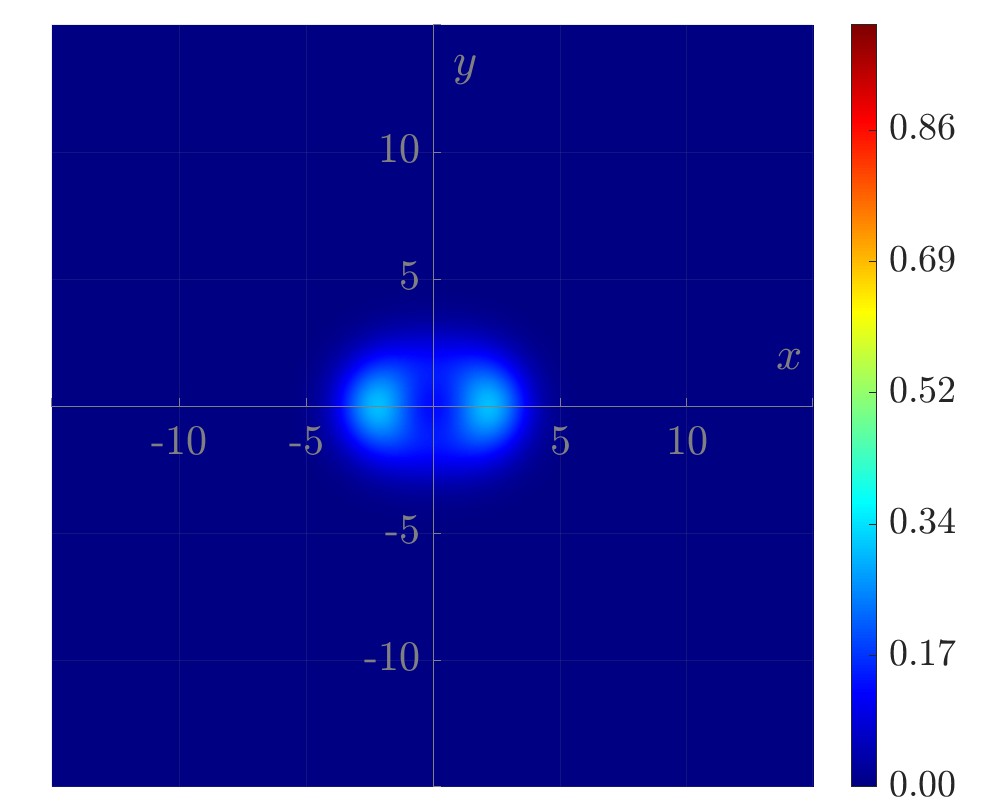}
\caption*{$t = 120$}
\end{minipage}
\begin{minipage}{0.24\textwidth}
\centering
\includegraphics[width=\linewidth]{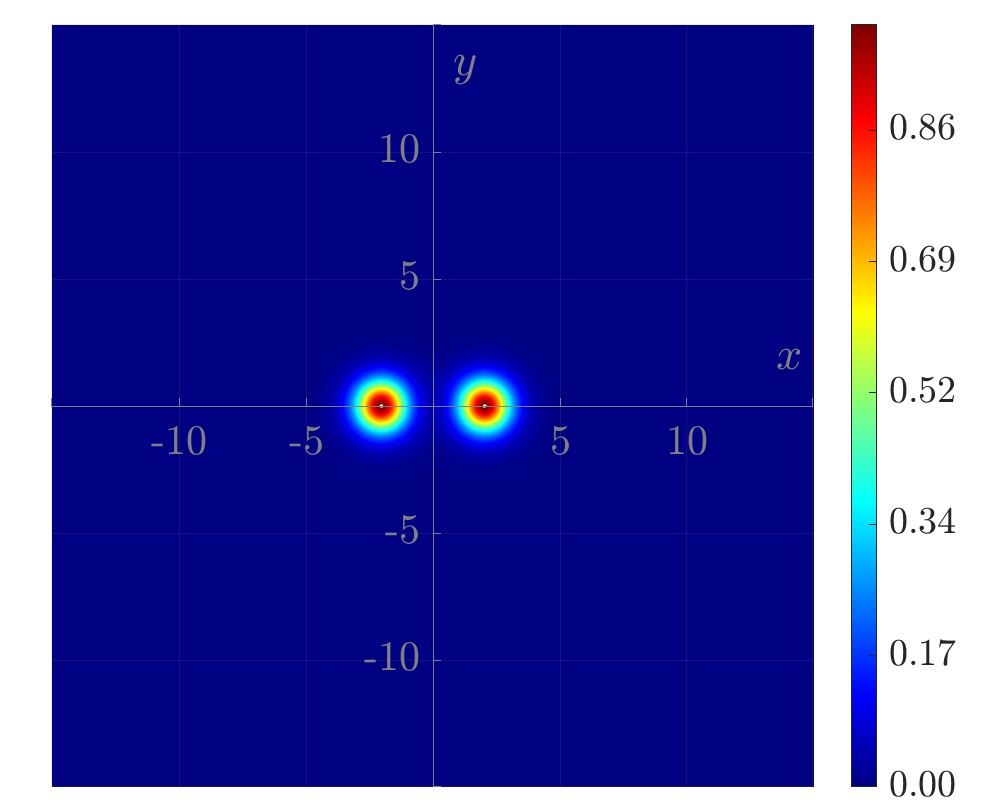}
\caption*{$t = 140$}
\end{minipage}
\begin{minipage}{0.24\textwidth}
\centering
\includegraphics[width=\linewidth]{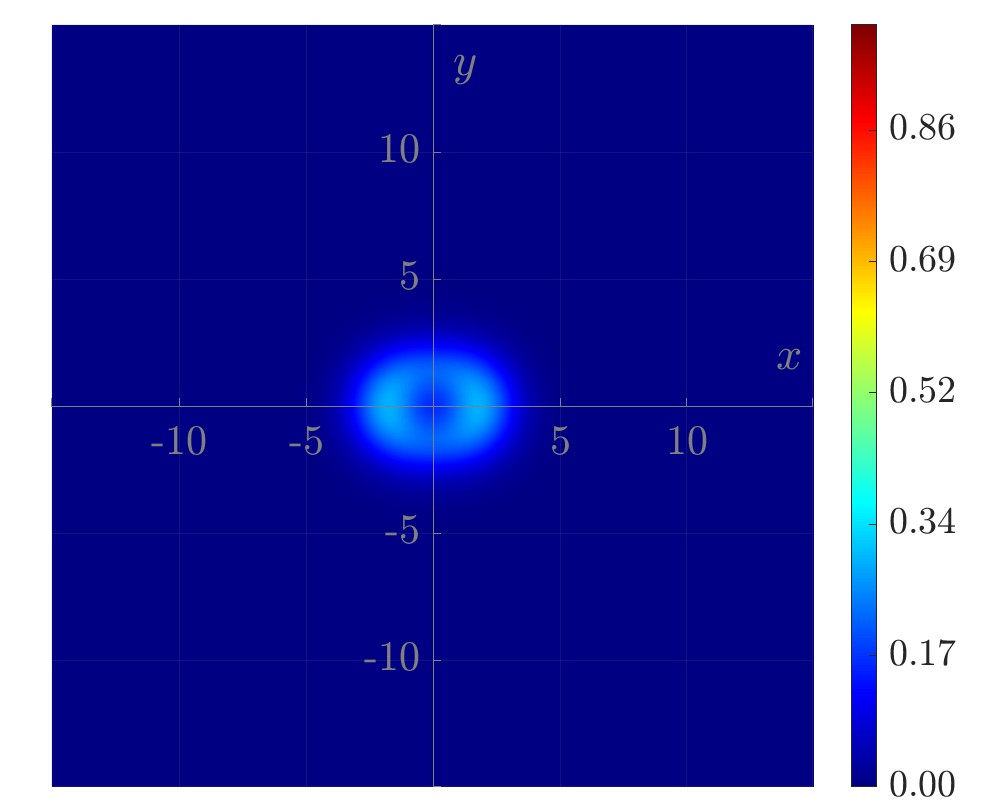}
\caption*{$t = 150$}
\end{minipage}
\begin{minipage}{0.24\textwidth}
\centering
\includegraphics[width=\linewidth]{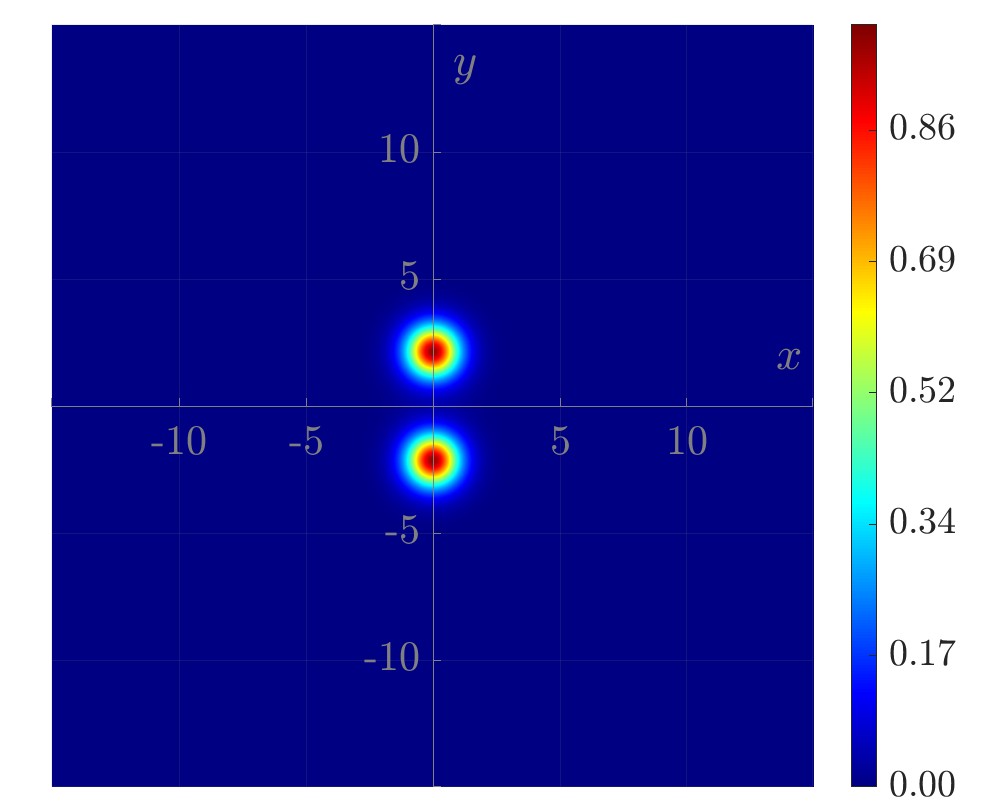}
\caption*{$t = 170$}
\end{minipage}
\begin{minipage}{0.24\textwidth}
\centering
\includegraphics[width=\linewidth]{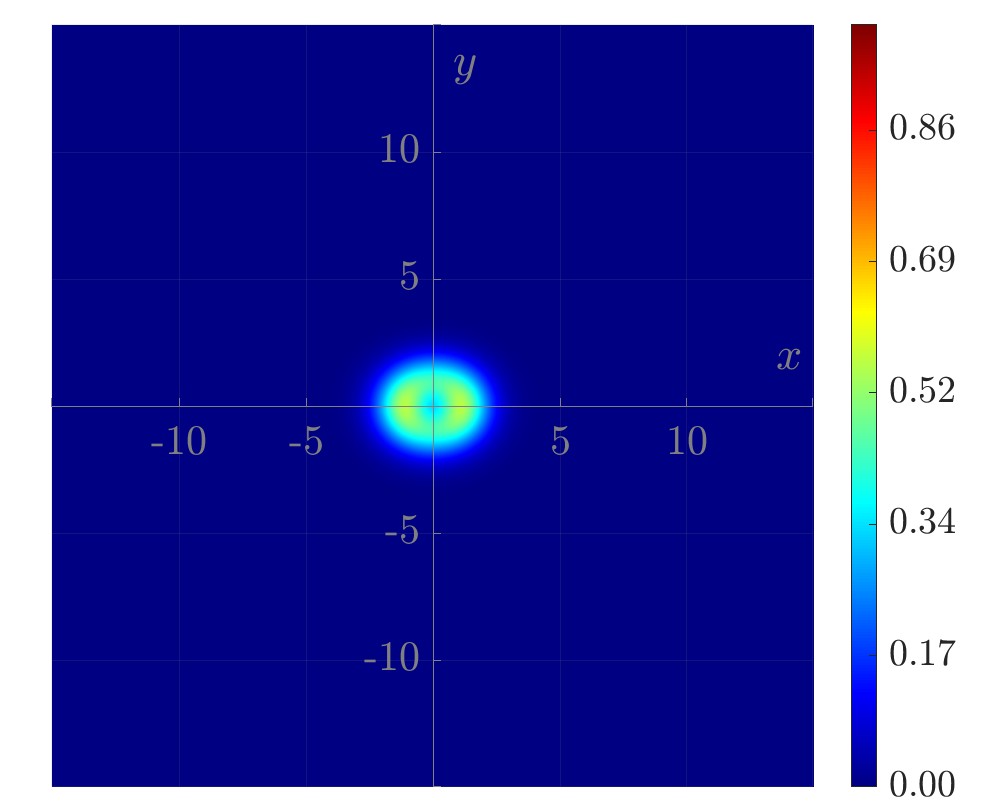}
\caption*{$t = 200$}
\end{minipage}
\begin{minipage}{0.24\textwidth}
\centering
\includegraphics[width=\linewidth]{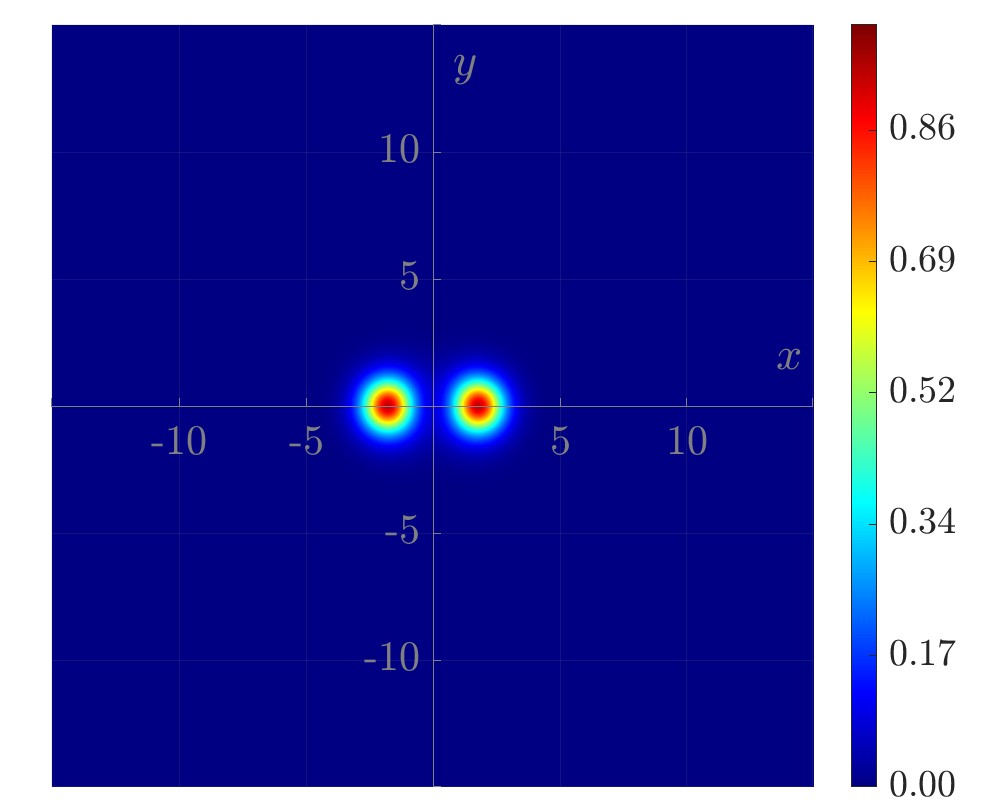}
\caption*{$t = 210$}
\end{minipage}
\begin{minipage}{0.24\textwidth}
\centering
\includegraphics[width=\linewidth]{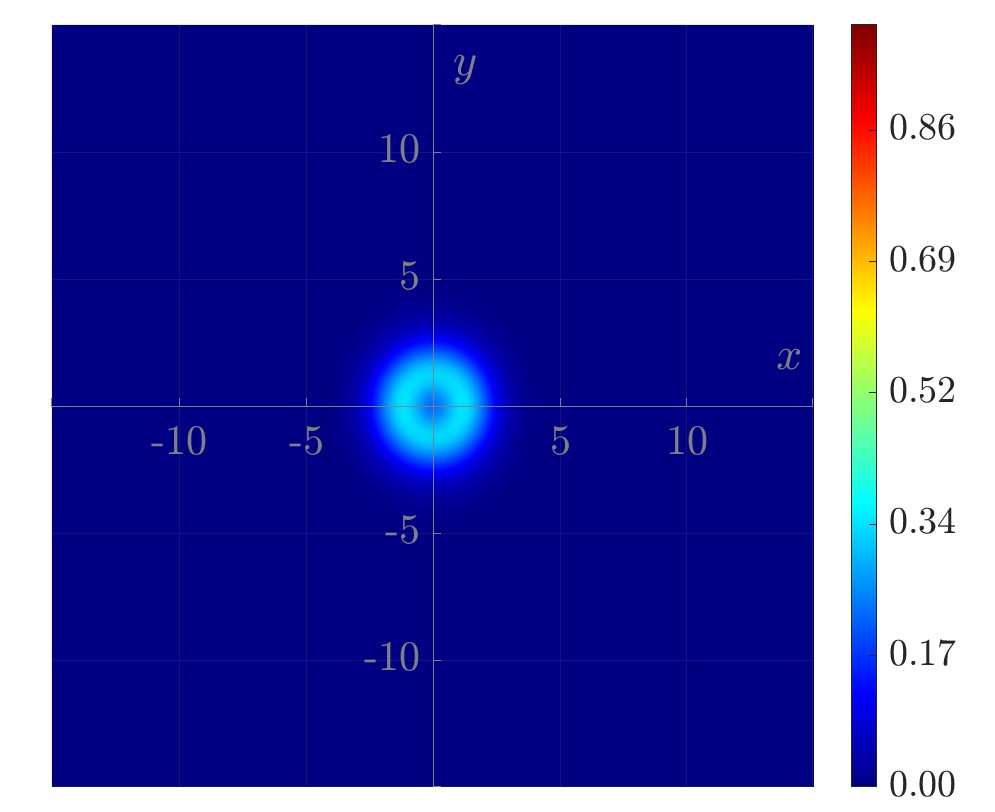}
\caption*{$t = 260$}
\end{minipage}
\begin{minipage}{0.24\textwidth}
\centering
\includegraphics[width=\linewidth]{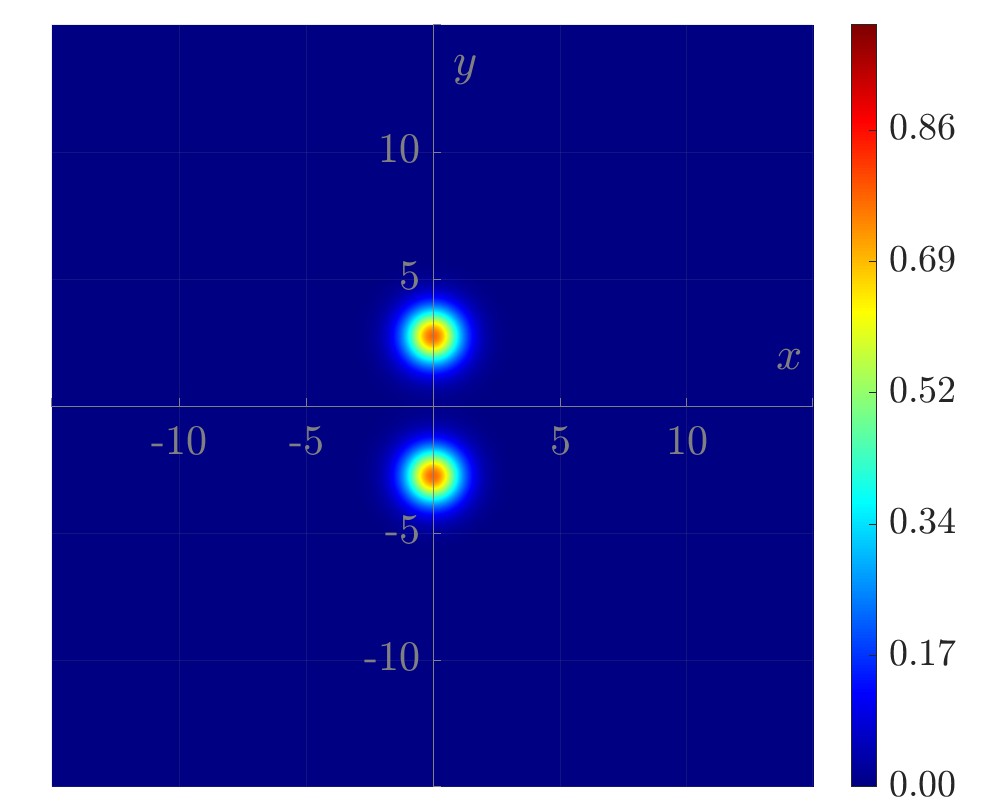}
\caption*{$t = 320$}
\end{minipage}
\begin{minipage}{0.24\textwidth}
\centering
\includegraphics[width=\linewidth]{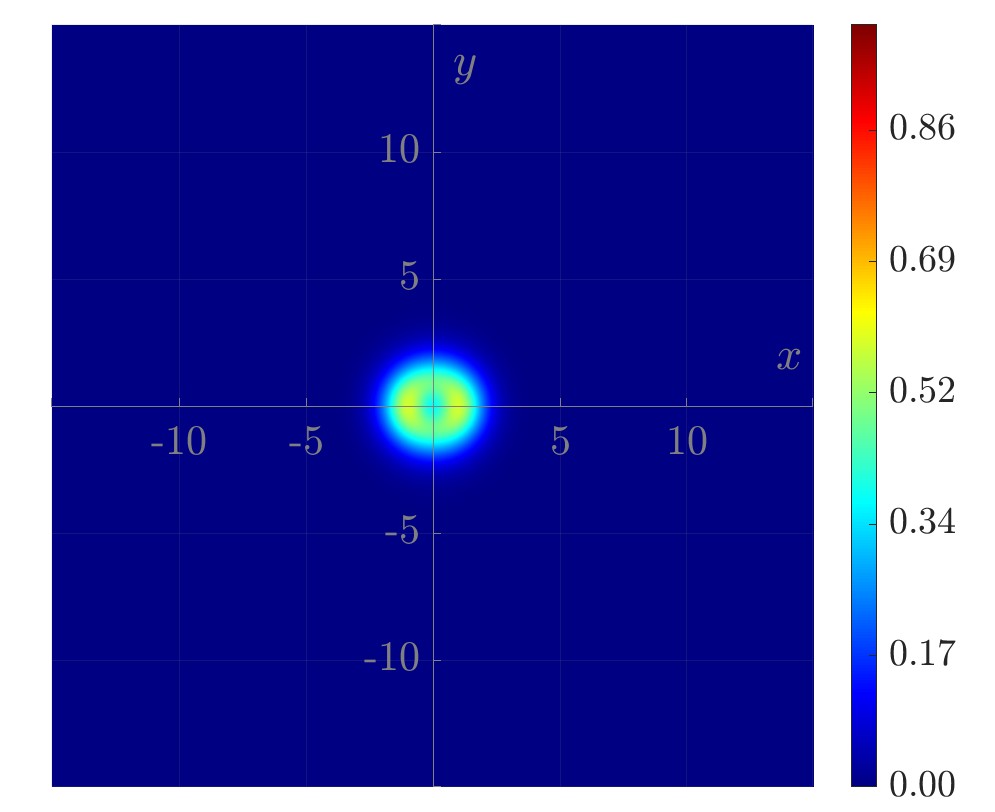}
\caption*{$t = 360$}
\end{minipage}
\begin{minipage}{0.24\textwidth}
\centering
\includegraphics[width=\linewidth]{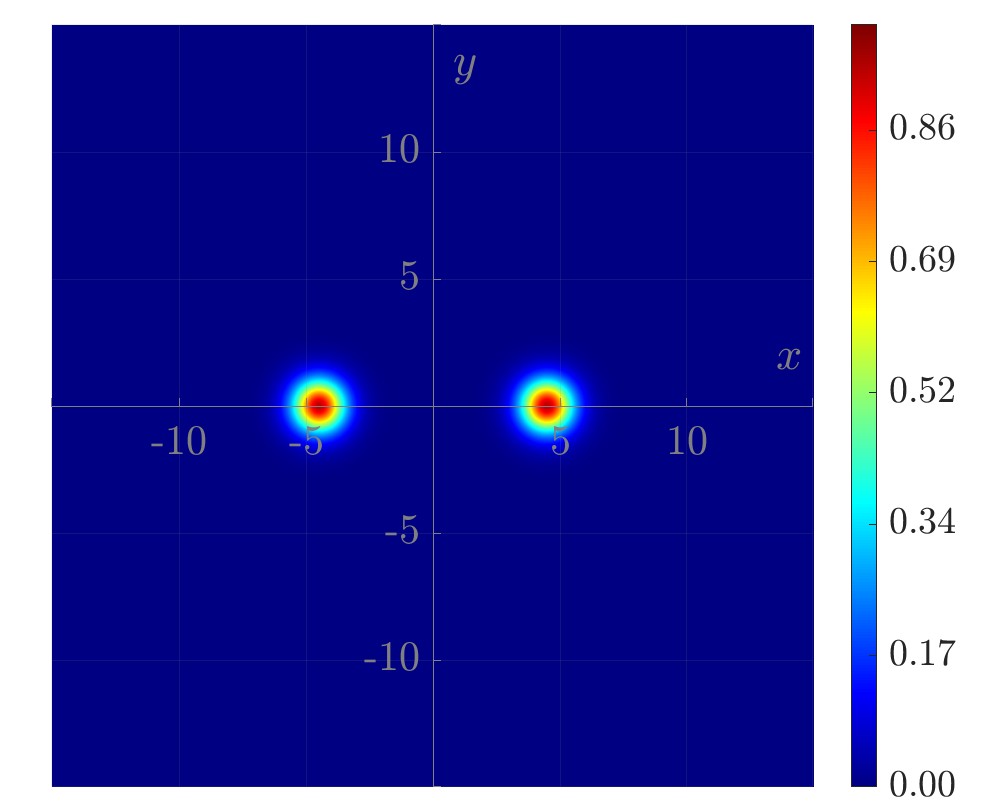}
\caption*{$t = 480$}
\end{minipage}
\begin{minipage}{0.24\textwidth}
\centering
\includegraphics[width=\linewidth]{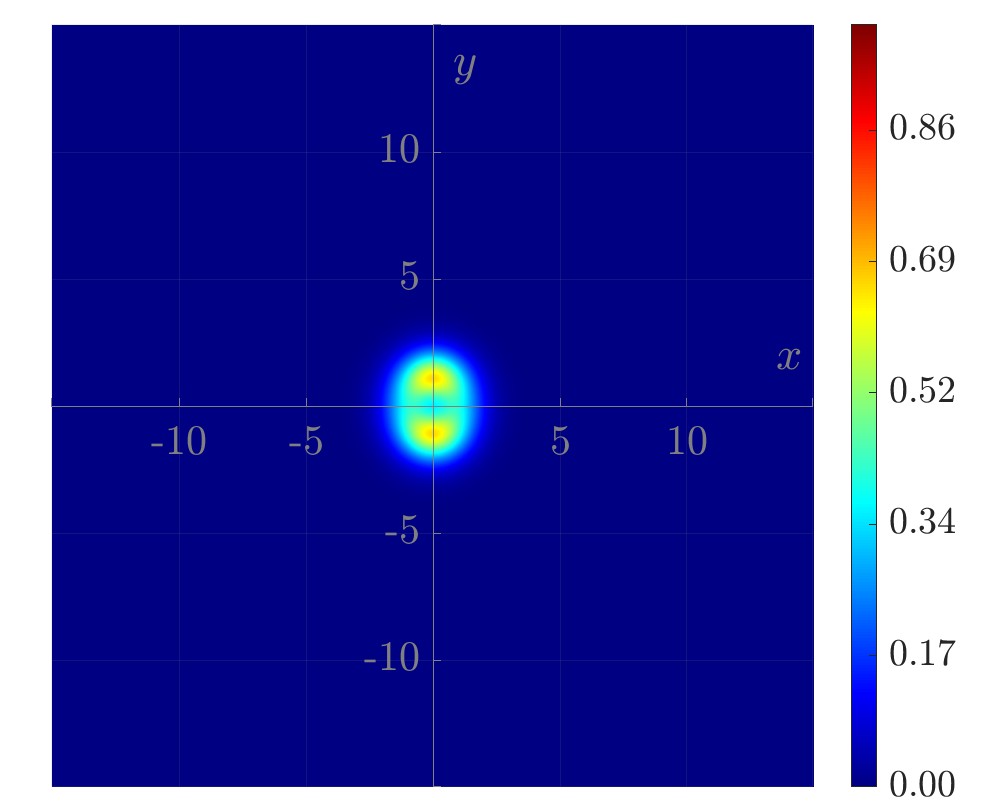}
\caption*{$t = 770$}
\end{minipage}
\begin{minipage}{0.24\textwidth}
\centering
\includegraphics[width=\linewidth]{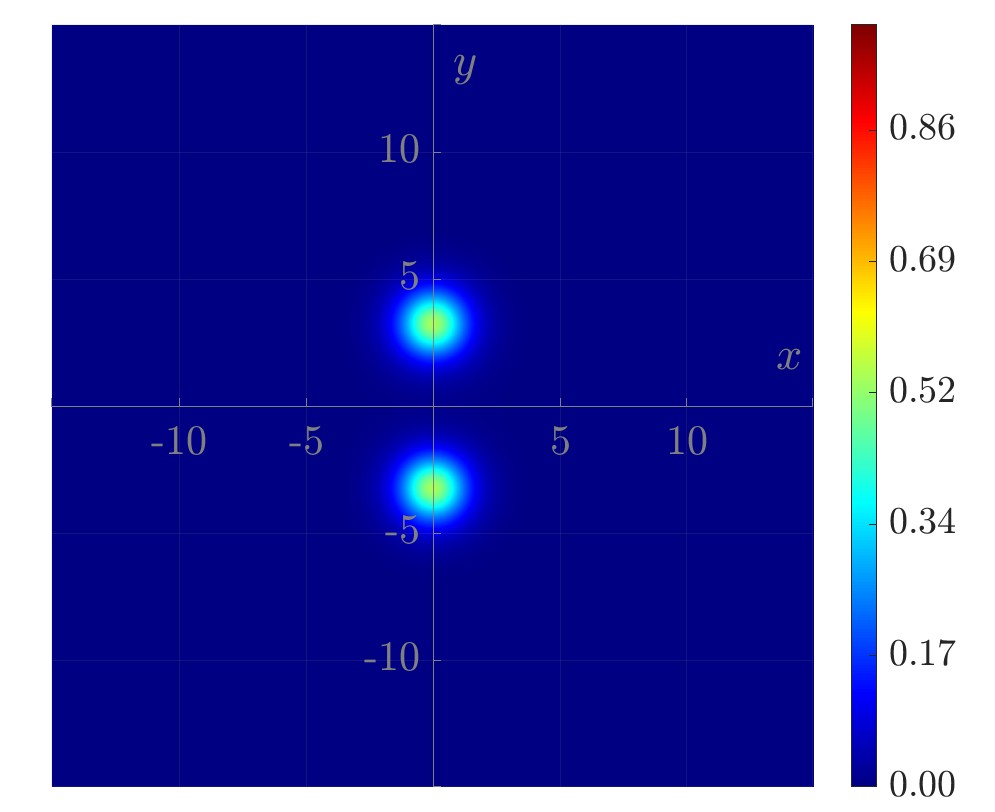}
\caption*{$t = 820$}
\end{minipage}
\begin{minipage}{0.24\textwidth}
\centering
\includegraphics[width=\linewidth]{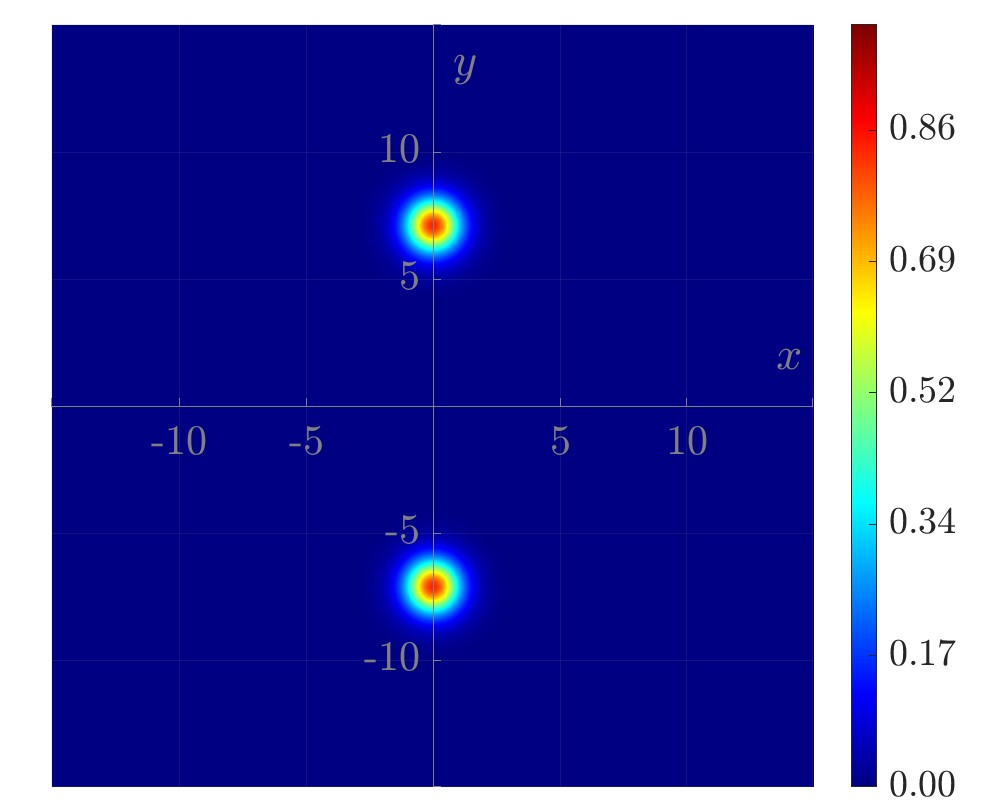}
\caption*{$t = 920$}
\end{minipage}
\caption{Snapshots of energy density for a two-vortex scattering ($\lambda = 1.1$, $v_{\rm in} = 0.1$, $I(0) = 0.7$), with in-phase $k=0$ shape modes, as seen in \cref{fig:l1.1_trajectories1}.}
\label{fig:l1.1_snapshot}
\end{figure}

We can explore some dynamical snapshots \cref{fig:l1.1_trajectories1} displaying a heat plot of the energy density. The simulation shown in \cref{fig:l1.1_snapshot} displays the scattering of two $N=1$ vortices with $\lambda = 1.1$. We have excited the $k = 0$ shape mode on each vortex to induce an attractive force, where $\omega_{1,0}^2 = 0.8352168$. We have chosen an initial intensity of $I(0) = 0.7$, where $I(0) = \frac{1}{2}(\epsilon \omega)^2$, and an initial velocity of $v_{\rm in} = 0.1$.

We can observe in \cref{fig:l1.1_snapshot} that the vortices are in phase with excited $k=0$ shape modes. We can see that the vortices move closer together, whilst oscillating in shape. We also observe the presence of a quasi-bound state where the vortices scatter multiple times. This is quite interesting because the static force in the \rom{2} regime is repulsive, and therefore the vortices naturally want to move apart.

\begin{figure}
\centering
\includegraphics[width=0.7\linewidth]{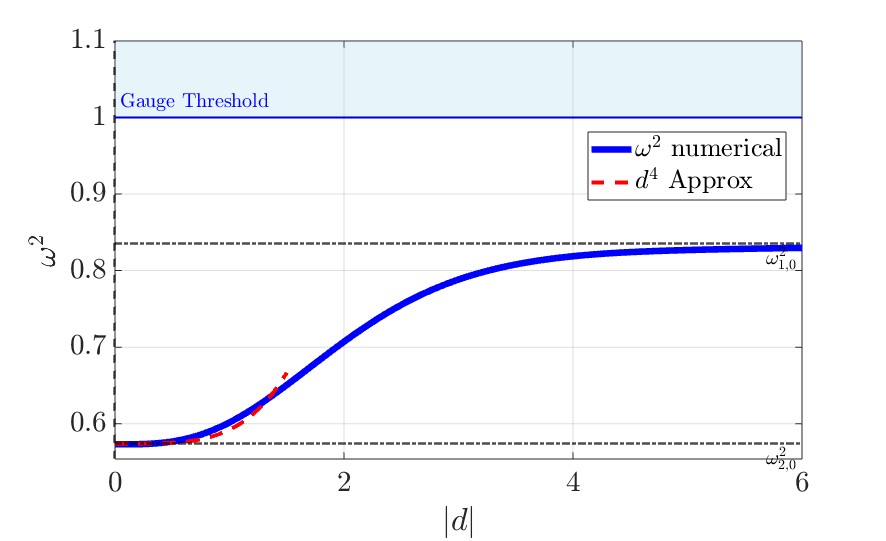}
\caption{Angular frequency flow for a two-vortex system ($\lambda = 1.1$) as a function of vortex separation $d$ from the origin. The blue area indicates the mass threshold, with a $d^4$ approximation (dashed red) for $d \to 0$.}
\label{fig:l11_sepFreq}
\end{figure}

To begin to understand this behaviour, we can measure the angular frequency from the dynamical simulation (see \cref{fig:l11_sepFreq}). We plot the $d^4$ approximation for $d \to 0$. We see that the frequency interpolates between the value of the coincident $N=2$ configuration, $\omega_{2,0}^2 = 0.5738714$, and the asymptotic value describing well separated $N=1$ vortices, $\omega_{1,0}^2 = 0.8352168$. This suggests an attractive intervortex force induced by the in-phase excitation.

\begin{figure}
\centering
\begin{minipage}{0.75\textwidth}
\includegraphics[width=\linewidth]{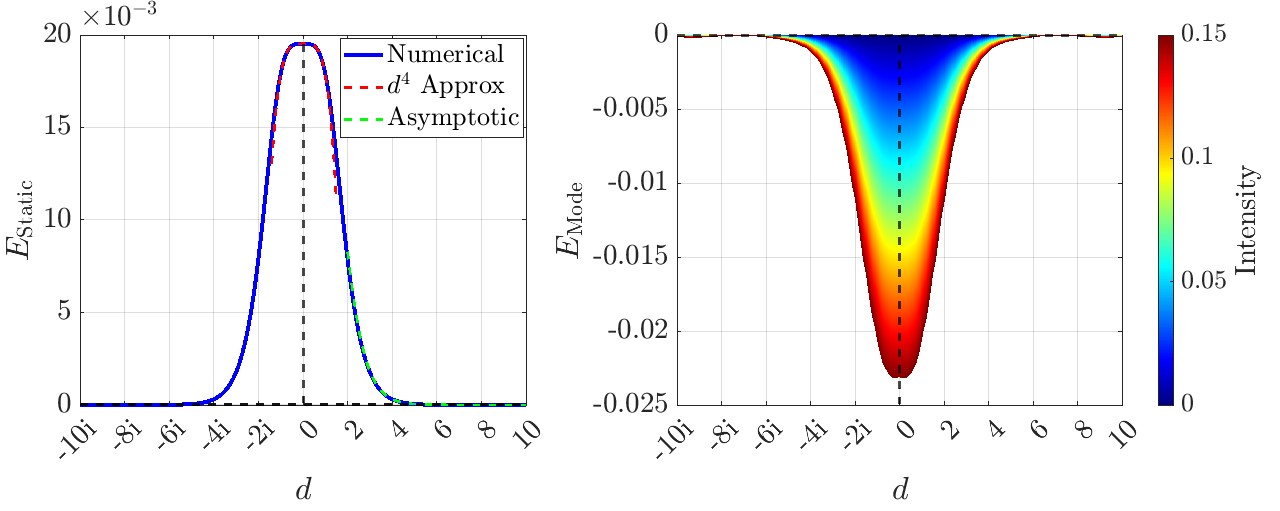}
\end{minipage}
\begin{minipage}{0.75\textwidth}
\includegraphics[width=\linewidth]{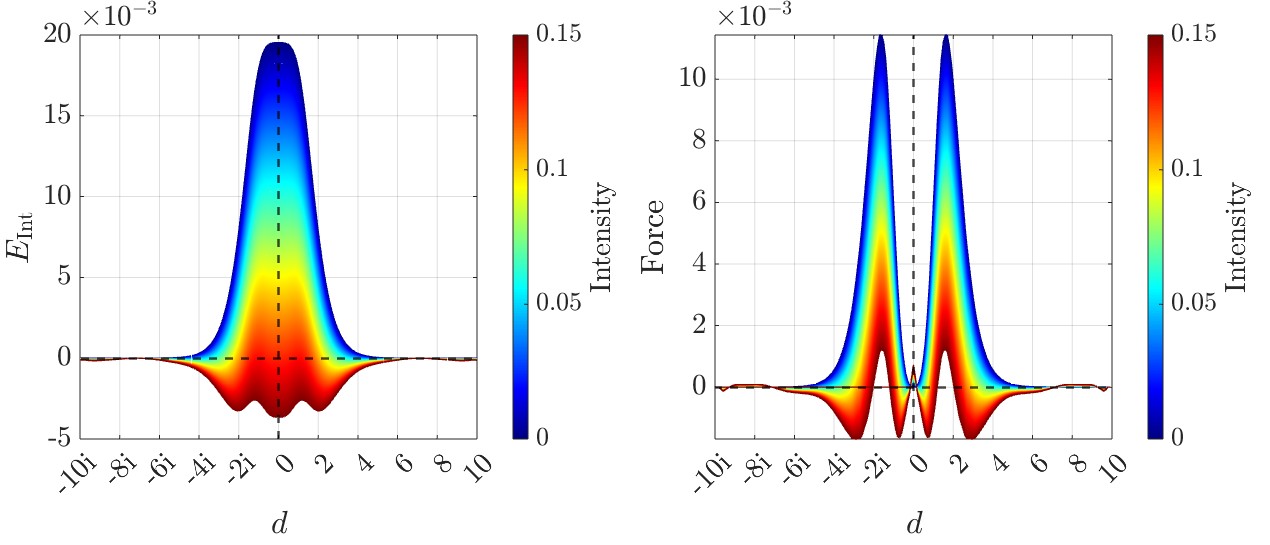}
\end{minipage}
\caption{Interaction energies and forces ($\lambda = 1.1$) as a function of vortex separation $d$. Top left: static force (blue) via \eqref{eq:Estatic}, with $d^4$ (dashed red, \eqref{eq:SRInt}) and asymptotic (dashed green, \eqref{eq:AsyInt}) approximations. Top right: mode interaction energy via \eqref{eq:Emode}. Bottom left: total interaction energy (static plus mode). Bottom right: total force, $F = -\frac{\partial E_{\rm int}}{\partial d}$.}
\label{fig:Int_11}
\end{figure}

We can then calculate the energy contribution from the excitation (\cref{eq:Emode}). This can be summed with the energy contribution of the static force, to give a space-dependent measure of the interaction energy for a type \rom{2} 2-vortex system with excited shape modes, see \cref{fig:Int_11}.

We interpret the interaction energy (see \cref{fig:Int_11}) as follows. For small $d$, the interaction energy behaves similarly to the static force as $d^4$, hence the 2 vortices will remain at the origin if the mode is dominating, i.e. the intensity of the excitation is large enough such that it is stronger than the static repulsion.

When the vortex separation is large, both the static force and the mode interaction asymptotically go to zero, hence there is no effect on vortex dynamics in this region. Alternatively, if we consider the scattering of vortices, when the intensity of the excitation is small, the vortices will back scatter due to the static force; however, if it is large enough, the vortices will be attracted towards each other.

\begin{figure}
\centering
\begin{minipage}{0.7\textwidth}
\includegraphics[width=\linewidth]{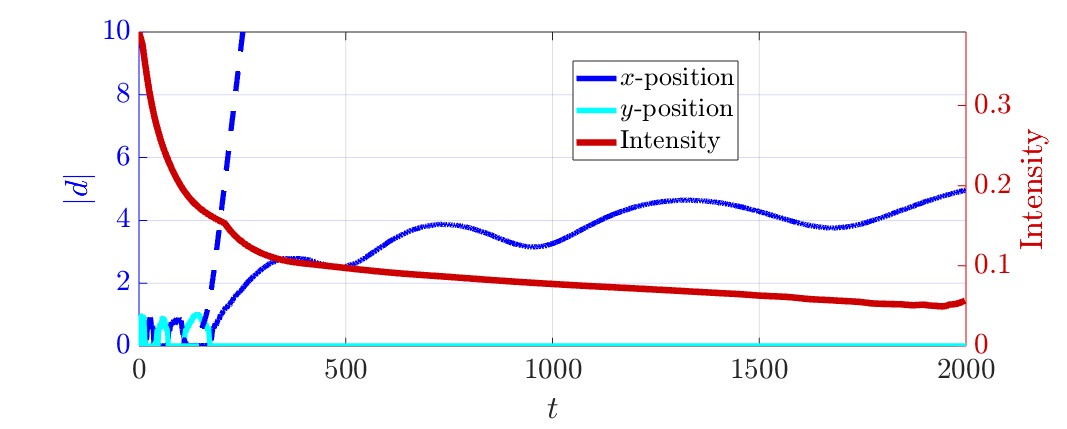}
\end{minipage}
\begin{minipage}{0.7\textwidth}
\includegraphics[width=\linewidth]{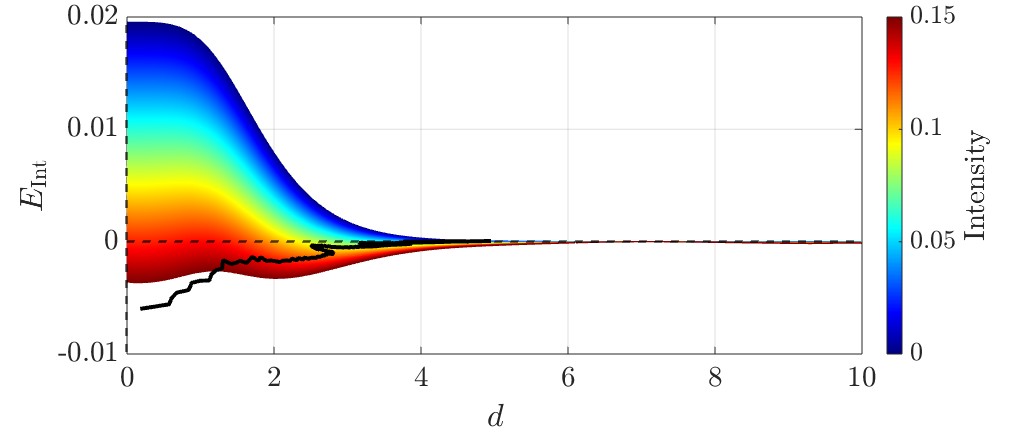}
\end{minipage}
\caption{Top: trajectories of a two-vortex system ($\lambda = 1.1$, $v_{\rm in} = 0$, $I(0) = 0.4$) coincident at $t=0$, where $I(0) = \frac{1}{2}(\epsilon \omega)^2$. The blue line shows the $x_1$-direction distance from the origin, cyan the $x_2$-direction, red the excitation intensity per vortex, and dashed blue the unexcited scattering in the $x_1$-direction. Bottom: interaction energy colour-map with black parametric curve showing the interaction energy of the simulation above.}
\label{fig:Int_A0_0.37}
\end{figure}

In-between these two regions, vortex dynamics become highly non-linear. There exists a local minimum in the interaction energy at $|d| \in (2,4)$ that depends on the intensity of the mode. This suggests that vortices can become stuck at a fixed separation where the net force is zero, resulting in a quasi-stationary state, which explains the latter part of the trajectories in \cref{fig:l1.1_trajectories1}. We can confirm the existence of this bound state by considering another dynamical solution, see \cref{fig:Int_A0_0.37}. The simulation in question begins with a saddle point solution, a radially symmetric $N=2$ vortex centred at the origin. We then perturb the solution by adding a linear combination of the $k=2$ splitting mode $(\omega_{2,2}^2 = -0.0107688)$ in the $x$ direction and the $k=0$ shape mode $(\omega_{2,0}^2 = 0.5738714)$.

We see from the blue line denoting the distance from the origin of the vortices in the $x_1$-plane (see \cref{fig:Int_A0_0.37}) that the vortex motion is slowed in the region of the local minimum of the interaction energy. In fact, we see that the vortices begin to oscillate in space within the region as the forces compete, until escaping as the intensity of the mode decays.

The black dots on the top of \cref{fig:Int_A0_0.37} can be followed to show how the intensity of the mode changes as the vortices evolve in time. It also shows that the vortices do indeed become trapped in this potential well. The colour-map shows the interaction energy assuming a fixed intensity, however, as seen in the top plot, the solid red line displays the intensity $I(t)$ as the vortices scatter, and it is not constant.

\begin{figure}
\centering
\includegraphics[width=0.7\linewidth]{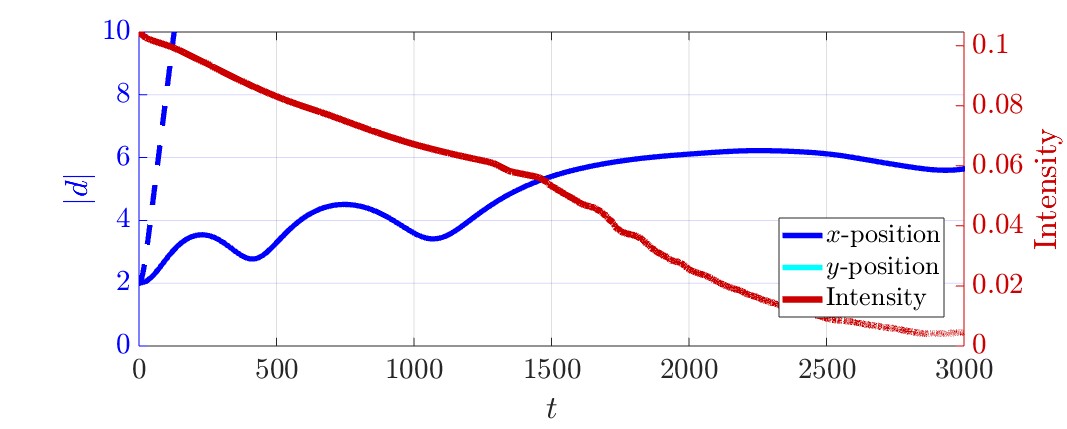}
\caption{Trajectories of a two-vortex system ($\lambda = 1.1$, $v_{\rm in} = 0$, $I(0) = 0.04$) near the local minimum, where $I(0) = \frac{1}{2}(\epsilon \omega)^2$. The blue line shows the $x_1$-direction distance from the origin, red the excitation intensity per vortex, and dashed blue the unexcited scattering in the $x_1$-direction.}
\label{fig:trajectories_l11_localMin}
\end{figure}

Similarly to \cref{fig:trajectories_localMax_l09}, we can situate the vortices near the local minimum found in \cref{fig:Int_11}, see \cref{fig:trajectories_l11_localMin}. Here, the vortices are initially positioned at $x_1 = \pm 2$, with initial intensity $I(0) = 0.04$. The blue line indicates the position of the vortices on the $x_1$-axis as a function of time. We also show the intensity $I(t)$ in red. We can see that the vortices oscillate in space, within the potential well shown in \cref{fig:Int_11}. As the intensity of the excitation decays, the position of the potential well shifts to the right; hence, we see that the turning points of the position move out also. This clearly confirms the presence of a quasi-stationary state, whereby the vortices become trapped in the potential well and stay in a fixed region away from the origin, as long as the excitation is large enough.

In \cref{fig:trajectories_l11_localMin}, the vortices are positioned initially at $x = \pm 2$, with initial intensity $I(0) = 0.04$. The blue line indicated the position of the vortices as a function of time. We also show the intensity $I(t)$ in red. We can see that the vortices oscillate in space, inside the potential well shown in \cref{fig:Int_11}. As the intensity of the excitation decays, the position of the potential well shifts to the right, hence we see that the turning points of the position move out also. This clearly confirms the presence of a quasi-stationary state, whereby the vortices become trapped in the potential well, and stay in a fixed region away from the origin, as long as the excitation is large enough.

It was found in \cite{SKMRTW,alonso2026bps} that BPS vortices with excited shape modes in the attractive channel exhibit a chaotic bound structure. Vortices will scatter more than once due to the energy transfer mechanism between the energy in the mode and the kinetic energy, akin to the resonant energy transfer originally observed in kink-antikink collisions in the $\phi^4$ model \cite{sug,CSW}. A closely related chaotic bounce window structure has recently been observed in unexcited vortex-antivortex scattering in the deep type \rom{2} regime, where the energy transfer is mediated by a Feshbach quasi-normal mode of the vortex \cite{bachmaier2026resonance}.
We can perform the same analysis here. Even though the static force of the type \rom{2} vortices makes these quasi-bound states less likely to exist, they can still occur, as observed in \cref{fig:l1.1_trajectories1}.

\begin{figure}
\centering
\begin{minipage}{0.45\textwidth}
\centering
\begin{adjustbox}{valign=c}
\includegraphics[width=\linewidth]{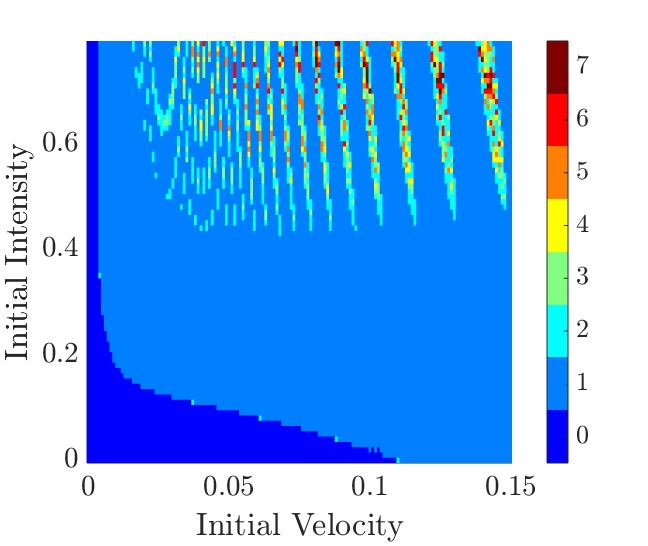}
\end{adjustbox}
\caption{Parameter space for a two-vortex system ($\lambda = 1.1$), varying $v_{\rm in}$ and $I(0)$. The colour indicates the number of times the vortices passed through the coincident configuration.}
\label{fig:l11_velAmpScat}
\end{minipage}
\hfill
\begin{minipage}{0.45\textwidth}
\centering
\begin{adjustbox}{valign=c}
\includegraphics[width=\linewidth]{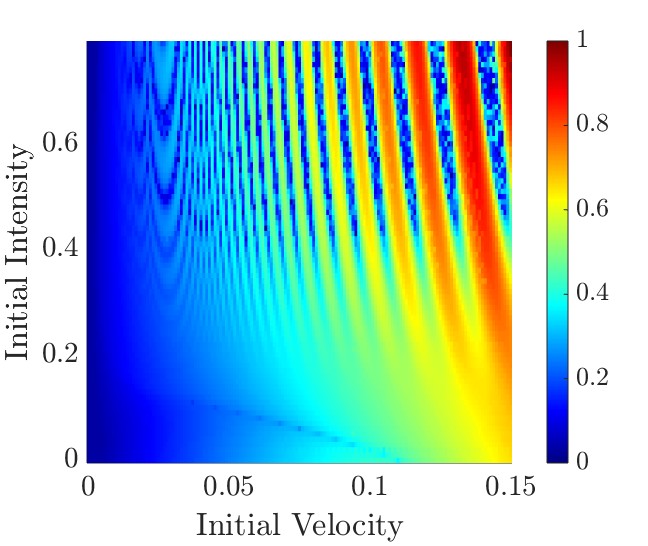}
\end{adjustbox}
\caption{Parameter space for a two-vortex system ($\lambda = 1.1$), varying $v_{\rm in}$ and $I(0)$. The colour indicates the normalised simulation length, $\frac{t_{\rm max}/t_{\rm end}}{\max(t_{\rm max}/t_{\rm end})}$, from 0 ($t_{\rm max} = 3000$) to 1 ($t_{\rm min}$).}
\label{fig:l11_velAmp}
\end{minipage}
\end{figure}

We can vary the initial intensity of the perturbation and initial velocity of the vortices to explore the chaotic nature of these multi-bounce solutions, see \cref{fig:l11_velAmpScat,fig:l11_velAmp}.

In \cref{fig:l11_velAmpScat}, we plot the number of times the vortices pass through the coincident configuration (bounces) against initial velocity and initial intensity of the excitation. We observe a critical velocity of $v_{\rm in} = 0.11$, whereby the vortices will always scatter at least once, i.e. the kinetic motion will overcome the static repulsive force.

We also observe a fractal structure of lines with more than one bounce, akin to that observed at critical coupling in \cite{SKMRTW} and to the well-known fractal pattern of bounce windows in kink-antikink scattering \cite{sug,CSW}. As the initial velocity increases, the lines of multi-bounce solutions increase in width, however we notice that for smaller intensities, these regions are cut off as the static force dominates the interaction, making it difficult for the vortices to form a quasi-bound state.

Interestingly, there are some solutions on the critical line separating solutions that do not pass the coincident configuration (zero bounces) and those that do, whereby we have 2 bounces, suggesting some fine tuned solutions were quasi-bound states at low initial intensities are possible.

To better understand the structure of the parameter space of solutions, we plot the normalised simulation length, $\frac{t_{\rm max}/t_{\rm end}}{\max(t_{\rm max}/t_{\rm end})}$, from 0 ($t_{\rm max} = 3000$) to 1 ($t_{\rm min}$). This allows us to see that the fractal structure does indeed continue for smaller initial intensities, as we see the existence of the bands of longer simulation length inside regions of shorter simulations comparatively.

We have calculated the interaction energy for a range of $\lambda$. It should be noted that the $N=1$ $k=0$ shape mode, which we assume to be the only shape mode excited, exists only in the discrete spectrum up to $\lambda \approx 1.5$ (see \cref{sec:Linearisation}). The plots are not shown here; however, they show the same behaviour as \cref{fig:Int_11}.

\begin{figure}
\centering
\includegraphics[width=0.7\linewidth]{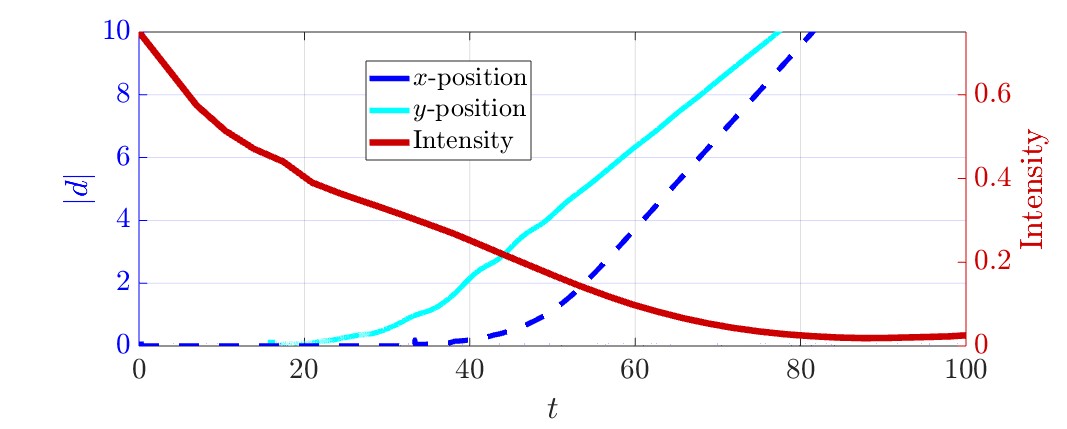}
\caption{Trajectories of a two-vortex system ($\lambda = 2$, $v_{\rm in} = 0$, $I(0) = 0.75$) as a function of time, where $I(0) = \frac{1}{2}(\epsilon \omega)^2$. The cyan line shows the $x_2$-direction distance from the origin, red the excitation intensity per vortex, and dashed blue the unexcited scattering in the $x_1$-direction.}
\label{fig:l2_trajectories}
\end{figure}

It is still of interest to consider the dynamics of vortices above this threshold $\lambda > 1.5$. For example, consider a $N=2$ vortex at $\lambda = 2$, centred at the origin with an excited $k=0$ shape mode, $\omega_{2,0}^2 = 0.8161198$. This mode is still in the discrete spectrum; however, the $N=1$ shape modes are not. As such we can consider the splitting of a $N=2$ vortex, and monitor the dynamics of the constituent $N=1$ vortices as they separate, see \cref{fig:l2_trajectories}. One might assume that as the vortices separate, their frequencies will increase, and at some fixed distance the angular frequency will reach the continuous spectrum. This is a criterion for the existence of spectral walls, so it could be possible that the motion of the vortices is affected.

\begin{figure}
\centering
\includegraphics[width=0.6\linewidth]{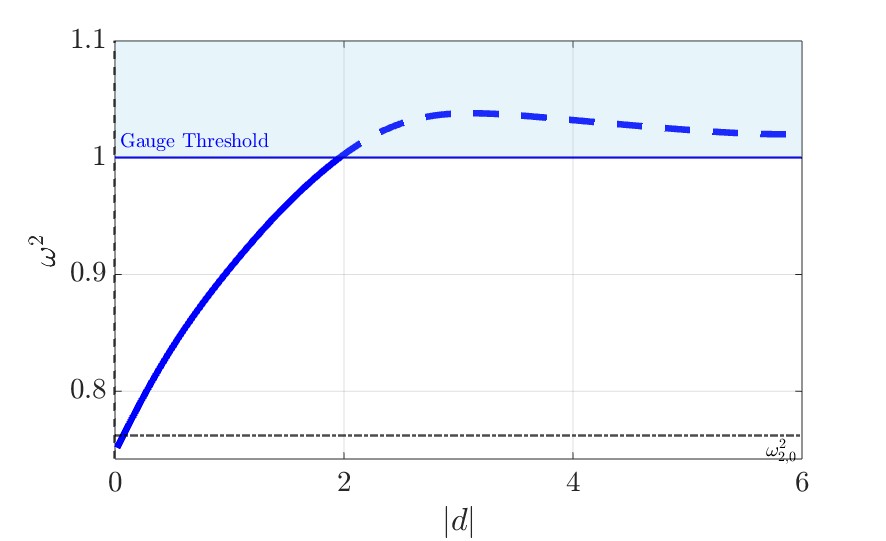}
\caption{Angular frequency flow for a two-vortex system ($\lambda = 2$) as a function of vortex separation $d$ from the origin. The blue area indicates the mass threshold.}
\label{fig:l2_sepFreq}
\end{figure}

We can see from \cref{fig:l2_trajectories} that there exist some irregularities in the position at $|d| \approx 2$. This could suggest the presence of a spectral wall. Note that due to the static repulsion, the vortices are not expected to form a quasi-stationary state, however, we do expect to see a change in velocity due to the presence of a spectral wall, which is what might be observed here. We can confirm that this irregularity is indeed an artefact of a spectral wall by tracking the angular frequency as the vortices separate, see \cref{fig:l2_sepFreq}.

We observe in \cref{fig:l2_sepFreq} that the frequency rapidly increases from the coincident configuration, $\omega_{2,0}^2 = 0.8161198$, as the vortices separate. We see that at $d\approx2$, the frequency enters the continuous spectrum. This confirms the hypothesis of the existence of a spectral wall.

\begin{figure}
\centering
\begin{minipage}{0.75\textwidth}
\includegraphics[width=\linewidth]{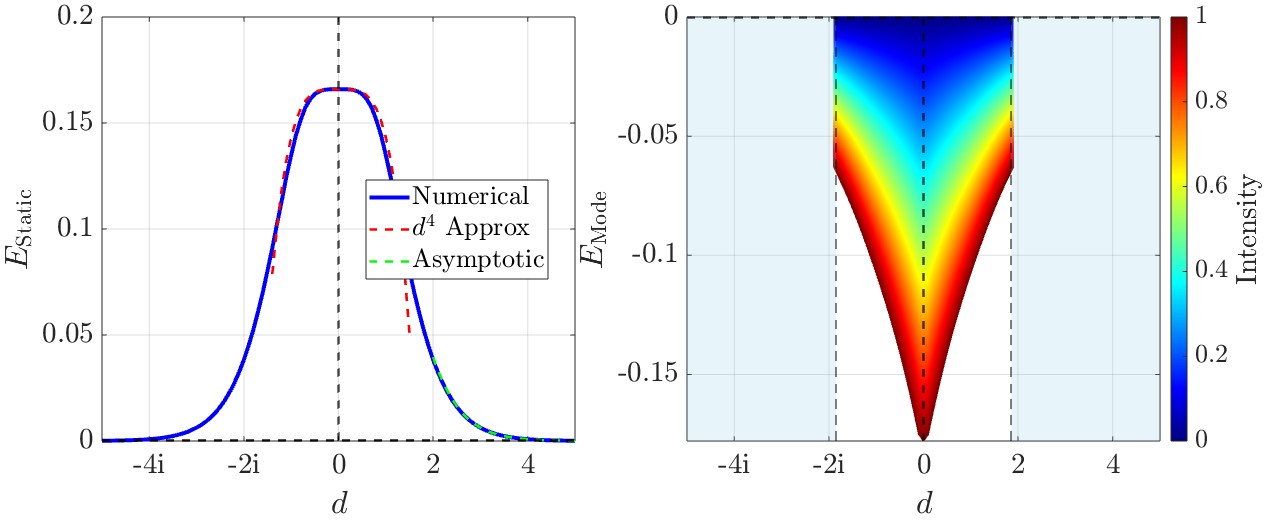}
\end{minipage}
\begin{minipage}{0.75\textwidth}
\includegraphics[width=\linewidth]{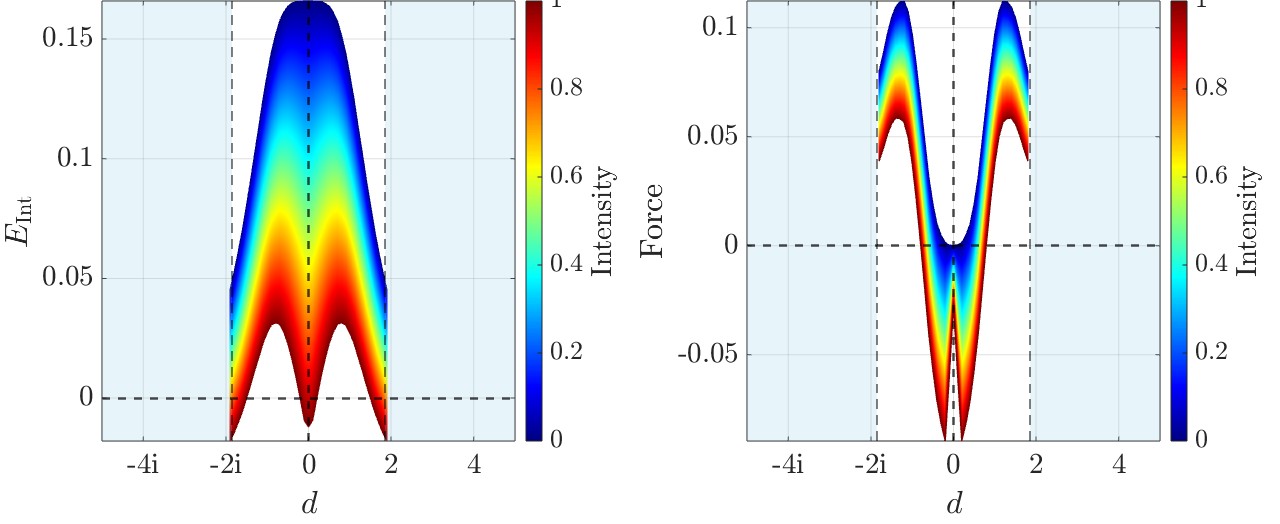}
\end{minipage}
\caption{Interaction energies and forces ($\lambda = 2$) as a function of vortex separation $d$. Top left: static force (blue) via \eqref{eq:Estatic}, with $d^4$ (dashed red, \eqref{eq:SRInt}) and asymptotic (dashed green, \eqref{eq:AsyInt}) approximations. Top right: mode interaction energy via \eqref{eq:Emode}. Bottom left: total interaction energy (static plus mode). Bottom right: total force, $F = -\frac{\partial E_{\rm int}}{\partial d}$.}
\label{fig:Int_2}
\end{figure}

Moreover, to fully understand the dynamics, we can again calculate the interaction energy, see \cref{fig:Int_2}. Firstly, that for $|d| \ll 1$ there is a critical point in the interaction energy, suggesting that the $N=2$ vortex can remain in the coincident configuration when excited. However, the excitation will quickly decay, and the vortices will separate. For well separated vortices, the frequency is in the continuous spectrum, hence all the energy from the excitation goes into the spectral wall, hence if the vortex is moving slow enough, it will bounce back towards the origin, or at least be slowed.

\section{Vortex Orbits}
\label{sec:VortexOrbits}
In this section, we discuss the full field theory dynamics for orbital vortex solutions, whereby the vortices undergo rotational motion around the origin. These orbits arise in various contexts. We expand on the previous section, whereby we will take advantage of the attractive static force in type \rom{1} superconductivity, and show that this attraction can lead to stable rotational states.

We will discuss the role of the tangential velocity in balancing the centrifugal force with the mutual attraction of the type \rom{1} vortices. Next, we will consider type \rom{2} vortices. Here, the static intervortex force is repulsive, hence we will rely on the local minimum found in previous section, and discuss solutions with rotational motion inside this potential well. 

We will again attempt to balance the attractive force with the repulsive centrifugal force to obtain stable vortex orbits. Finally, it is known that there is a strong attractive force for vortex-antivortex pairs. We will follow a similar procedure to that for type \rom{1} vortices and show that we can achieve stable orbits.

Consider a 2-vortex system with mass $m = V_1^\lambda$ per vortex, where $V_1^\lambda$ is the static energy of a single $N=1$ vortex, and the vortices have separation $s = |r_1 - r_2|$. We define the relative separation as $r = r_1 - r_2$. The reduced mass is thus $\mu = \frac{m}{2} = \frac{V_1^\lambda}{2}$.
In polar coordinates, we have that $r = (s \cos \theta, s \sin \theta)$. The relative kinetic energy is then
$$    T_{\text{rel}} = \frac{m}{4} (\dot{s}^2 + s^2 \dot{\theta}^2).$$
If we include a radial potential, then the reduced Lagrangian is then,
\begin{equation}
    L_{\rm red} = \frac{m}{4} (\dot{s}^2 + s^2 \dot{\theta}^2) - V(s).
\label{eq:Lred}
\end{equation}
The angular coordinate $ \theta $ is cyclic. Therefore, the angular momentum defined as
$$    L_z = \frac{m}{4} s^2 \dot{\theta}
    $$
is conserved. Note that $L_z = \frac{V_1^\lambda}{4} s v,$    
where $v = s \dot{\theta}$.
Varying \eqref{eq:Lred} with respect to $s$ gives,
$$    \frac{m}{4} \ddot{s} = \frac{m}{4} s \dot{\theta}^2 - \frac{dV}{ds}.$$
hence we have that,
$$    m \ddot{s} = \frac{16 L_z^2}{m s^3} - 4 \frac{dV}{ds}.$$
The centrifugal force is thus,
$$    F_{\rm centrifugal} = \frac{16 L_z^2}{V_1^\lambda s^3}.$$
We then have that the centrifugal part of the interaction energy per vortex is
$$    E_{\rm centrifugal} = \frac{L_z^2}{V_1^\lambda d^2},$$
where $s = 2d$, and $L_z = \frac{V_1^{\lambda}}{2}dv$

\subsection{Vortex Orbits at Critical Coupling}
\begin{figure}
\centering
\includegraphics[width=0.95\linewidth]{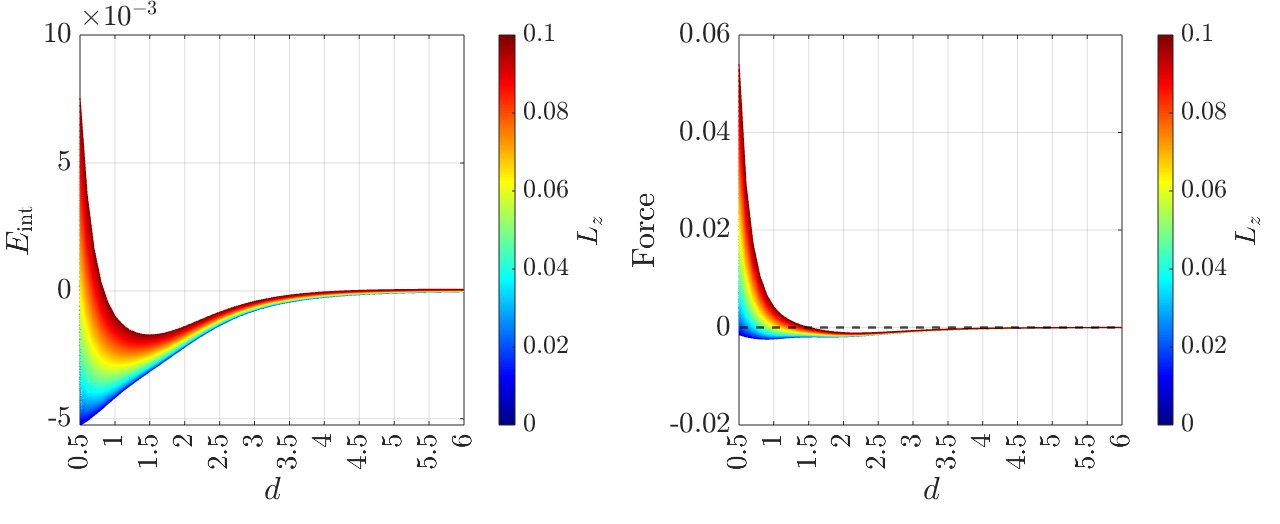}
\caption{Interaction energy (left) and total force (right) for a two-vortex system ($\lambda = 1$) with initial intensity $I(0) = 0.025$ and angular momentum $L_z = \frac{V_1^\lambda}{2} d v_{\rm in}$, as a function of orbit radius $d$.}
\label{fig:interaction_orbit_l1}
\end{figure}

We can gain some intuition for suitable orbits by studying the interaction energy, and hence the total force of the system. At critical coupling, the static force is zero, so the only competing forces are the attractive mode induced force, and the repulsive centrifugal force. We assume a fixed intensity of the excitation, and calculate the contributions to the interaction energy, see \cref{fig:interaction_orbit_l1}, where we have chosen a fixed intensity of $I(0) = 0.025$.

\begin{figure}
\centering
\includegraphics[width=0.75\linewidth]{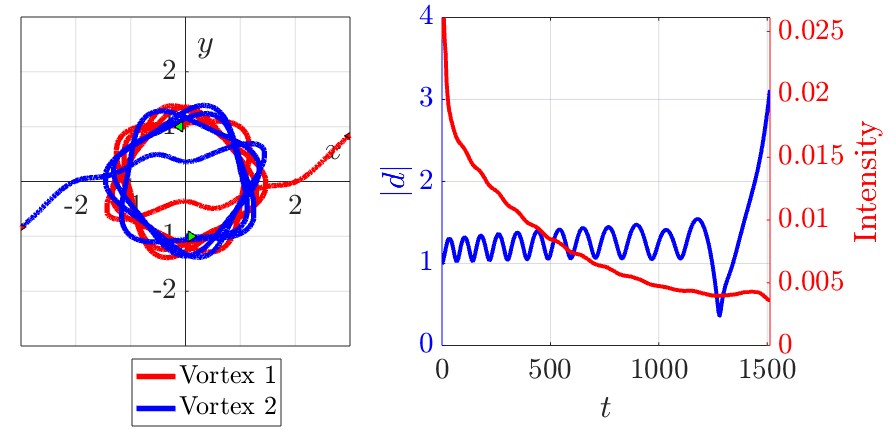}
\caption{Left: trajectories of a two-vortex system ($\lambda = 1$, $v_{\rm in} = 0.05$, $I(0) = 0.025$) at $x_2 = \pm 1$. The blue line shows the $(x_1, x_2)$ position of one vortex, red the other. Right: intensity of the excitation per vortex as a function of time (red), and distance $|d|$ of the vortices from the origin as a function of time.}
\label{fig:l1_orbit}
\end{figure}

We see from \cref{fig:interaction_orbit_l1} 
that for $I(0) = 0.025$, that the force crosses the $x$-axis at $|d| \approx 1$, i.e. there is a local minimum in the interaction energy (see left of \cref{fig:interaction_orbit_l1}). Choosing $L_z = 0.078$, we find that $v_{\rm in} \approx 0.05$. We can test these parameters in \cref{fig:l1_orbit}.

We see in \cref{fig:l1_orbit} that the vortices orbit the origin 6 times. The left plot shows the positions of the vortices in the $x_1,x_2$ plane, where blue shows the path of one vortex, and red the other vortex. The right plot shows the distance of the vortices from the origin as a function of time in blue, and the intensity of the excitation per vortex as a function of time in red. We can see from the distance of the vortices from the origin in the right plot (blue) that as the intensity of the excitation decreases, the size of the orbit increases, which is expected by observing \cref{fig:interaction_orbit_l1}. It is trivial that as the intensity of the excitation decreases, the attractive force induced by the mode also decreases, hence since the centrifugal force stays roughly the same (assuming a circular orbit), hence the local minimum where the interaction energy is zero, i.e. the force crosses the $x_1$-axis, moves out ($|d|$ increases).

We can show snapshots of the energy density for the simulation shown in \cref{fig:l1_orbit}.
\begin{figure}
\centering
\begin{minipage}{0.24\textwidth}
\centering
\includegraphics[width=\linewidth]{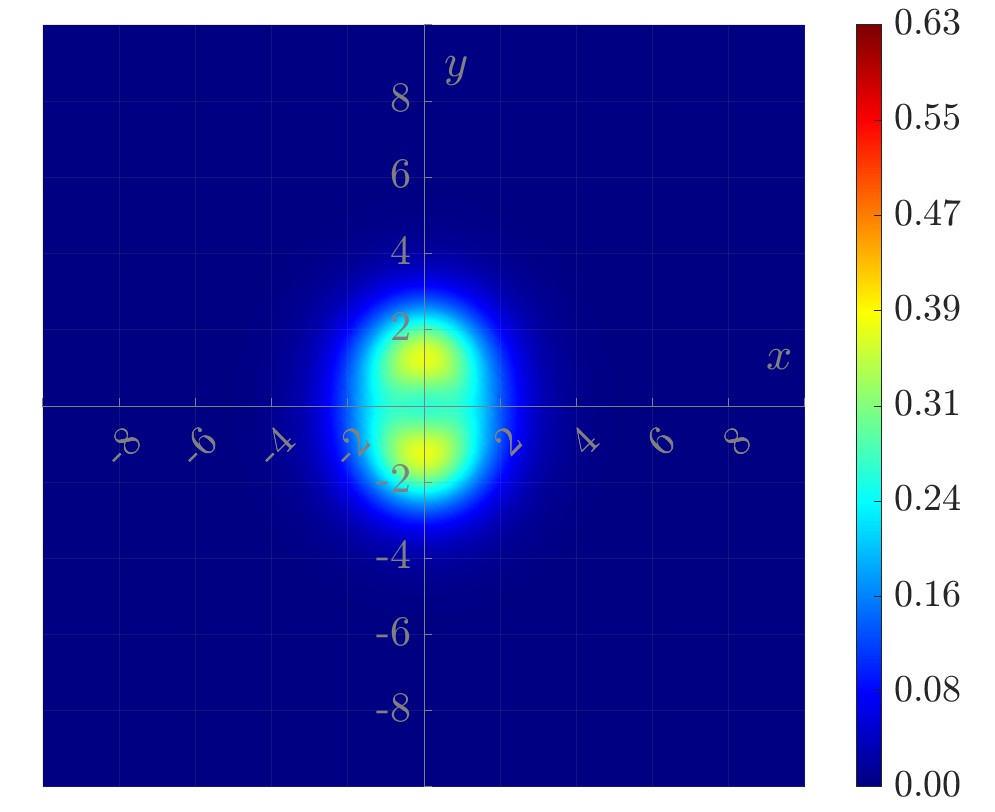}
\caption*{$t = 0$}
\end{minipage}
\begin{minipage}{0.24\textwidth}
\centering
\includegraphics[width=\linewidth]{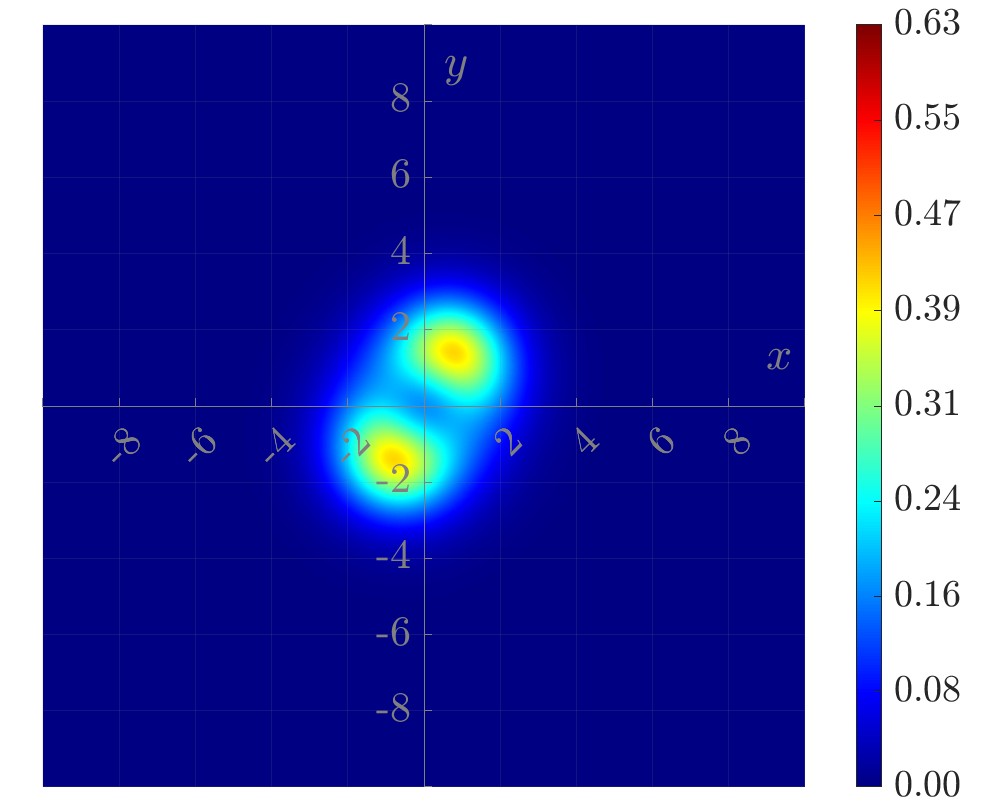}
\caption*{$t = 100$}
\end{minipage}
\begin{minipage}{0.24\textwidth}
\centering
\includegraphics[width=\linewidth]{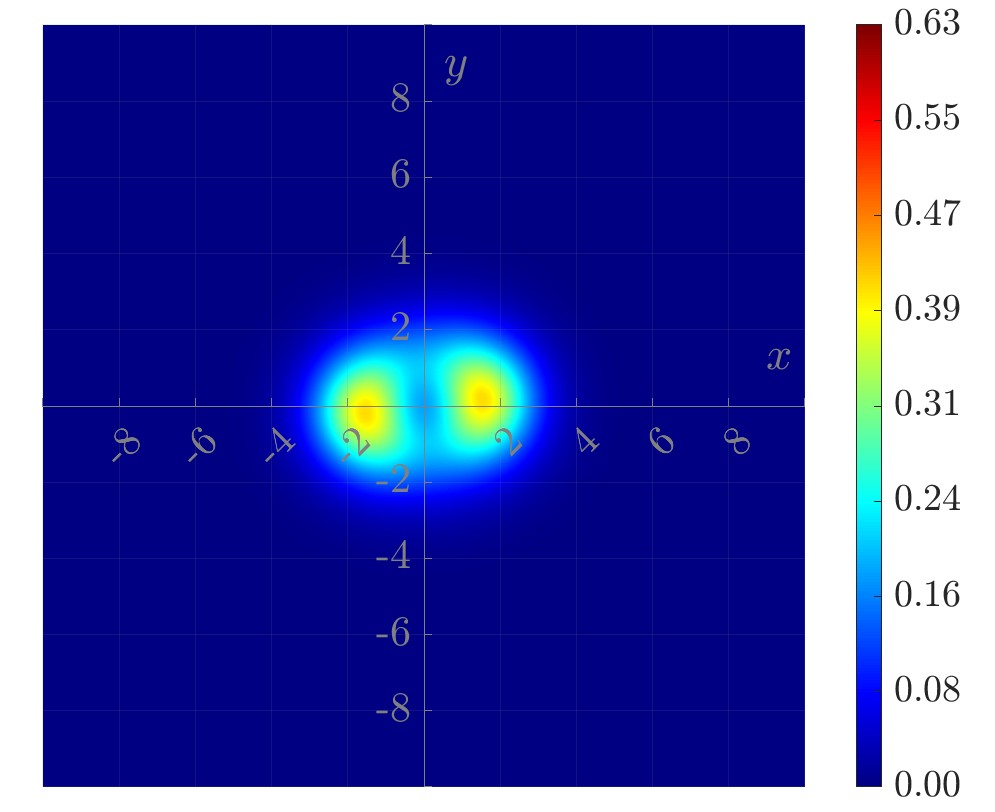}
\caption*{$t = 200$}
\end{minipage}
\begin{minipage}{0.24\textwidth}
\centering
\includegraphics[width=\linewidth]{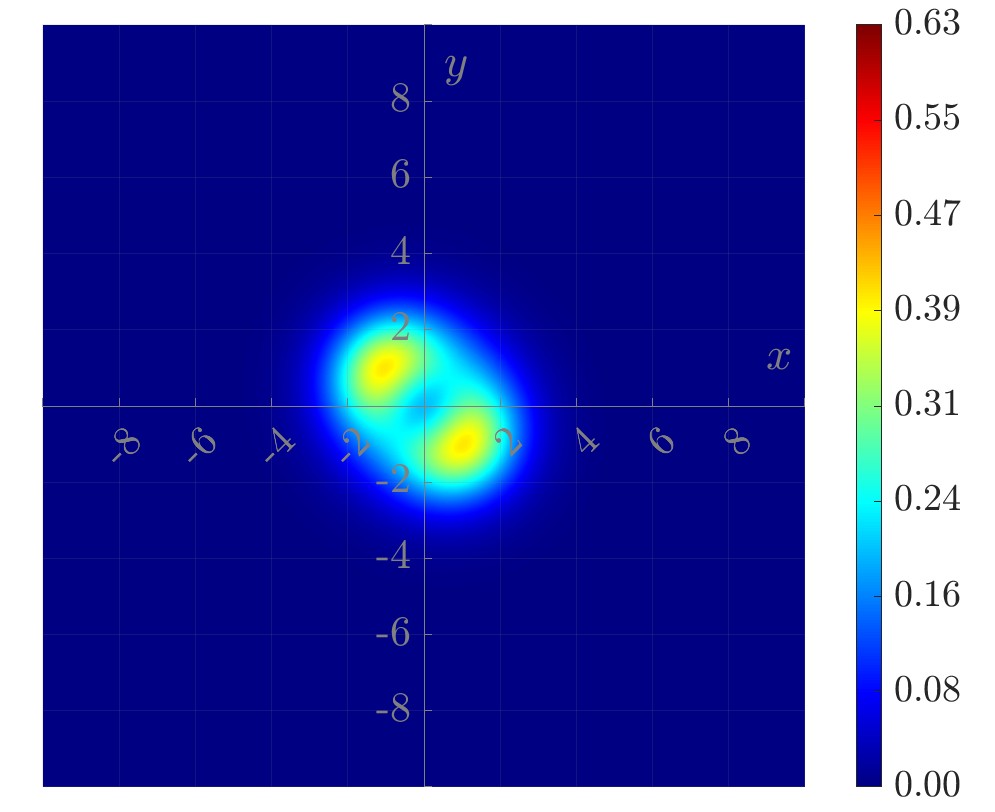}
\caption*{$t = 300$}
\end{minipage}
\begin{minipage}{0.24\textwidth}
\centering
\includegraphics[width=\linewidth]{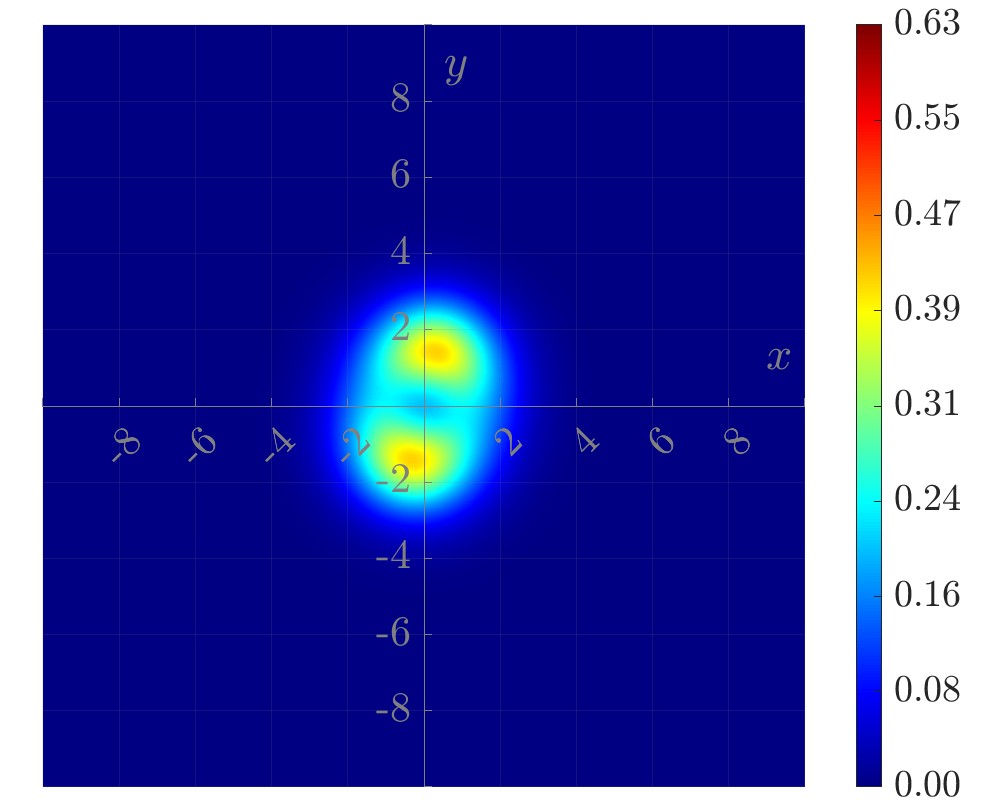}
\caption*{$t = 400$}
\end{minipage}
\begin{minipage}{0.24\textwidth}
\centering
\includegraphics[width=\linewidth]{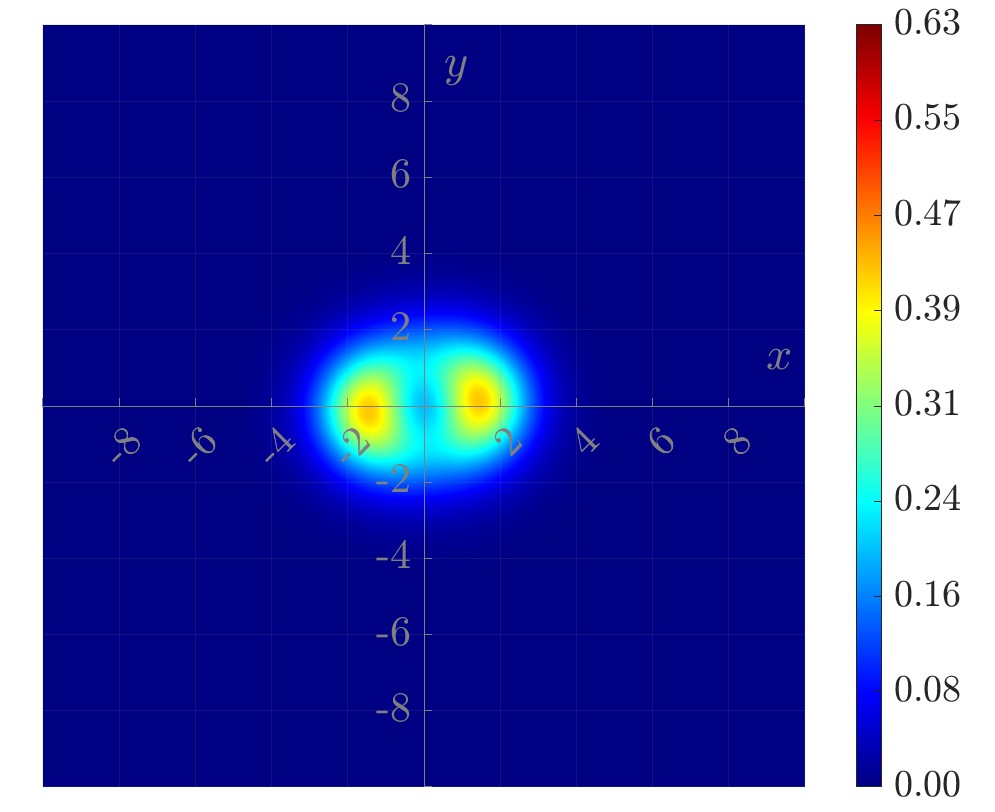}
\caption*{$t = 500$}
\end{minipage}
\begin{minipage}{0.24\textwidth}
\centering
\includegraphics[width=\linewidth]{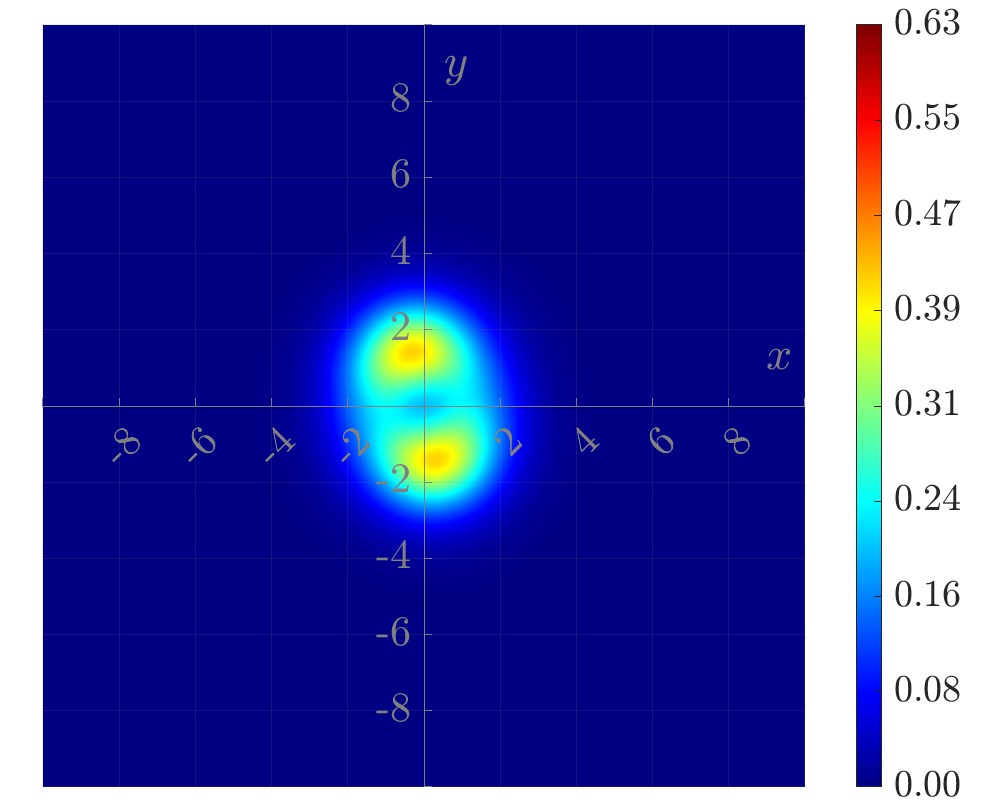}
\caption*{$t = 600$}
\end{minipage}
\begin{minipage}{0.24\textwidth}
\centering
\includegraphics[width=\linewidth]{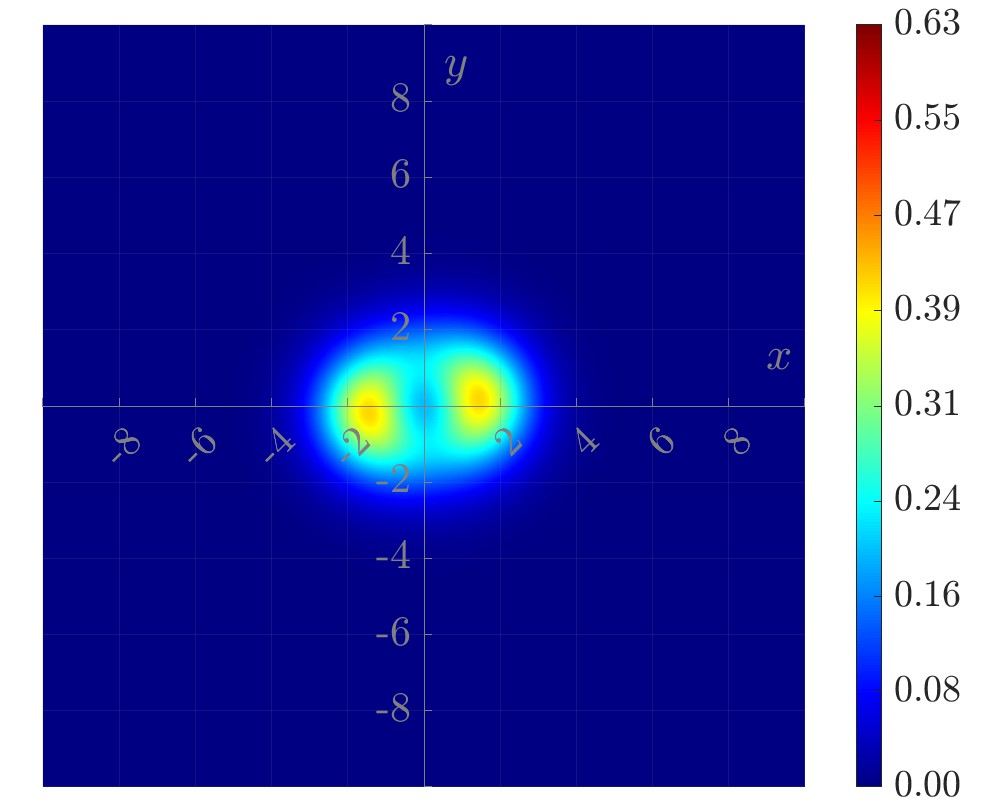}
\caption*{$t = 700$}
\end{minipage}
\begin{minipage}{0.24\textwidth}
\centering
\includegraphics[width=\linewidth]{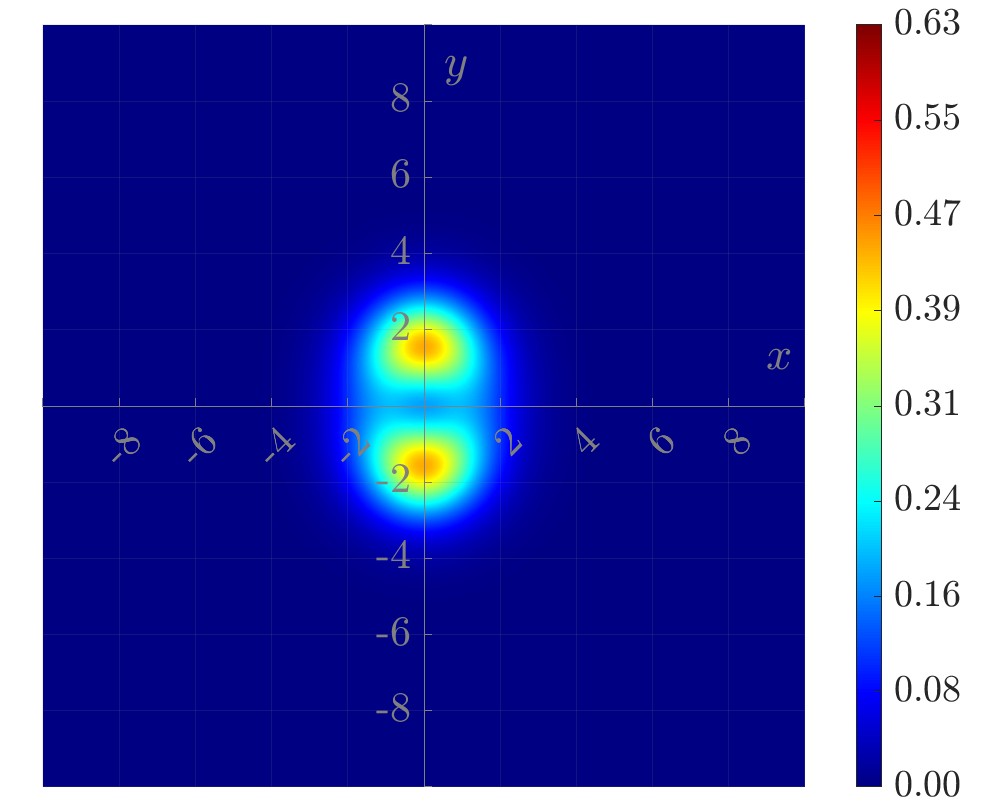}
\caption*{$t = 800$}
\end{minipage}
\begin{minipage}{0.24\textwidth}
\centering
\includegraphics[width=\linewidth]{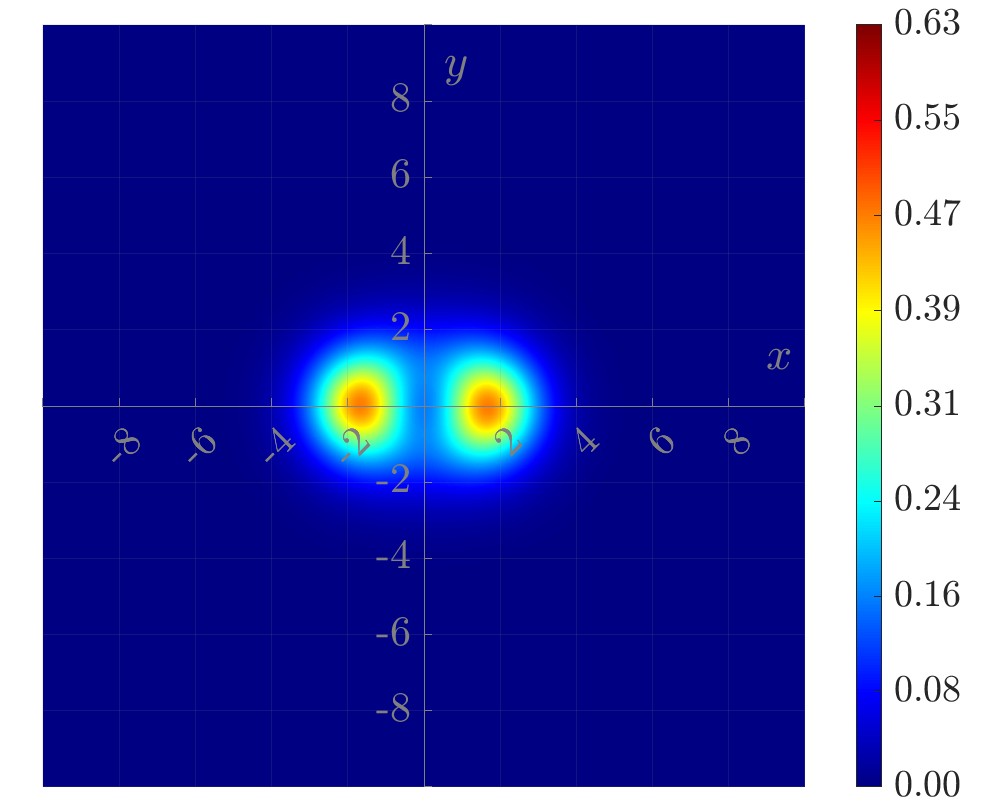}
\caption*{$t = 900$}
\end{minipage}
\begin{minipage}{0.24\textwidth}
\centering
\includegraphics[width=\linewidth]{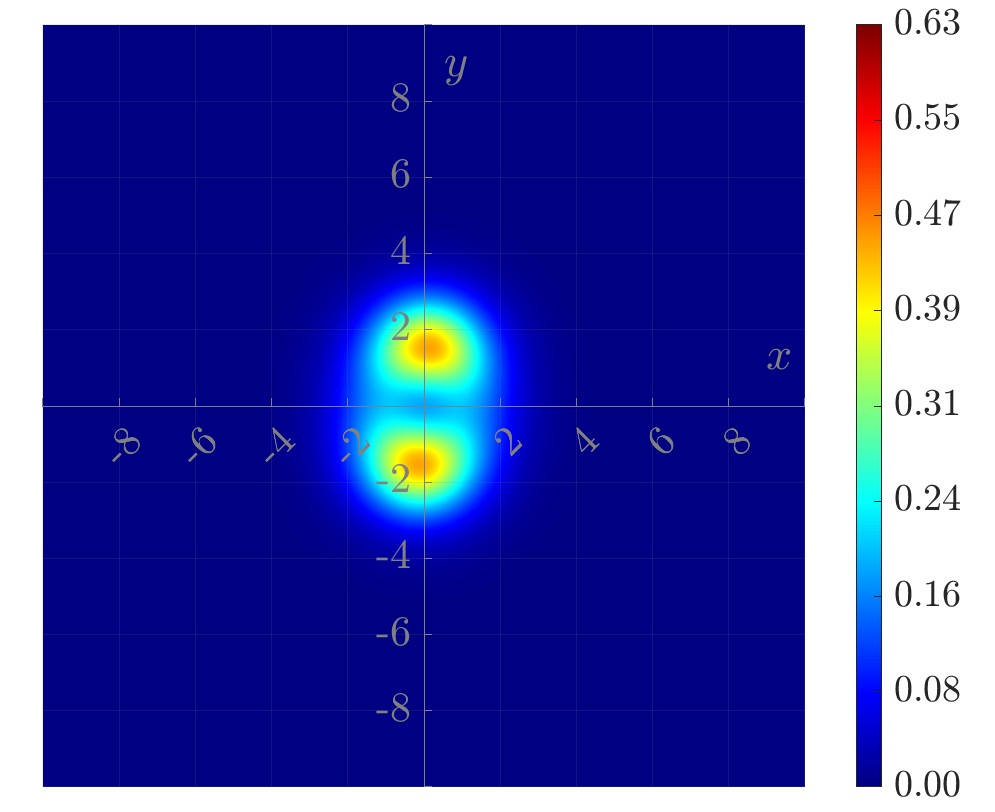}
\caption*{$t = 1000$}
\end{minipage}
\begin{minipage}{0.24\textwidth}
\centering
\includegraphics[width=\linewidth]{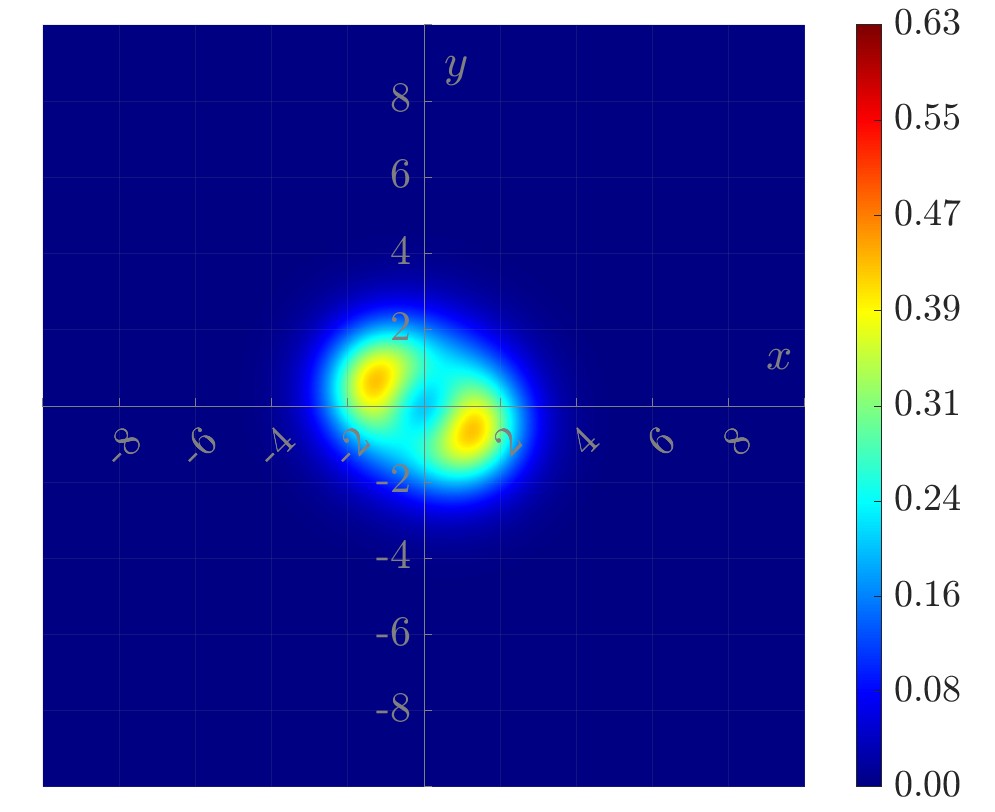}
\caption*{$t = 1100$}
\end{minipage}
\begin{minipage}{0.24\textwidth}
\centering
\includegraphics[width=\linewidth]{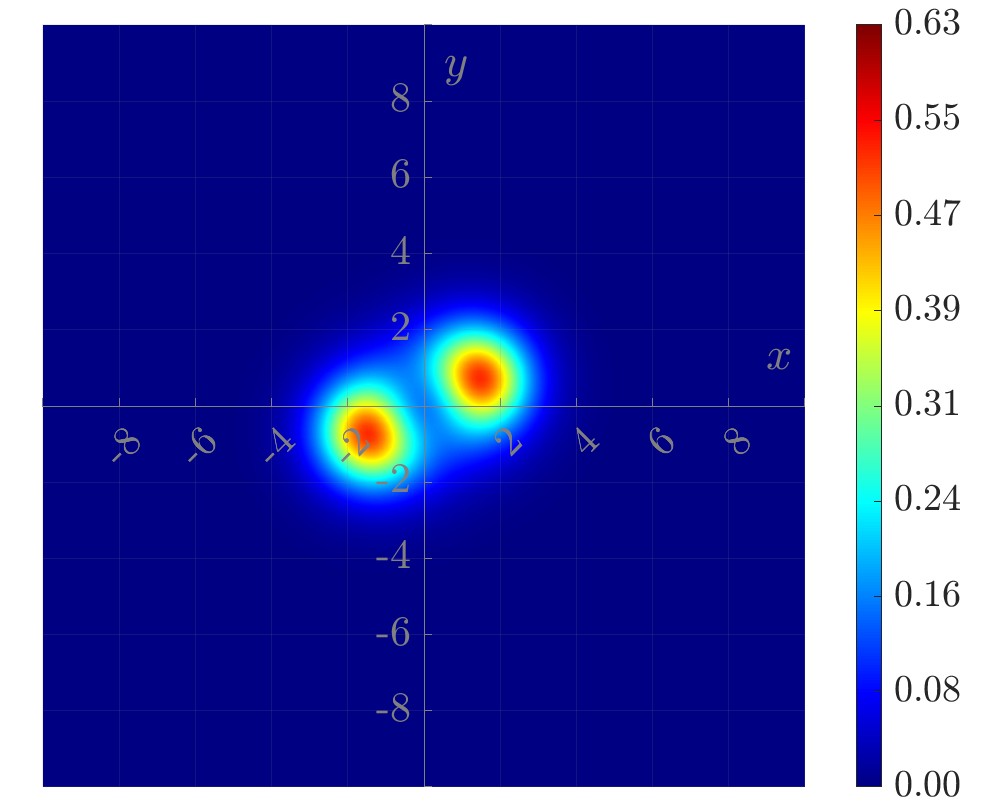}
\caption*{$t = 1200$}
\end{minipage}
\begin{minipage}{0.24\textwidth}
\centering
\includegraphics[width=\linewidth]{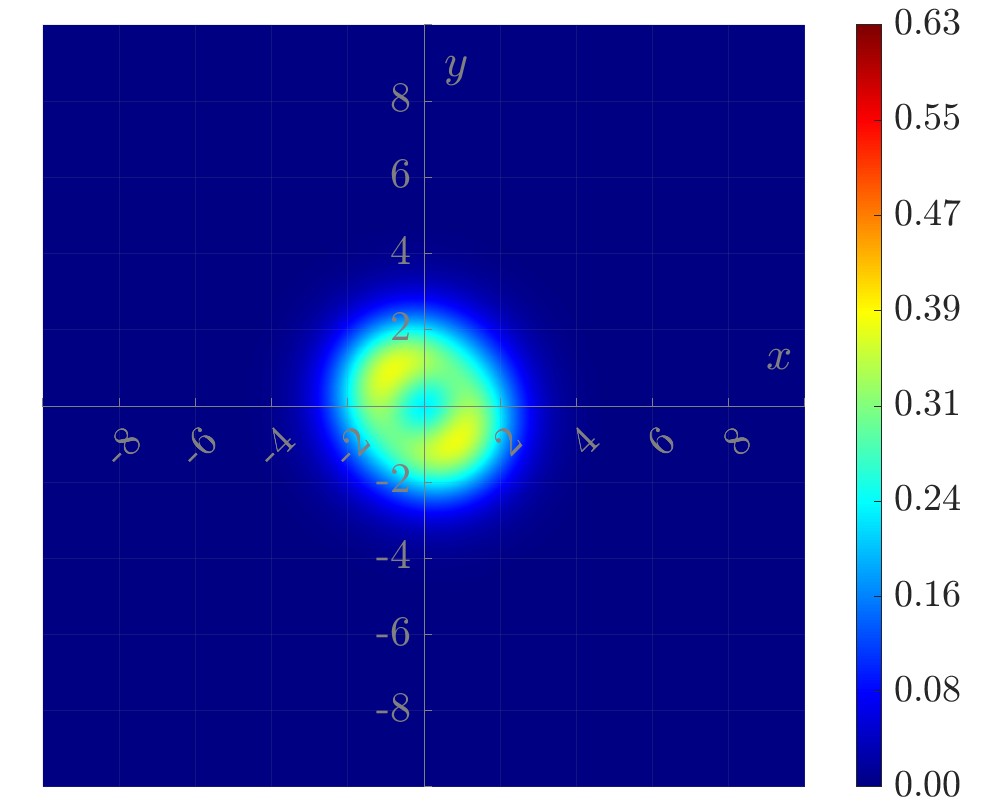}
\caption*{$t = 1300$}
\end{minipage}
\begin{minipage}{0.24\textwidth}
\centering
\includegraphics[width=\linewidth]{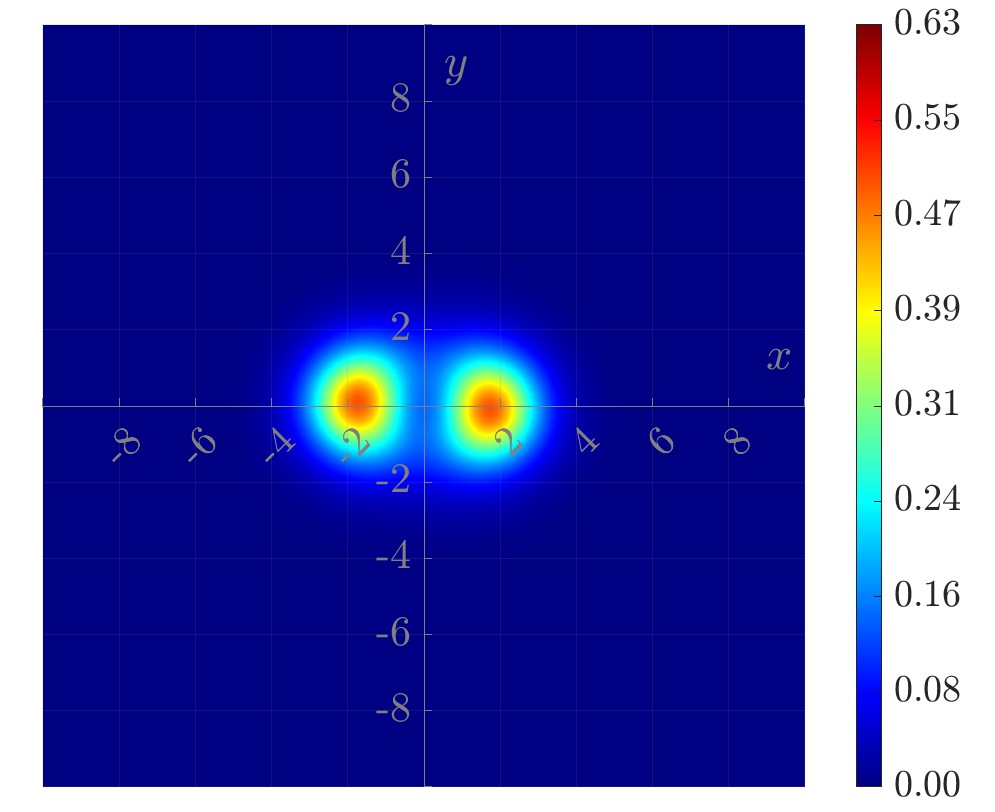}
\caption*{$t = 1400$}
\end{minipage}
\begin{minipage}{0.24\textwidth}
\centering
\includegraphics[width=\linewidth]{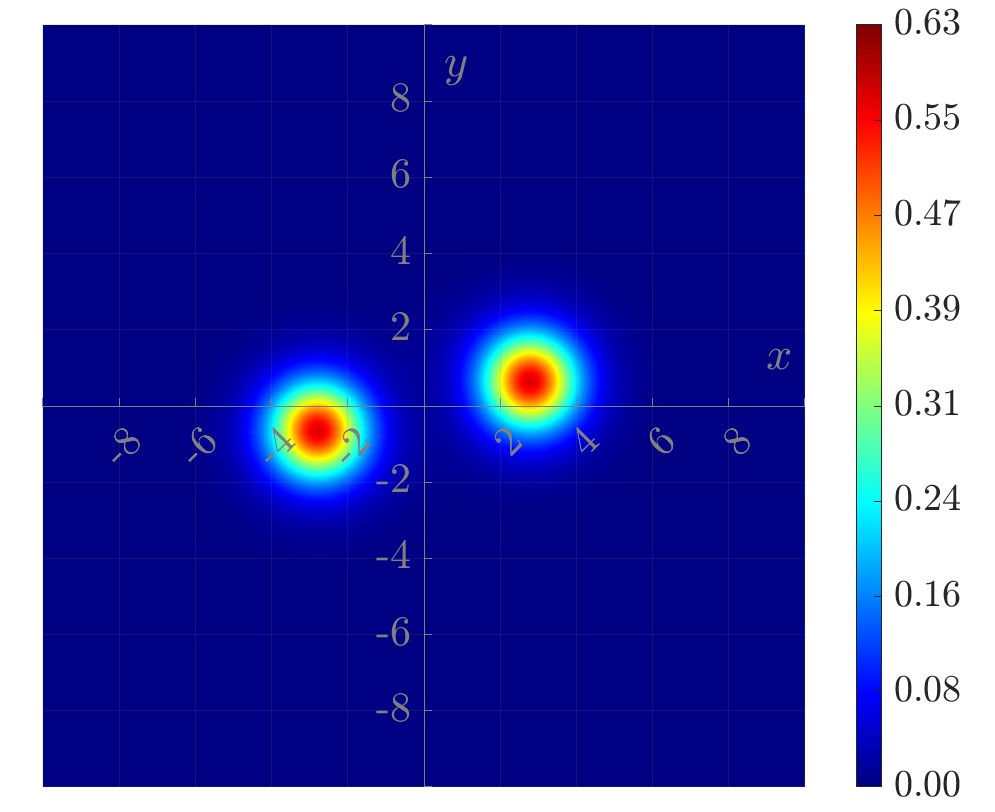}
\caption*{$t = 1500$}
\end{minipage}
\caption{Snapshots of energy density for a two-vortex scattering ($\lambda = 1$, $v_{\rm in} = 0.05$, $I(0) = 0.05$) at $d = \pm 1$, as seen in \cref{fig:l1_orbit}.}
\label{fig:l1_Orbitsnapshot}
\end{figure}

We see in \cref{fig:l1_Orbitsnapshot} that the vortices orbit the origin, whilst oscillating in shape, as well as oscillating about the local minimum in the interaction energy. The vortices orbit for a significantly long time, up to $t = 1500$, at which point the vortices escape after the orbit becomes unstable and the vortices pass close to the origin.

\subsection{Type \texorpdfstring{\rom{1}}{I} Vortex Orbits}
\label{sec:TypeIVortexOrbits}
When considering orbital vortices, it is first natural to consider the case of type \rom{1} vortices, without excitation. Here, the only competing forces are the attractive static force and the repulsive centrifugal force. If the tangential velocity is small, then the vortices will be drawn towards the centre. If the velocity is large, the kinetic energy of the vortices will dominate the interaction and they will escape to infinity. For a fine-tuned velocity, dependent on the orbit size, the vortices may form a stable circular orbit.

\begin{figure}
\centering
\includegraphics[width=0.9\linewidth]{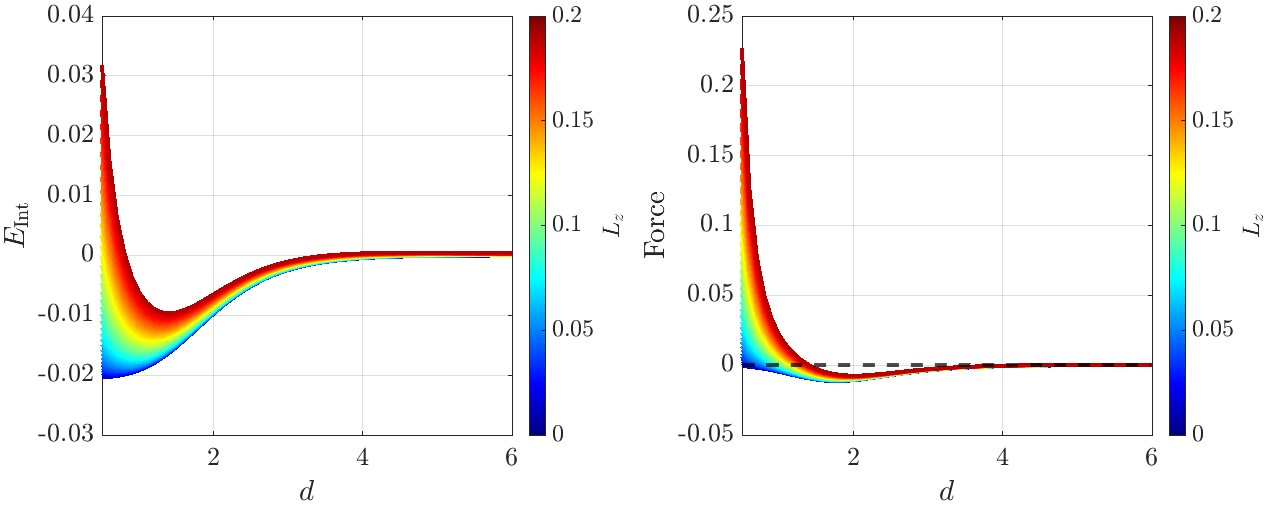}
\caption{Sum of static and centrifugal energies (left) and total force (right) as a function of vortex separation $d$ for varying velocities ($\lambda = 0.9$).}
\label{fig:l09_static_centrifugal_zoom}
\end{figure}

We can plot the interaction energy in this case, taking into account only that static force and centrifugal force, see \cref{fig:l09_static_centrifugal_zoom}.

We can see from \cref{fig:l09_static_centrifugal_zoom} that for $d > 4$, the net force is zero, where the force is the gradient of the interaction energy. Furthermore, we observe a critical point of the interaction energy for $d\in[0.5,1.5]$. We can see from the force that we should be able to obtain vortex orbits for small $d$ where the net force is zero. If the velocity is small, the static force will dominate and the vortices will accelerate towards the origin.

\begin{figure}
\centering
\begin{minipage}{0.4\textwidth}
\vspace{1.1em}
\includegraphics[width=\linewidth]{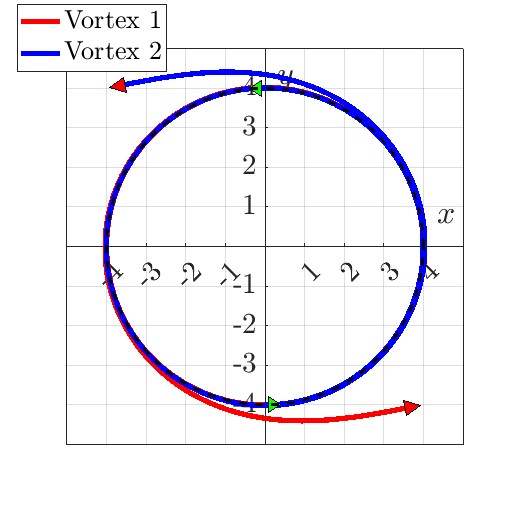}
\caption{Trajectories of a two-vortex system ($\lambda = 0.9$, $v_{\rm in} = 0.03230815$, $I(0) = 0$) orbiting at $d = \pm 4$. The blue line shows the $(x_1, x_2)$ position of one vortex, and red the other.}
\label{fig:l0.9_orbit2}
\end{minipage}
\hfill
\begin{minipage}{0.4\textwidth}
\includegraphics[width=\linewidth]{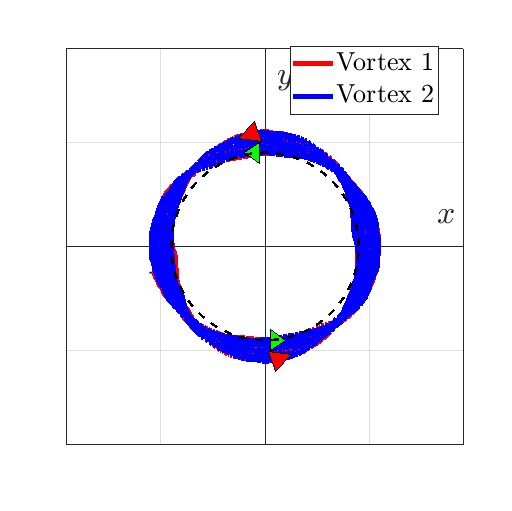}
\caption{Trajectories of a two-vortex system ($\lambda = 0.9$, $v_{\rm in} = 0.045$, $I(0) = 0$) orbiting at $d = \pm 0.9$. The blue line shows the $(x_1, x_2)$ position of one vortex, and red the other.}
\label{fig:l0.9_orbit3}
\end{minipage}
\end{figure}

We can predict the initial velocity required for the vortices to attempt to achieve a more circular orbit; see \cref{fig:l0.9_orbit2}. We can calculate the expected tangential velocity of $v = 0.038246$, at a fixed distance of $d = 4$.

\begin{figure}
\centering
\begin{minipage}{0.24\textwidth}
\centering
\includegraphics[width=\linewidth]{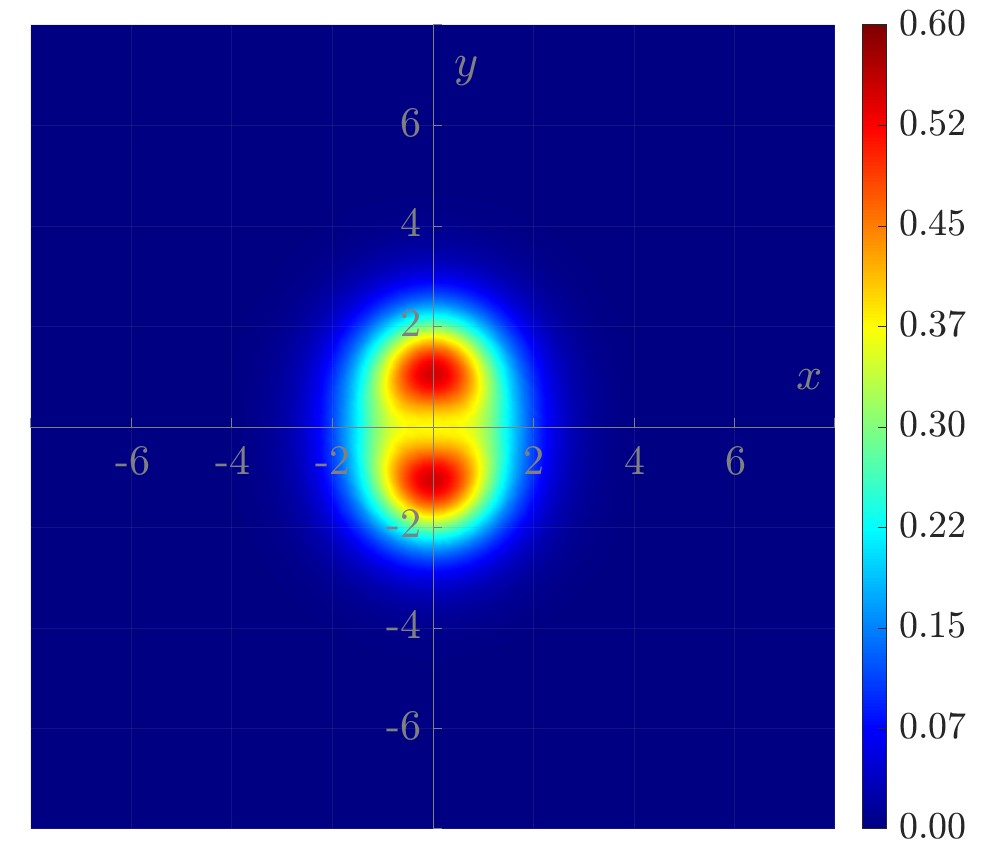}
\caption*{$t = 0$}
\end{minipage}
\begin{minipage}{0.24\textwidth}
\centering
\includegraphics[width=\linewidth]{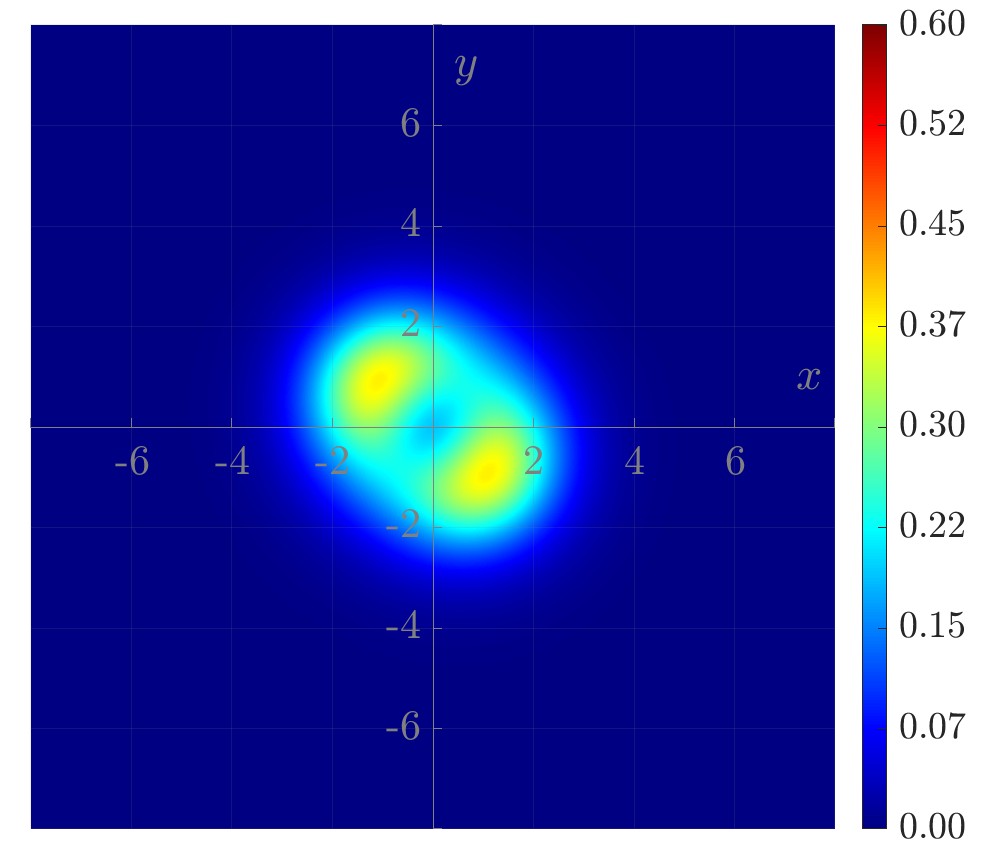}
\caption*{$t = 15$}
\end{minipage}
\begin{minipage}{0.24\textwidth}
\centering
\includegraphics[width=\linewidth]{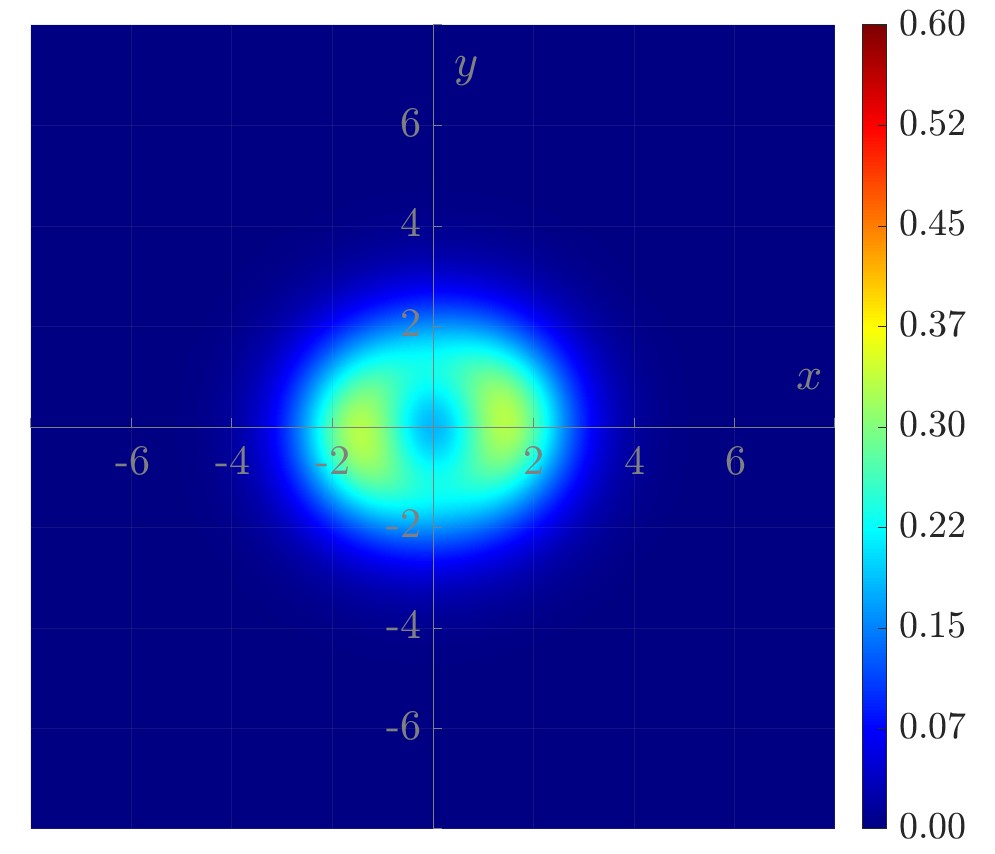}
\caption*{$t = 30$}
\end{minipage}
\begin{minipage}{0.24\textwidth}
\centering
\includegraphics[width=\linewidth]{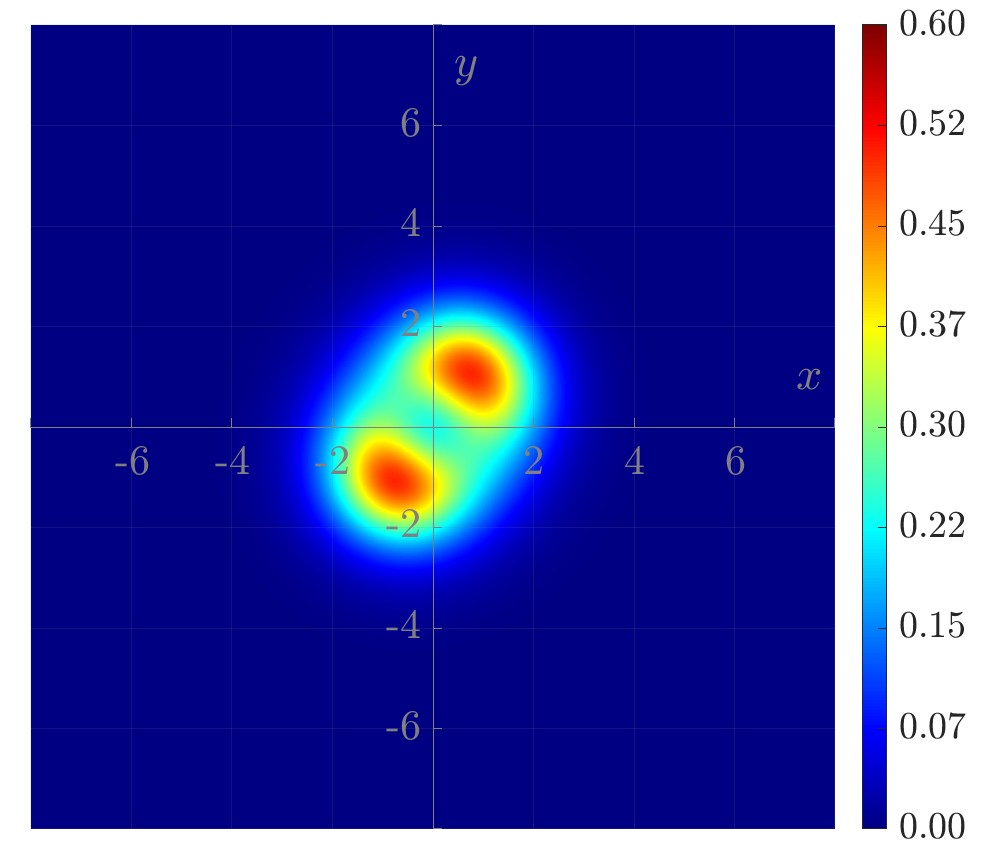}
\caption*{$t = 45$}
\end{minipage}
\begin{minipage}{0.24\textwidth}
\centering
\includegraphics[width=\linewidth]{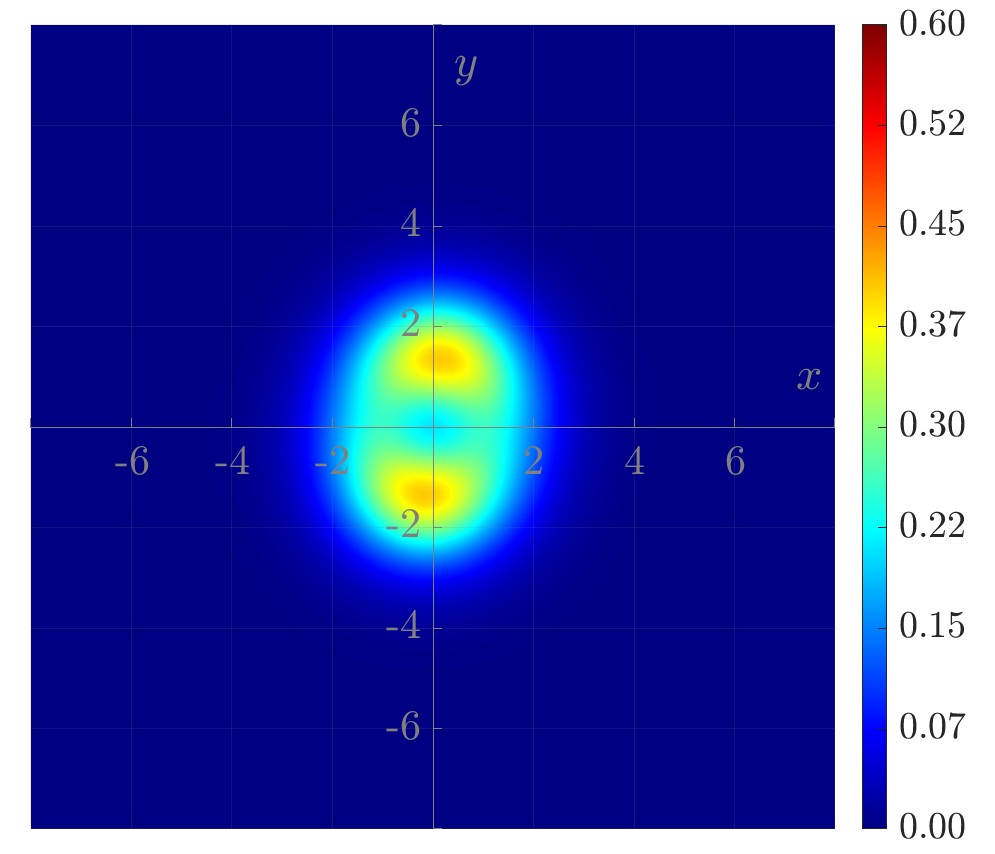}
\caption*{$t = 55$}
\end{minipage}
\begin{minipage}{0.24\textwidth}
\centering
\includegraphics[width=\linewidth]{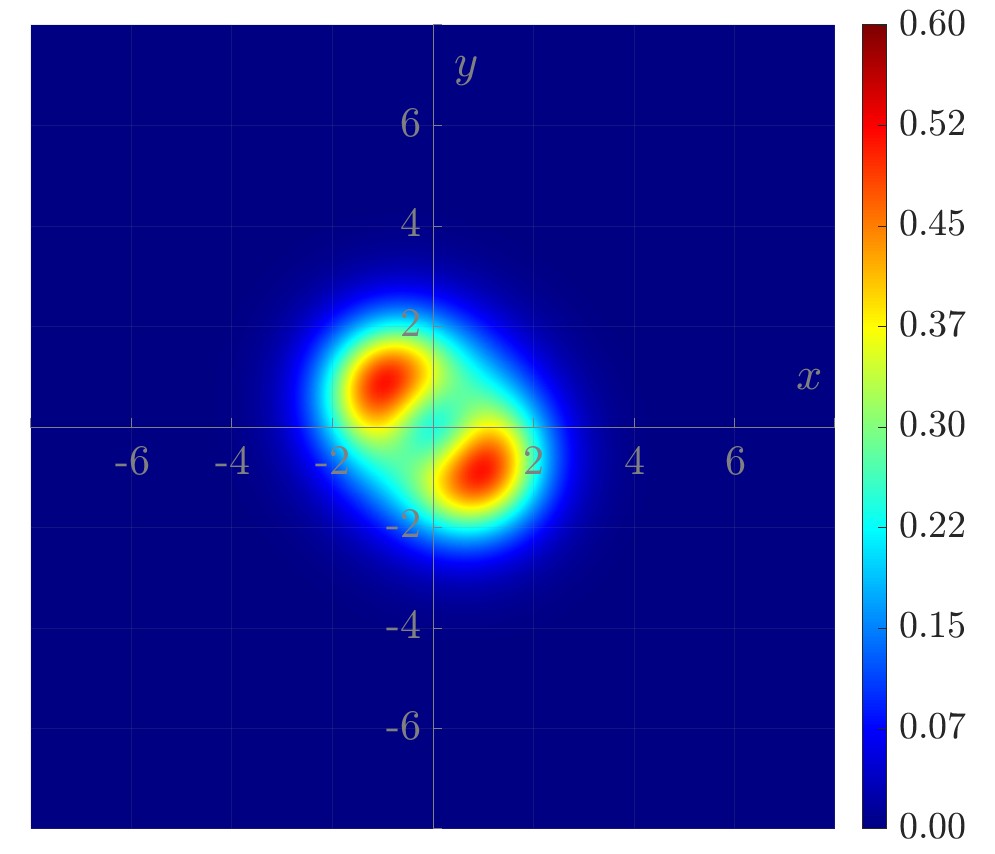}
\caption*{$t = 70$}
\end{minipage}
\begin{minipage}{0.24\textwidth}
\centering
\includegraphics[width=\linewidth]{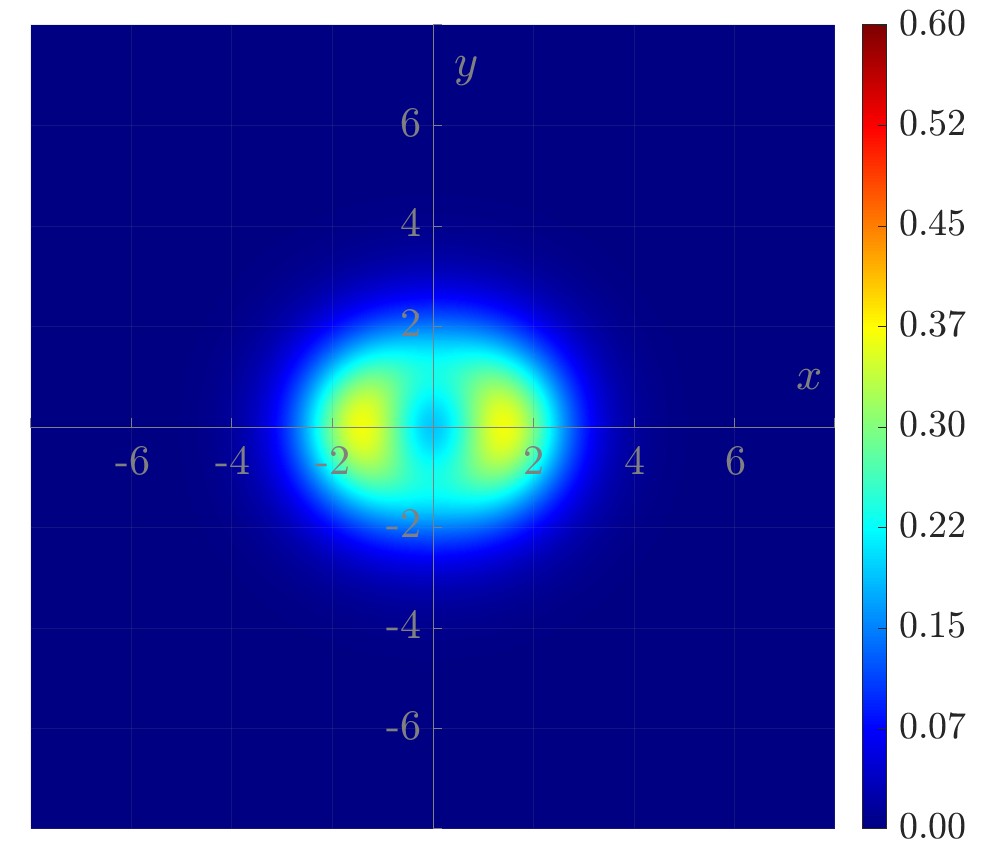}
\caption*{$t = 85$}
\end{minipage}
\begin{minipage}{0.24\textwidth}
\centering
\includegraphics[width=\linewidth]{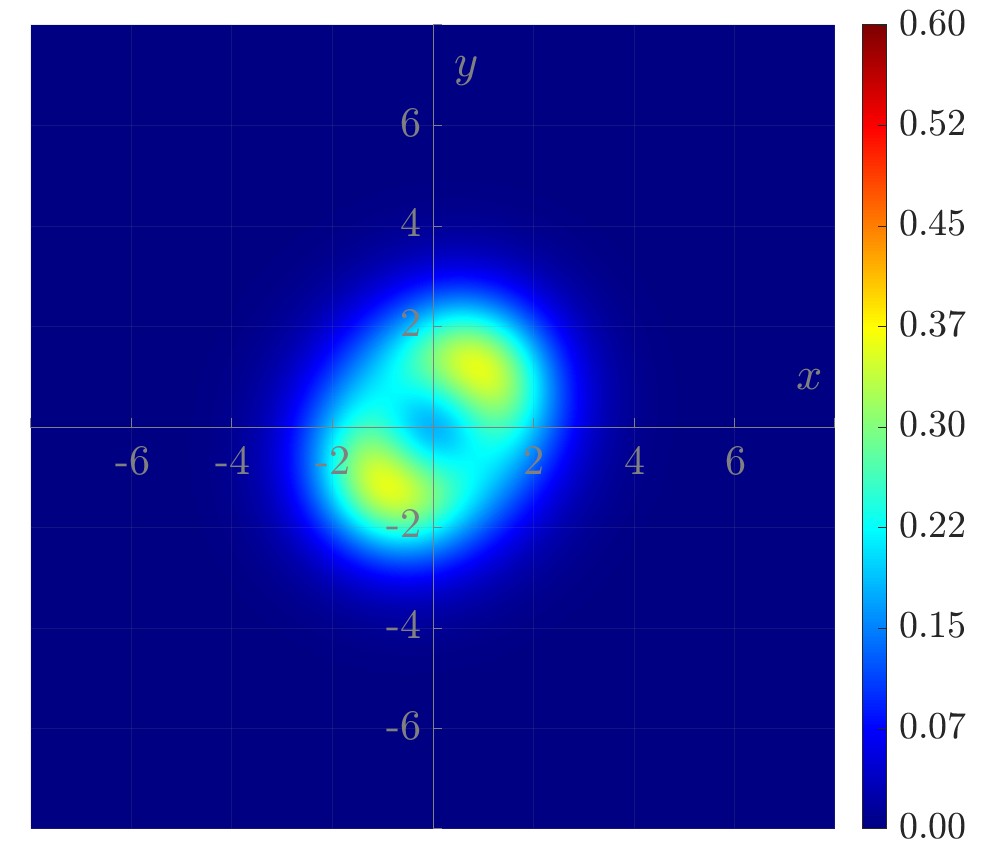}
\caption*{$t = 100$}
\end{minipage}
\begin{minipage}{0.24\textwidth}
\centering
\includegraphics[width=\linewidth]{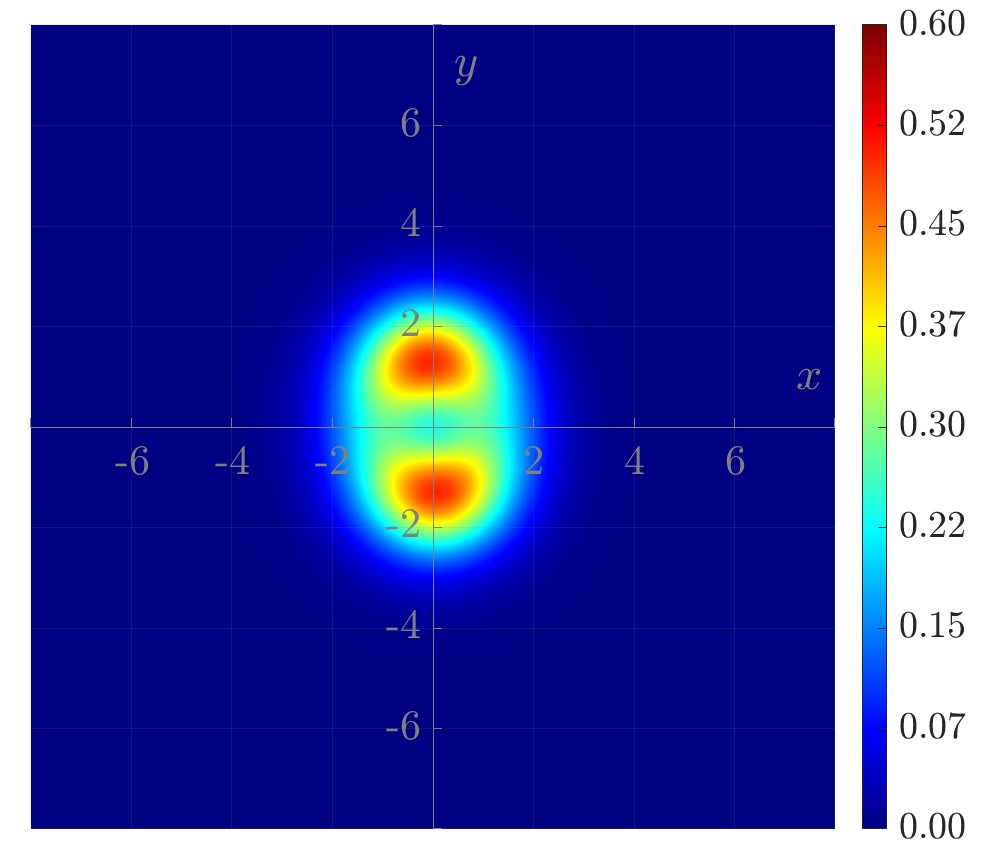}
\caption*{$t = 115$}
\end{minipage}
\begin{minipage}{0.24\textwidth}
\centering
\includegraphics[width=\linewidth]{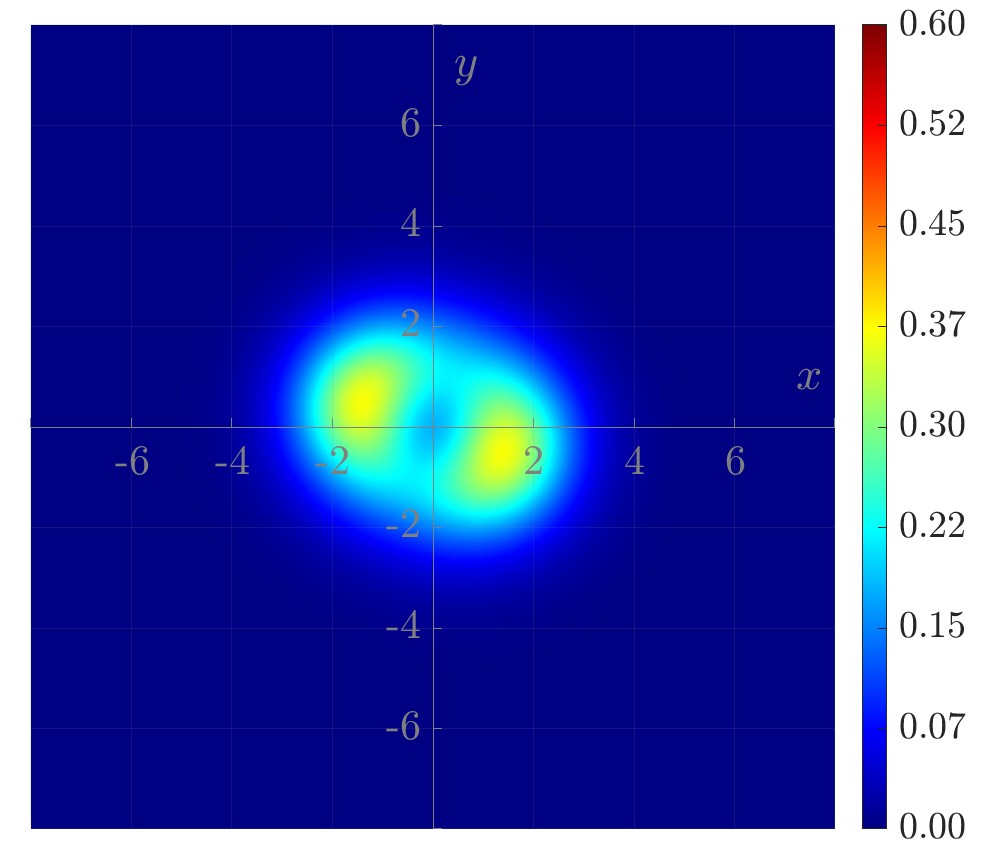}
\caption*{$t = 135$}
\end{minipage}
\begin{minipage}{0.24\textwidth}
\centering
\includegraphics[width=\linewidth]{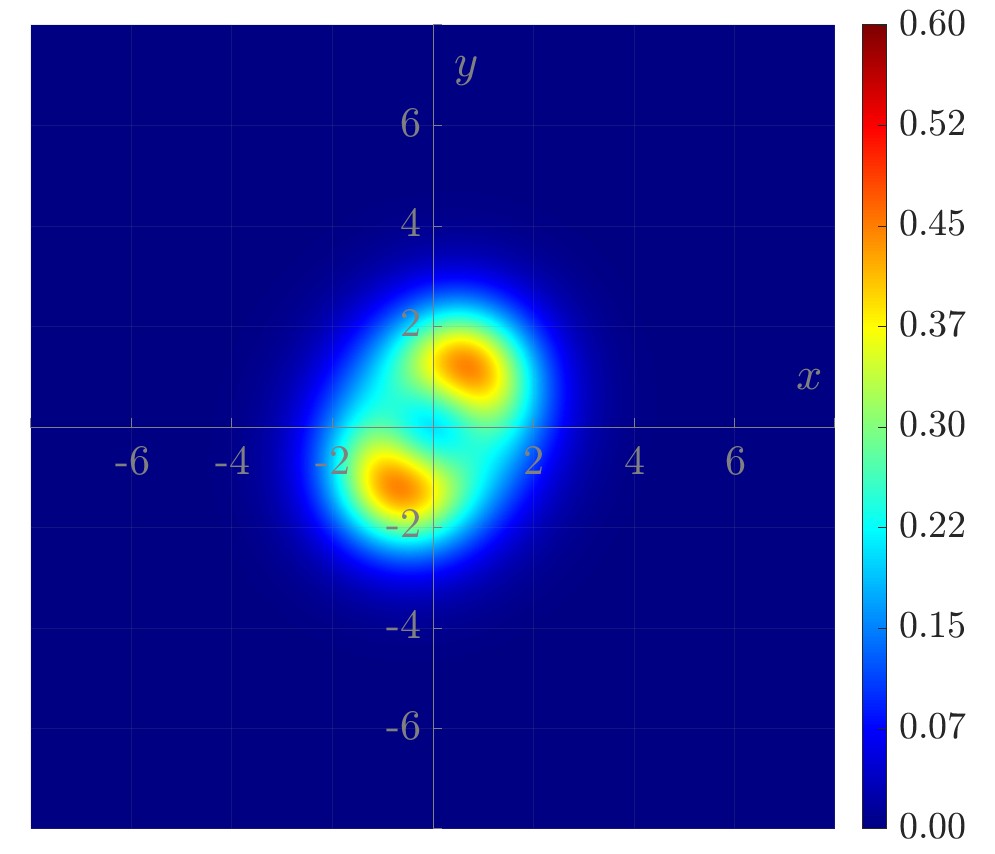}
\caption*{$t = 165$}
\end{minipage}
\begin{minipage}{0.24\textwidth}
\centering
\includegraphics[width=\linewidth]{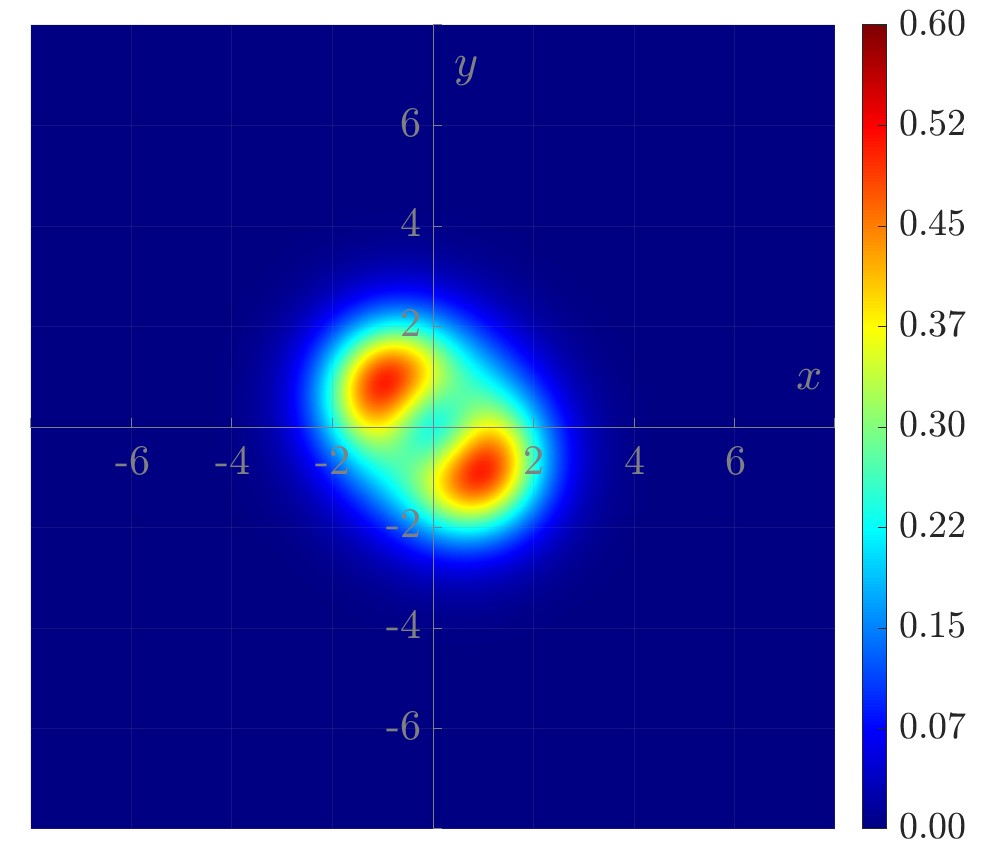}
\caption*{$t = 190$}
\end{minipage}
\caption{Snapshots of energy density for a two-vortex scattering ($\lambda = 0.9$, $v_{\rm in} = 0.045$, $I(0) = 0$) at $d = \pm 0.9$, as seen in \cref{fig:l0.9_orbit3}.}
\label{fig:l0.9_Orbitsnapshot2}
\end{figure}

We see in \cref{fig:l0.9_orbit2} a stable circular orbit in which the vortices orbit the origin multiple times before escaping to infinity.
Furthermore, we can observe vortex orbits at the critical point in the interaction energy, where $d\in[0.5,1.5]$, see \cref{fig:l0.9_orbit3}. We see that the vortices seem to orbit each other more than ten times (up to when the numerics fail). The orbit is not perfectly circular. This could be due to the vortex oscillating in the potential well around the critical point in the interaction energy. Additionally, it could be a numerical artefact as a result of the vortices being initially positioned close together, such that the initial configuration is not a perfect approximation to the field theory. We can plot dynamical snapshots of the same simulation \cref{fig:l0.9_orbit3}, displaying the Higgs field $|\phi|^2$ as a heat plot; see \cref{fig:l0.9_Orbitsnapshot2}.

\subsection{Type \texorpdfstring{\rom{2}}{II} Vortex Orbits}
\label{sec:TypeIIVortexOrbits}
We have shown in \cref{fig:Int_11} that there exists a local minimum whereby the vortices can form a quasi-stationary state at a fixed distance away from the origin. This motivates us to consider type \rom{2} vortices that orbit the origin.

The interaction energy will differ slightly, since not only do we still have the repulsive static force and the attractive mode interaction, but we will also have the centrifugal force. We can add the centrifugal force to the interaction energy, and approximate the ideal velocity for a circular orbit, given a fixed intensity.

\begin{figure}
\centering
\begin{minipage}{0.4\textwidth}
\centering
\includegraphics[width=\linewidth]{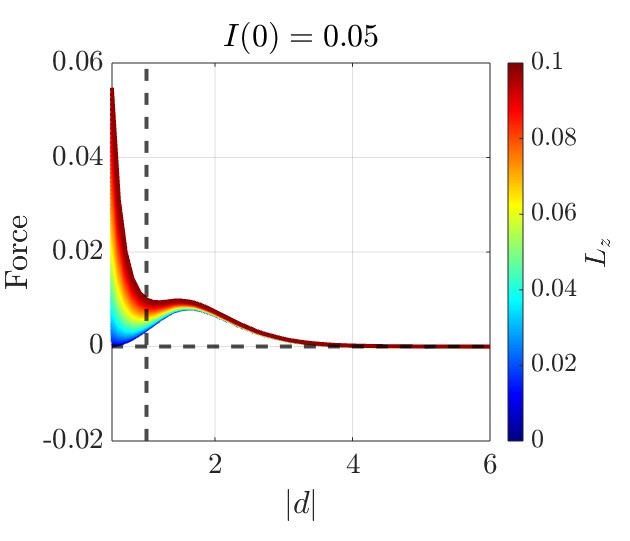}
\end{minipage}
\begin{minipage}{0.4\textwidth}
\centering
\includegraphics[width=\linewidth]{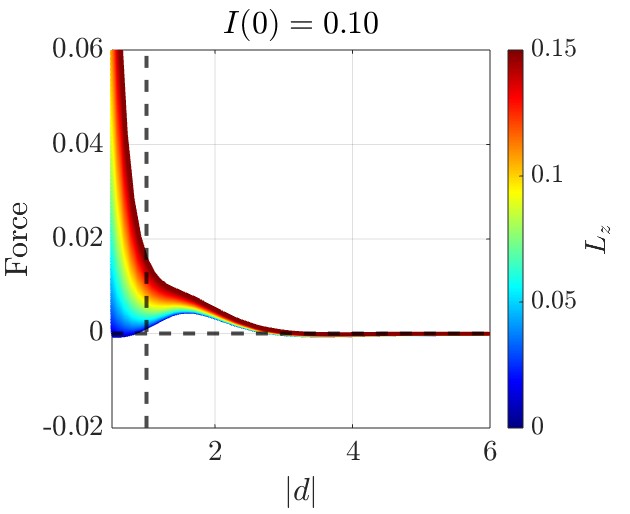}
\end{minipage}
\begin{minipage}{0.4\textwidth}
\centering
\includegraphics[width=\linewidth]{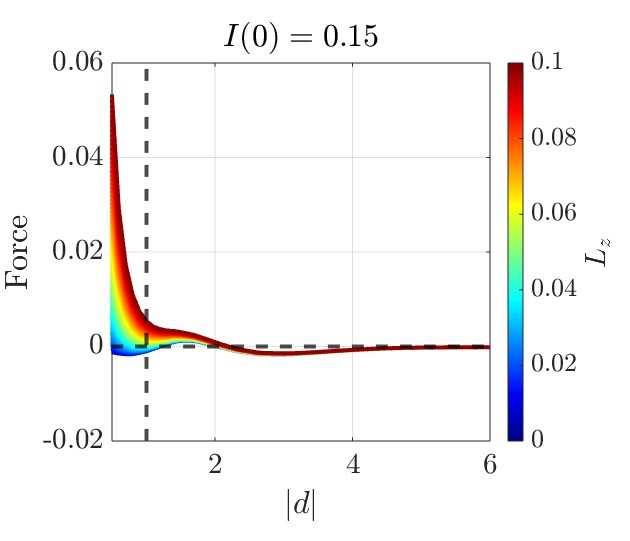}
\end{minipage}
\begin{minipage}{0.4\textwidth}
\centering
\includegraphics[width=\linewidth]{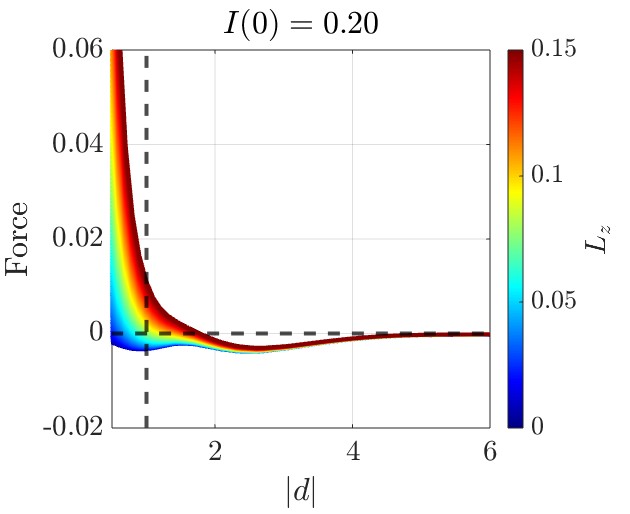}
\end{minipage}
\caption{Total force for a two-vortex system ($\lambda = 1.1$) with intensities $I(0) = 0.1, 0.2, 0.3, 0.4$ and angular momentum $L_z = \frac{V_1^\lambda}{2} d v_{\rm in}$, as a function of orbit radius $d$.}
\label{fig:interaction_orbit}
\end{figure}

In our dynamical simulations, the vortices are not initially well separated, and hence the initial condition is not perfect. As such, the intensity will drop as the solution flows to the correct solution, which is why we have chosen a sample of intensities $I(0) = [0.05, 0.1, 0.15, 0.2]$.

We see in \cref{fig:interaction_orbit} that for $I(0) = 0.2$, the net force is zero at $d \approx 1$ and also asymptotically, i.e., $d > 5$. The zero force at $d \approx 1$ indicates that stable vortex orbits might be possible in this regime. We know from \cref{fig:Int_11} that this zero force point corresponds to a local minimum in the interaction energy. As we decrease the intensity of the excitation, this minimum moves up, and hence the force is no longer zero at $d \approx 1$. As such, as the intensity decays in the dynamical simulation, the stability of the orbit will fail, and the vortices will escape.

\begin{figure}
\centering
\includegraphics[width=0.7\linewidth]{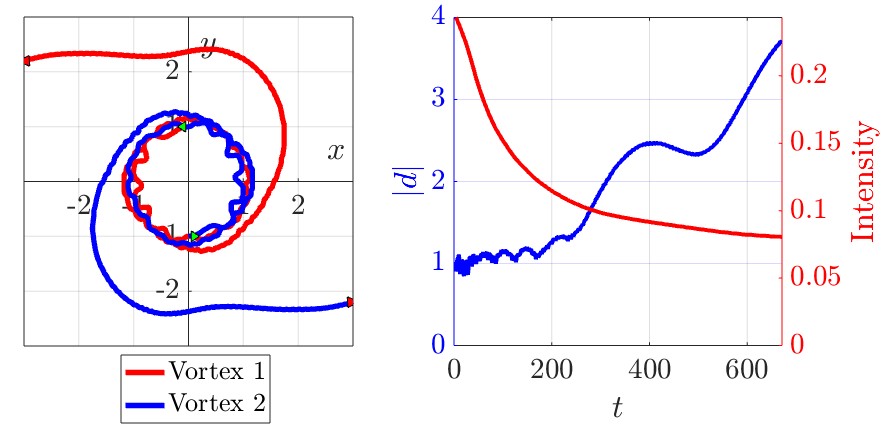}
\caption{Left: trajectories of a two-vortex system ($\lambda = 1.1$, $v_{\rm in} = 0.04$, $I(0) = 0.2$) at $x_2 = \pm 1$. The blue line shows the $(x_1, x_2)$ position of one vortex, red the other. Right: intensity of the excitation per vortex as a function of time (red), and distance $|d|$ of the vortices from the origin as a function of time.}
\label{fig:l1.1_orbit2}
\end{figure}

For $\lambda = 1.1$, $V_1^{\lambda} = 3.204508502$, so for a stable orbit at $d = 1$, for $I(0) = 0.2$, choosing $L_z = 0.07$, corresponding to the orange line in \cref{fig:interaction_orbit}, we would require an angular velocity of $v_{\rm in} = 0.043688.$ We see an example of this in \cref{fig:l1.1_orbit2}.

\cref{fig:l1.1_orbit2} shows a semi-stable orbit for type \rom{2} vortices. The blue line indicates the path of one vortex and the red line the other. The dashed black line displays the circle $x_1^2 + x_2^2 = d_0^2$, where $d_0 = 1$.

We see in \cref{fig:l1.1_orbit2} that the vortices orbit twice around the origin before escaping. This is due to the intensity of the excitation decaying, as seen in the right of \cref{fig:l1.1_orbit2} with the magenta line displaying the intensity of the excitation per vortex as a function of time, and as such the static force and centrifugal force dominate the interaction, and the vortices escape to infinity.

\begin{figure}
\centering
\begin{minipage}{0.24\textwidth}
\centering
\includegraphics[width=\linewidth]{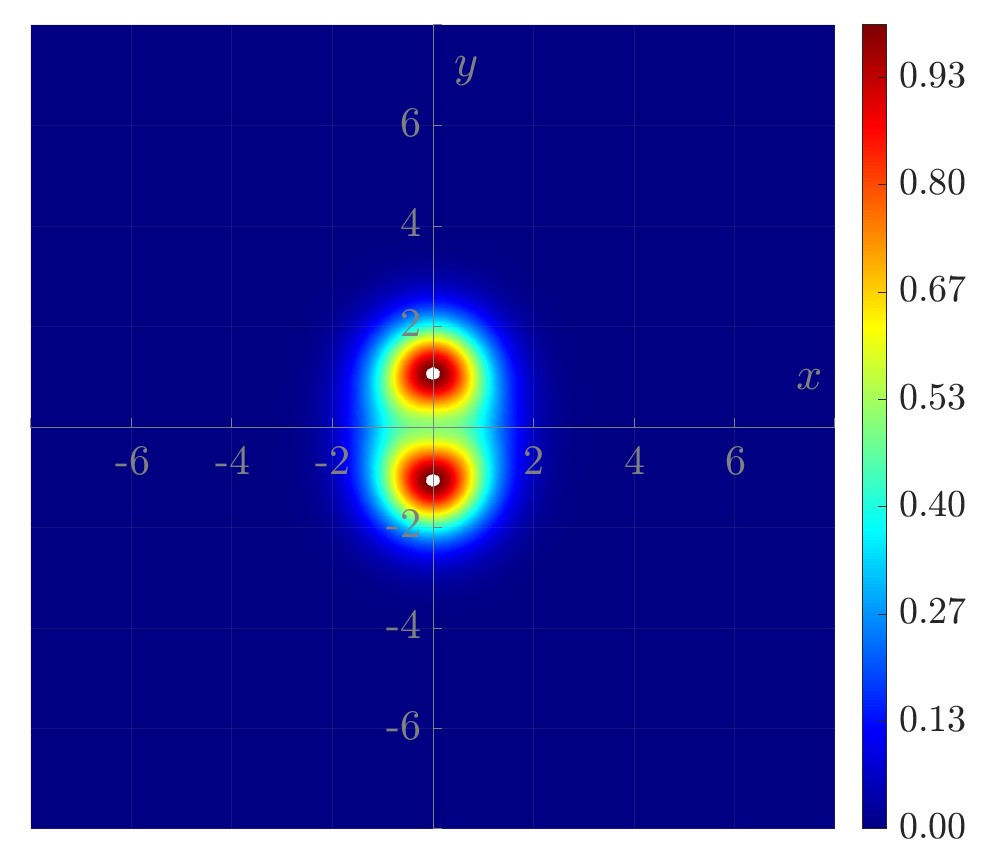}
\caption*{$t = 0$}
\end{minipage}
\begin{minipage}{0.24\textwidth}
\centering
\includegraphics[width=\linewidth]{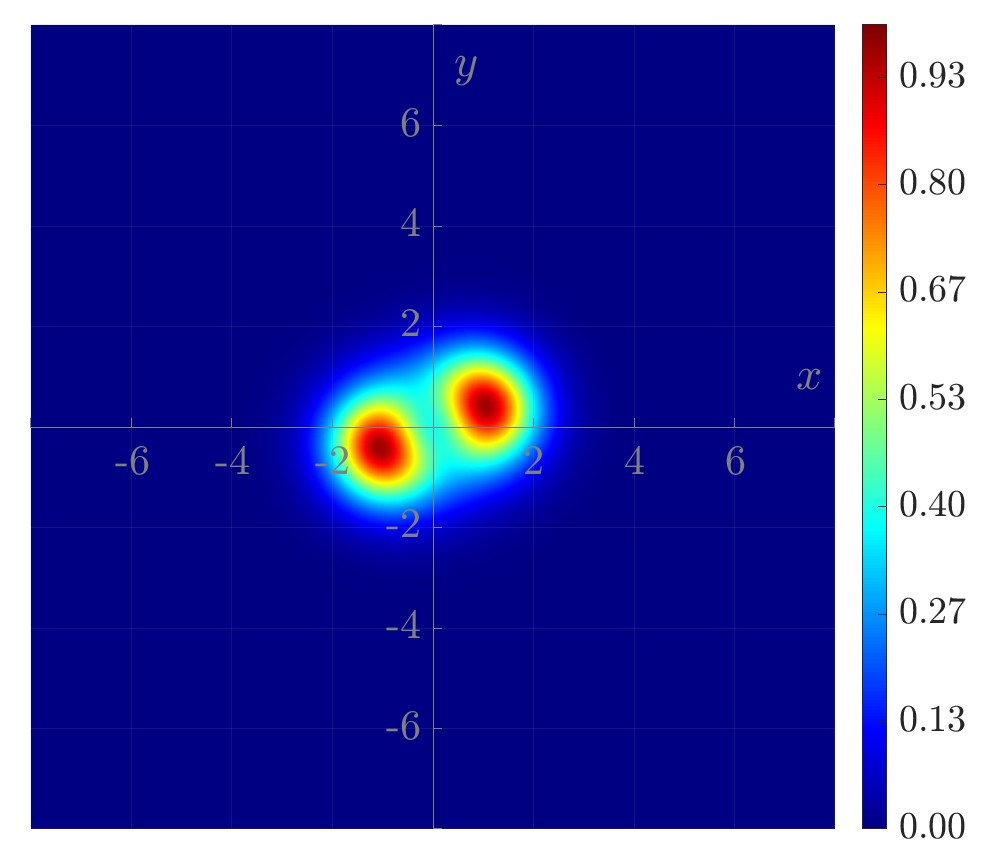}
\caption*{$t = 25$}
\end{minipage}
\begin{minipage}{0.24\textwidth}
\centering
\includegraphics[width=\linewidth]{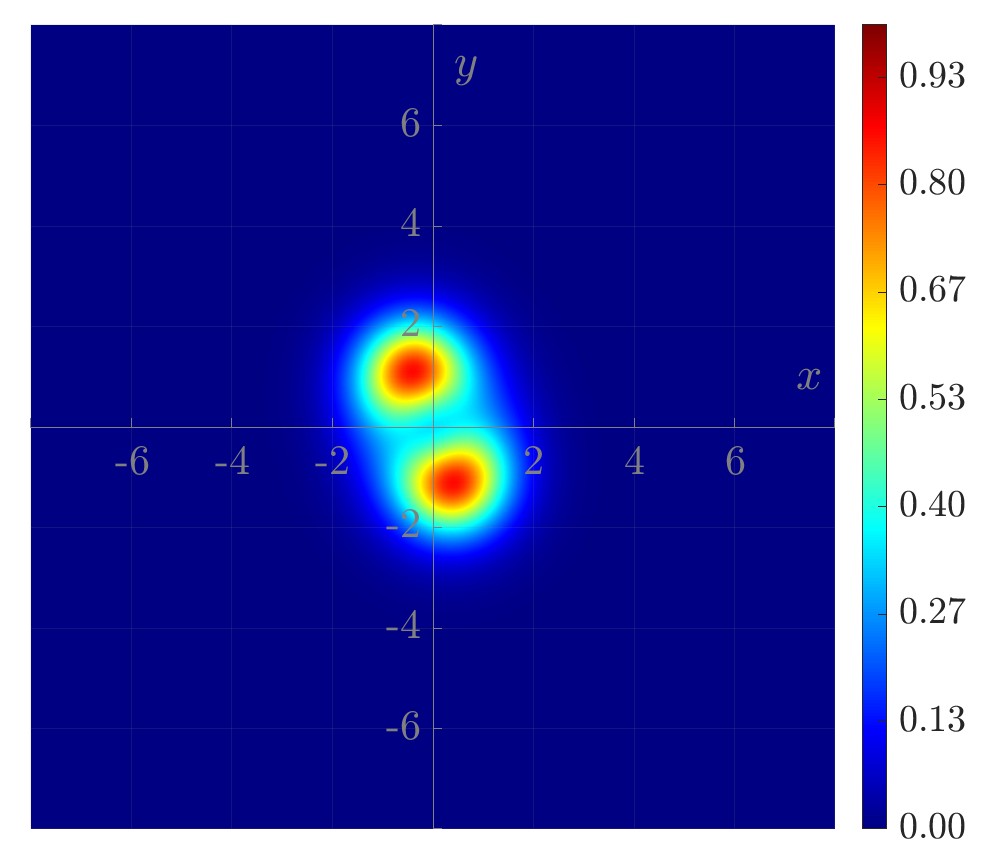}
\caption*{$t = 50$}
\end{minipage}
\begin{minipage}{0.24\textwidth}
\centering
\includegraphics[width=\linewidth]{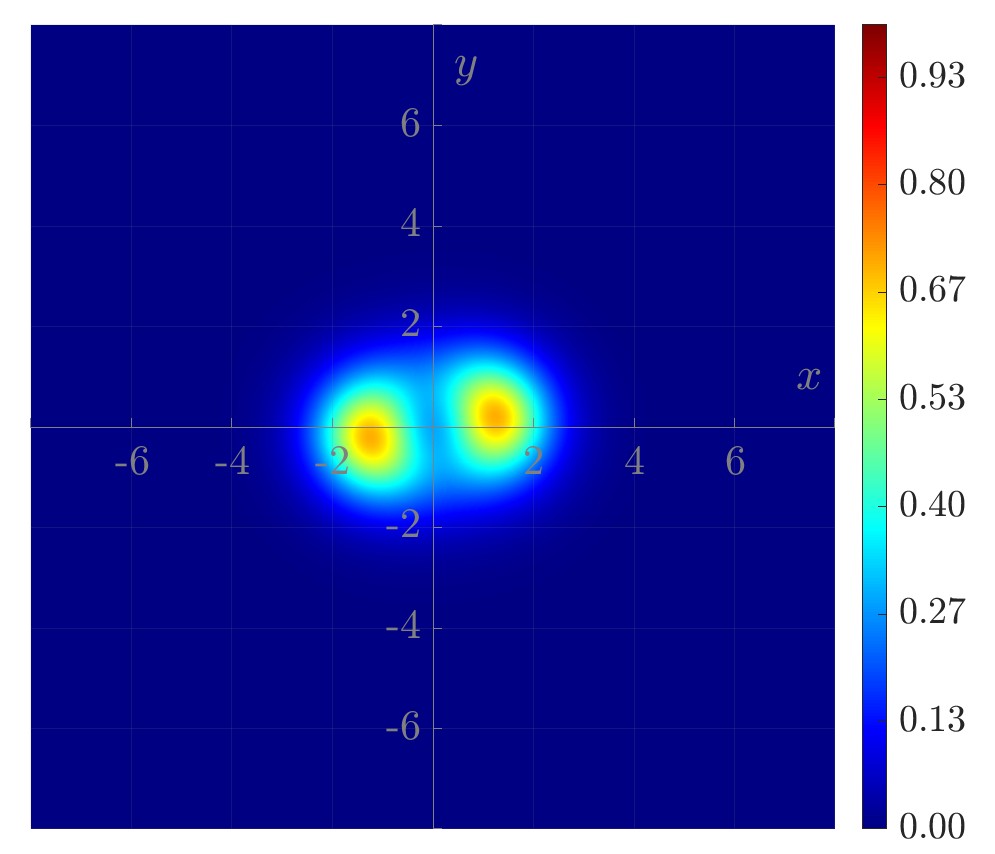}
\caption*{$t = 75$}
\end{minipage}
\begin{minipage}{0.24\textwidth}
\centering
\includegraphics[width=\linewidth]{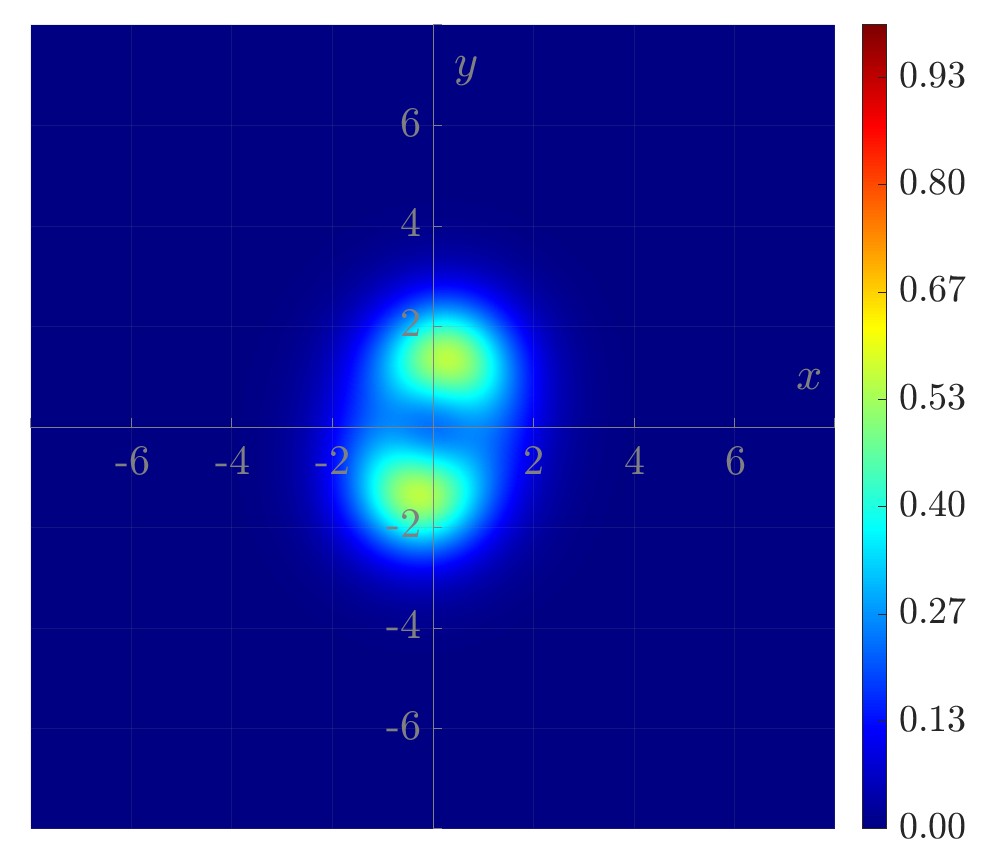}
\caption*{$t = 100$}
\end{minipage}
\begin{minipage}{0.24\textwidth}
\centering
\includegraphics[width=\linewidth]{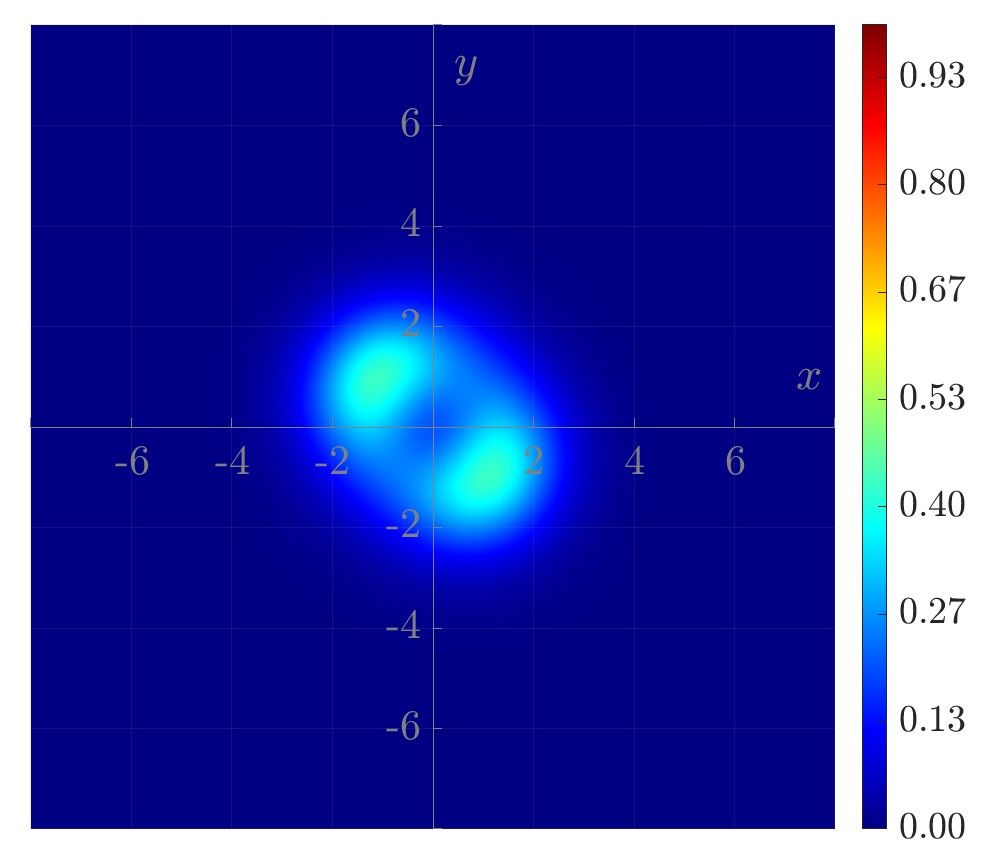}
\caption*{$t = 125$}
\end{minipage}
\begin{minipage}{0.24\textwidth}
\centering
\includegraphics[width=\linewidth]{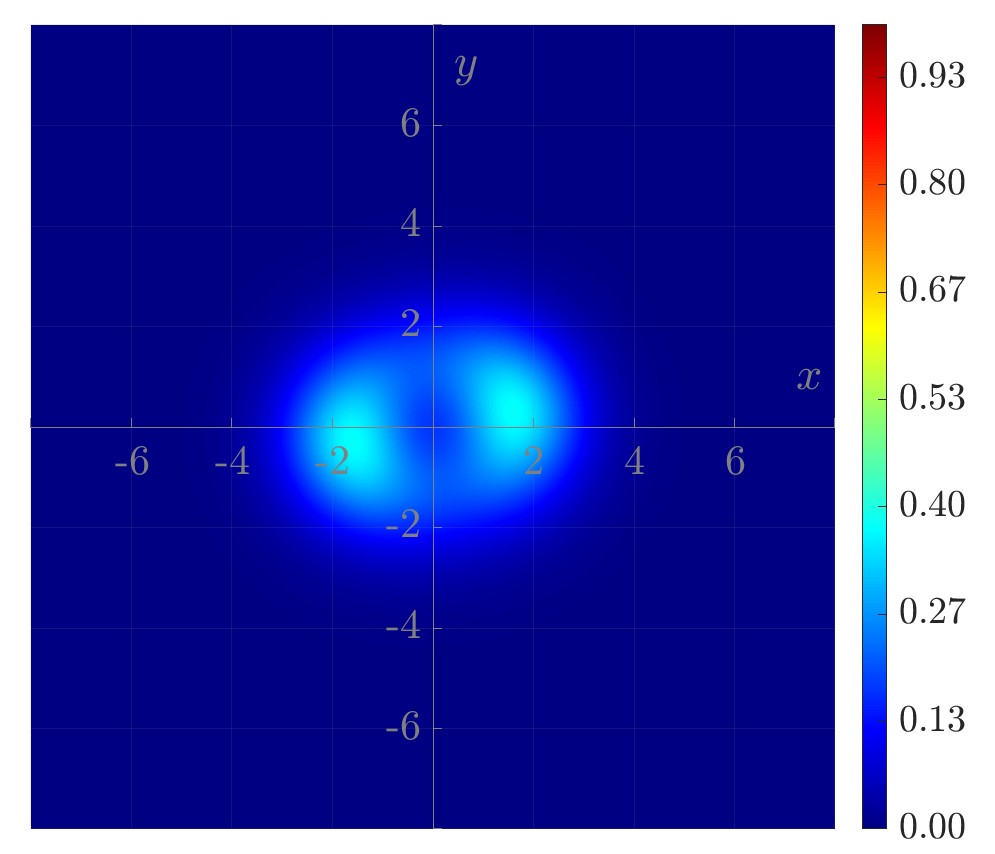}
\caption*{$t = 150$}
\end{minipage}
\begin{minipage}{0.24\textwidth}
\centering
\includegraphics[width=\linewidth]{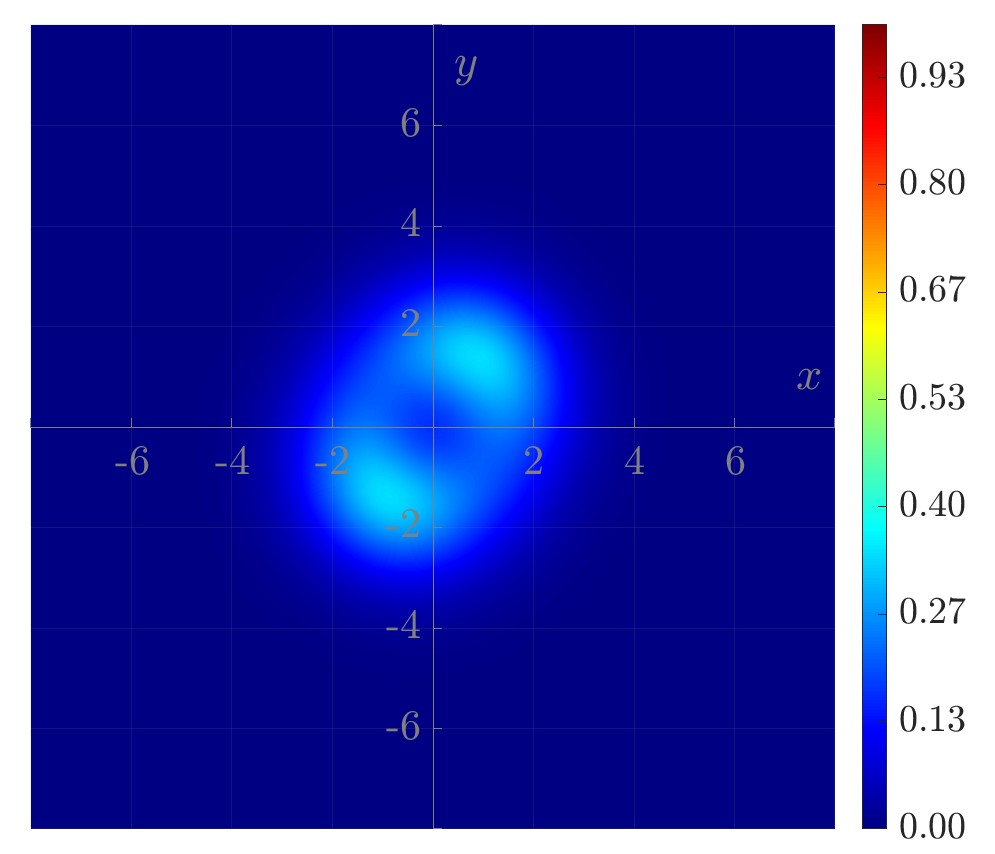}
\caption*{$t = 175$}
\end{minipage}
\begin{minipage}{0.24\textwidth}
\centering
\includegraphics[width=\linewidth]{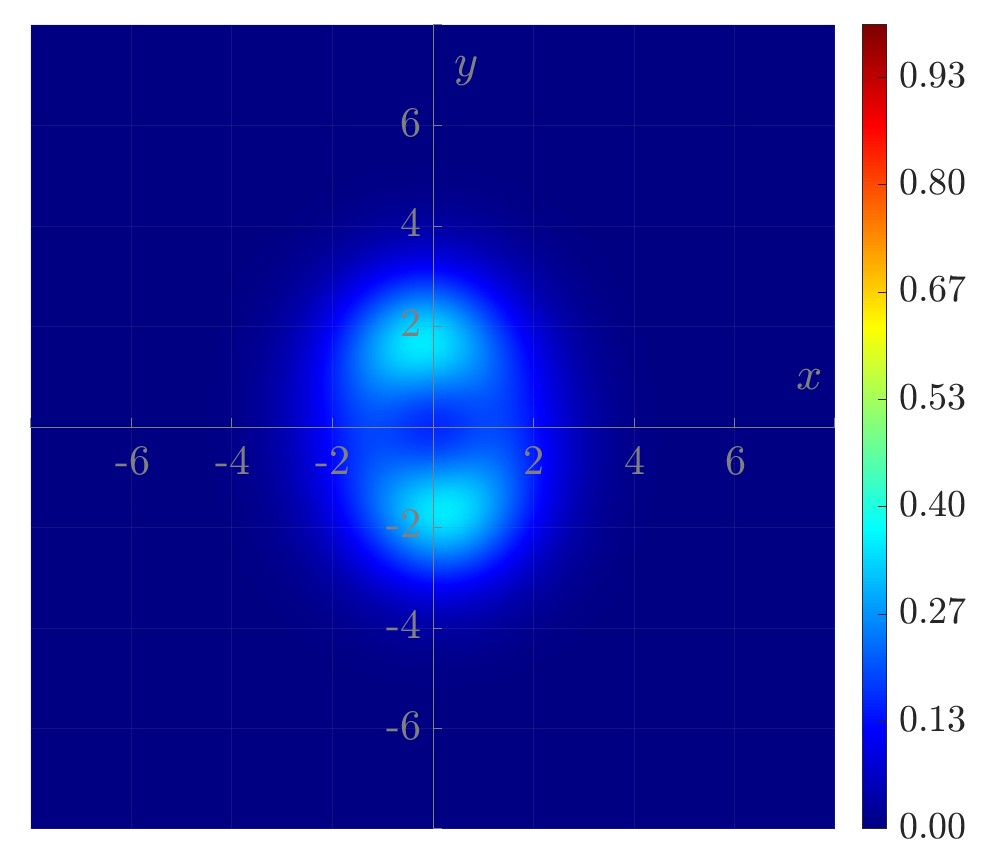}
\caption*{$t = 200$}
\end{minipage}
\begin{minipage}{0.24\textwidth}
\centering
\includegraphics[width=\linewidth]{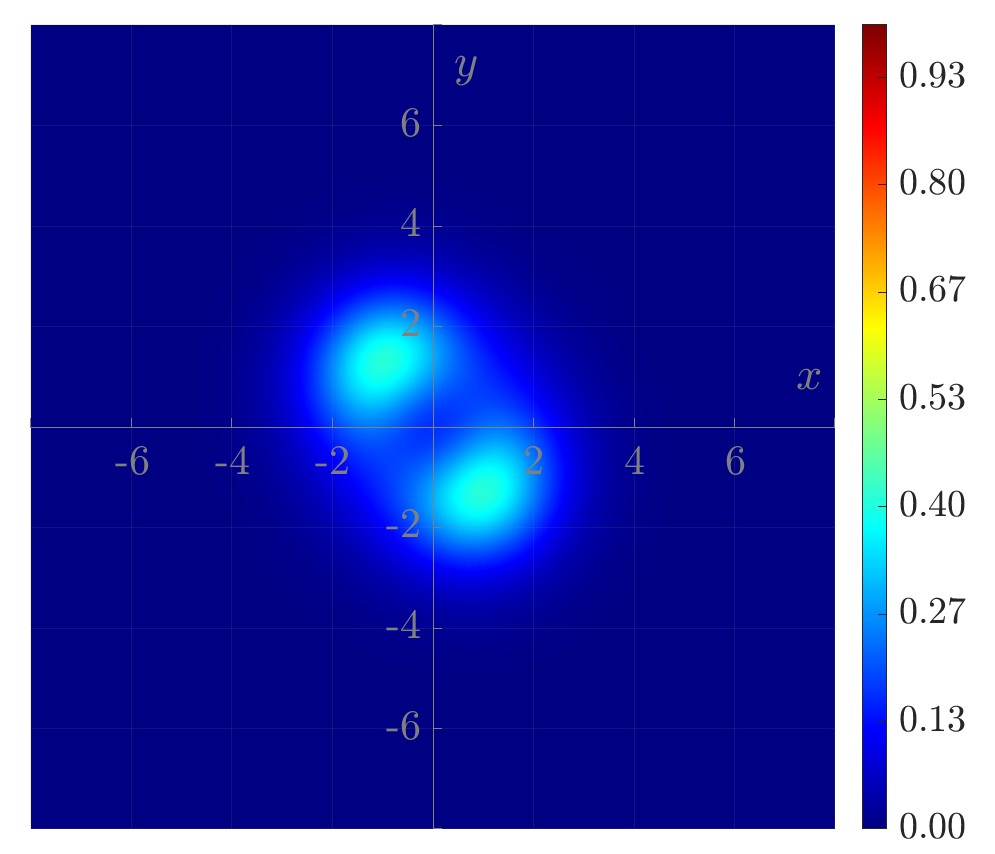}
\caption*{$t = 225$}
\end{minipage}
\begin{minipage}{0.24\textwidth}
\centering
\includegraphics[width=\linewidth]{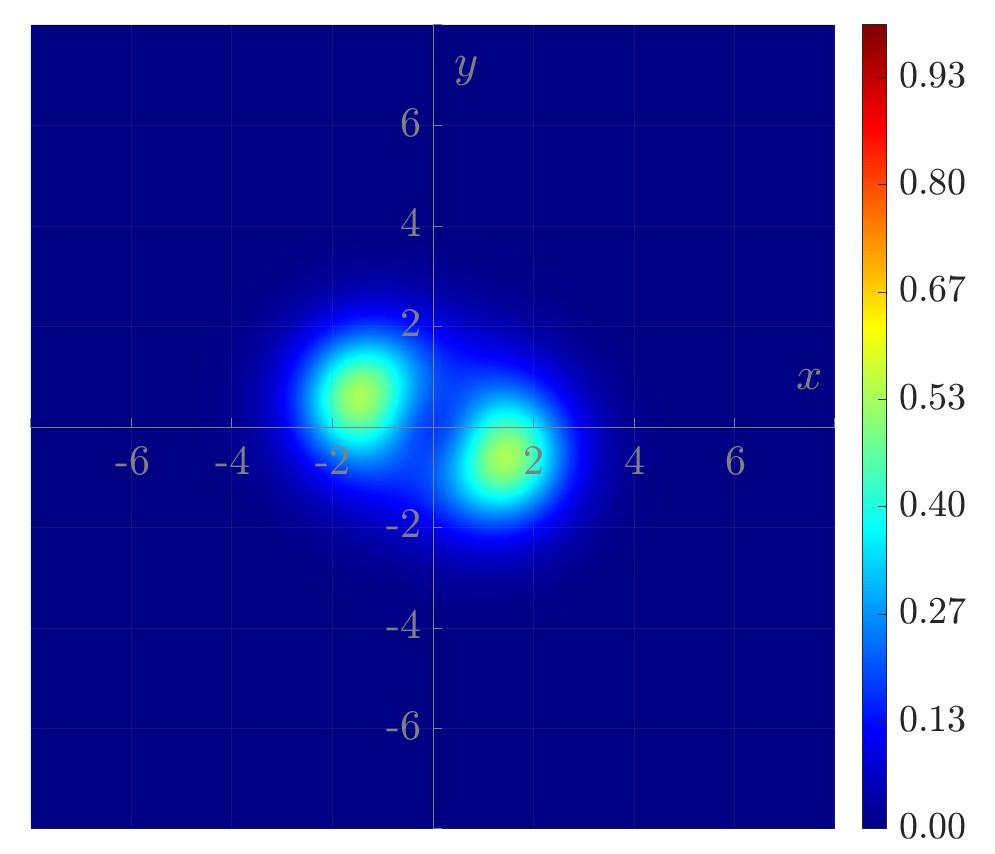}
\caption*{$t = 250$}
\end{minipage}
\begin{minipage}{0.24\textwidth}
\centering
\includegraphics[width=\linewidth]{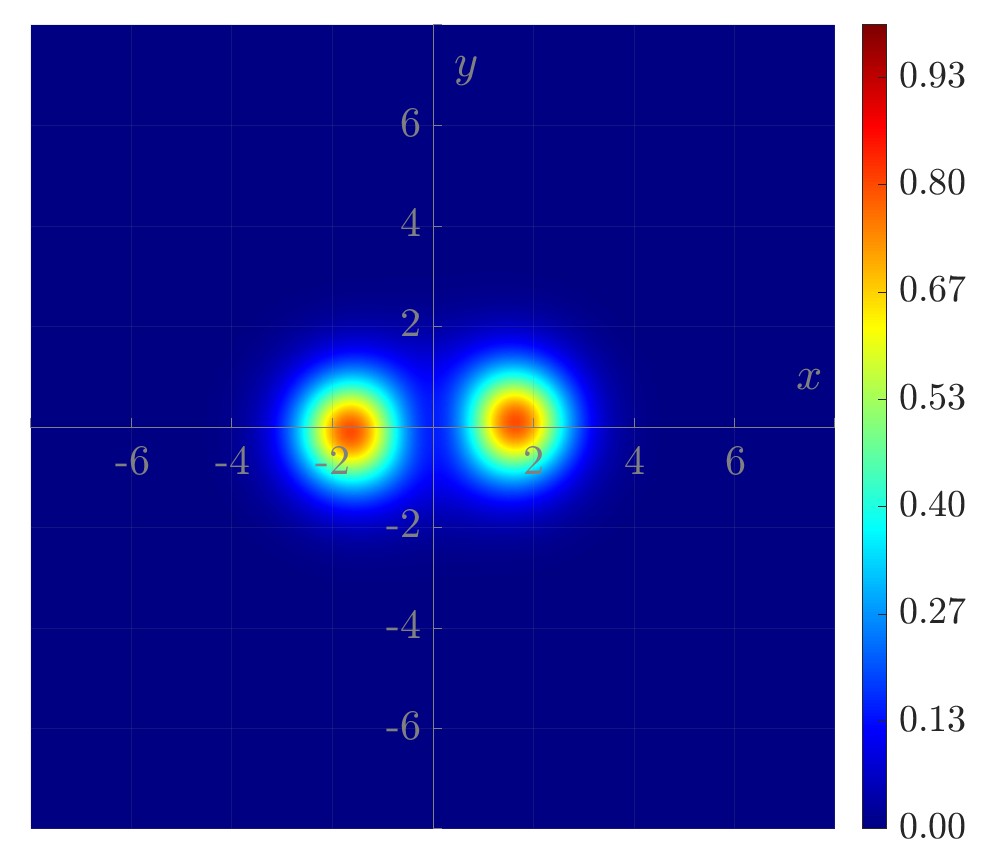}
\caption*{$t = 275$}
\end{minipage}
\begin{minipage}{0.24\textwidth}
\centering
\includegraphics[width=\linewidth]{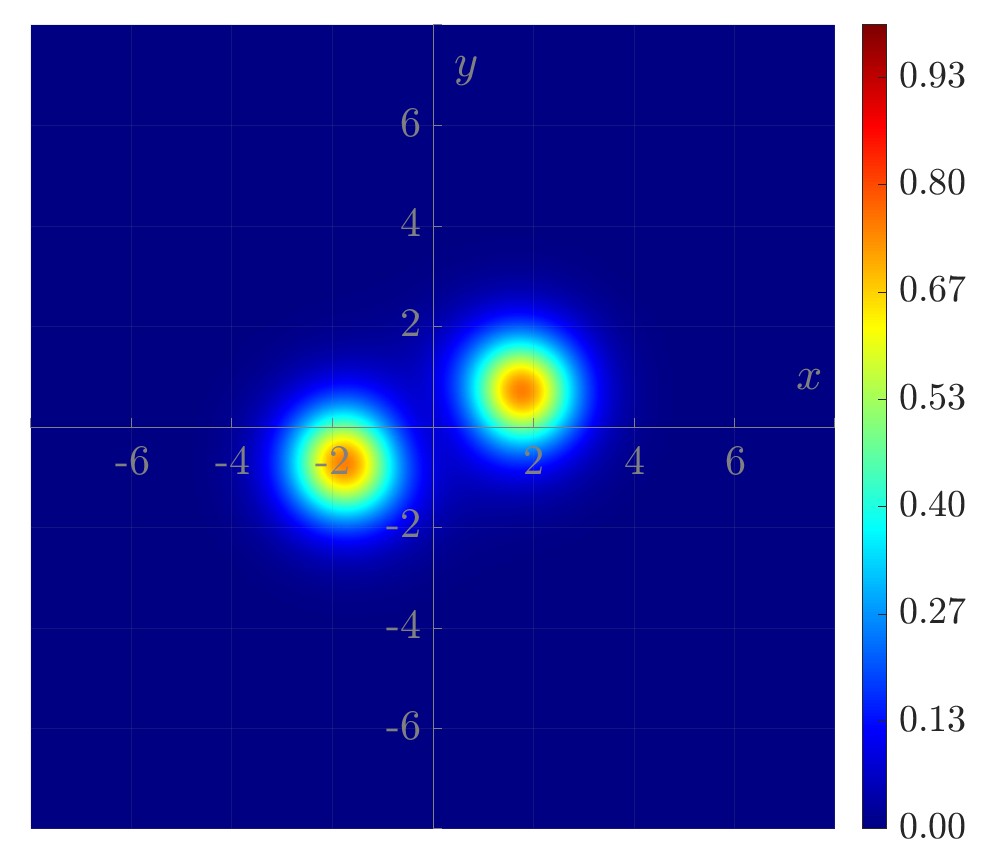}
\caption*{$t = 300$}
\end{minipage}
\begin{minipage}{0.24\textwidth}
\centering
\includegraphics[width=\linewidth]{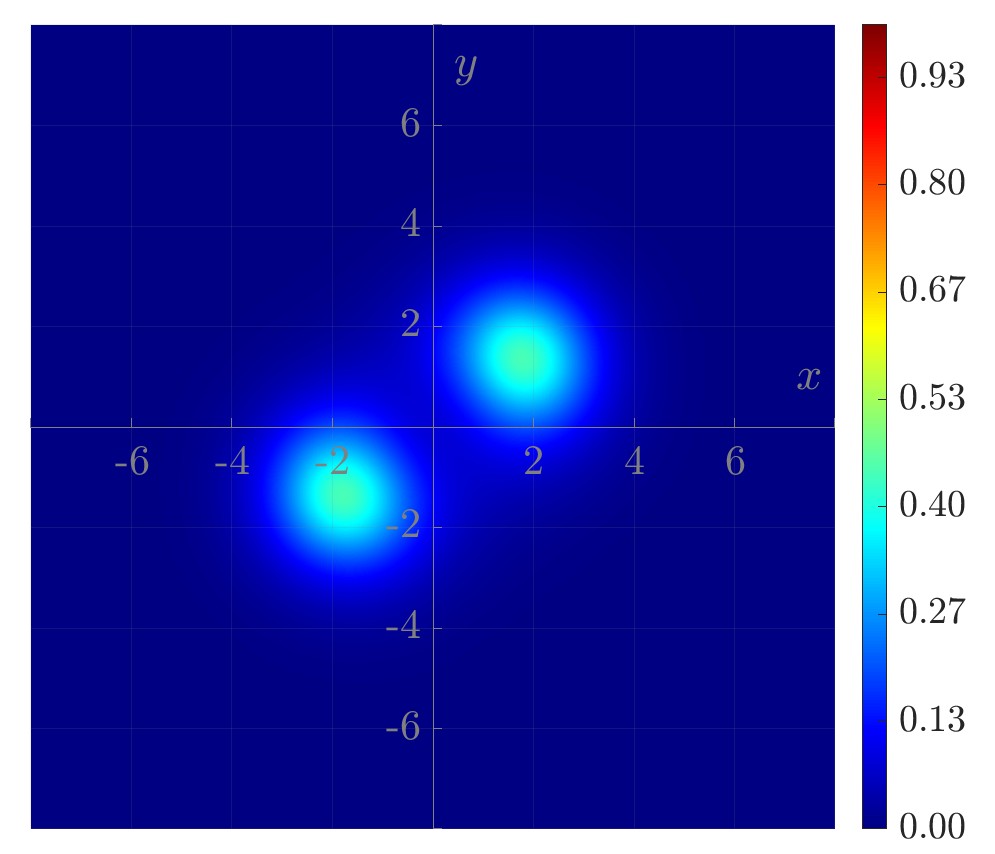}
\caption*{$t = 325$}
\end{minipage}
\begin{minipage}{0.24\textwidth}
\centering
\includegraphics[width=\linewidth]{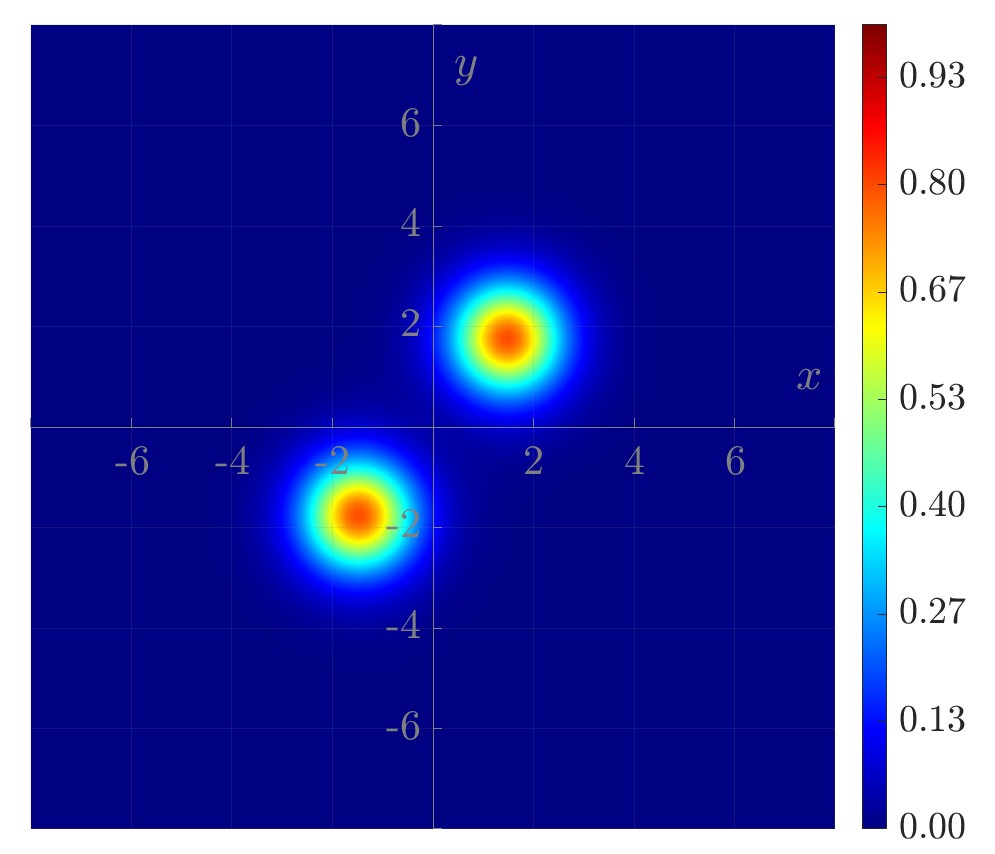}
\caption*{$t = 350$}
\end{minipage}
\begin{minipage}{0.24\textwidth}
\centering
\includegraphics[width=\linewidth]{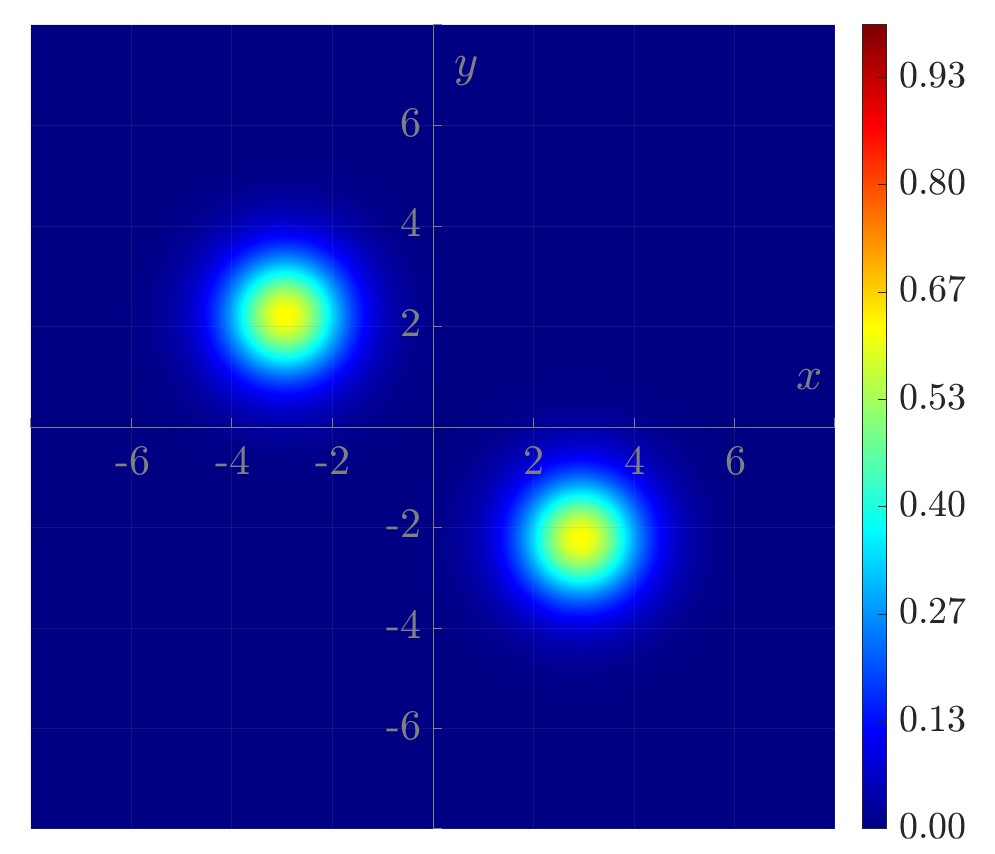}
\caption*{$t = 675$}
\end{minipage}
\caption{Snapshots of energy density for a two-vortex scattering ($\lambda = 1.1$, $v_{\rm in} = 0.04$, $I(0) = 0.2$) at $d = \pm 1$, as seen in \cref{fig:l1.1_orbit2}.}
\label{fig:l1.1_Orbitsnapshot}
\end{figure}

\subsection{Vortex Anti-Vortex Orbits}
\label{sec:VortexAntiVortexOrbits}
\begin{figure}
\centering
\begin{minipage}{0.45\textwidth}
    \includegraphics[width=\linewidth]{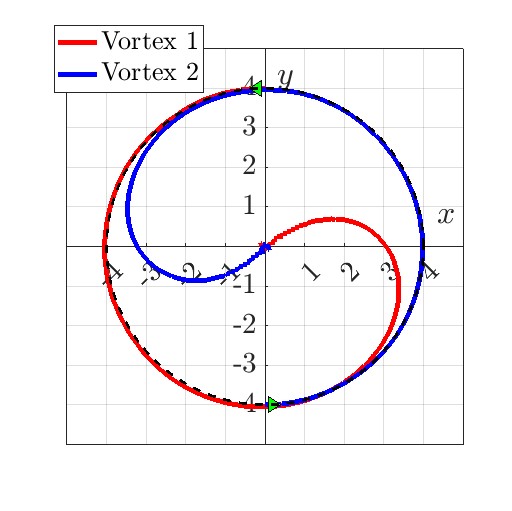}
\caption{Trajectories of a vortex-antivortex pair ($v_{\rm in} = 0.08$). The red line shows the $(x_1, x_2)$ position of the vortex, and blue the antivortex, completing a half-orbit before annihilation.}
\label{fig:V-AV-orbit1}
\end{minipage}
\hfill
\begin{minipage}{0.45\textwidth}
    \includegraphics[width=\linewidth]{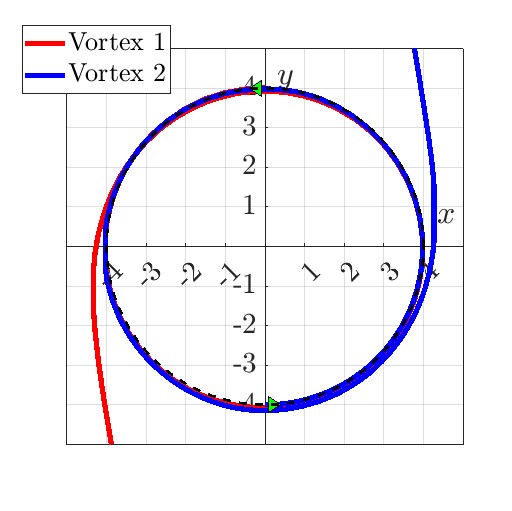}
\caption{Trajectories of a vortex-antivortex pair ($v_{\rm in} = 0.08327$). The red line shows the $(x_1, x_2)$ position of the vortex, and blue the antivortex, completing one orbit before escaping to infinity.}
\label{fig:V-AV-orbit2}
\end{minipage}
\end{figure}

Finally, we consider the case of vortex-antivortex orbits. The static force for a vortex-antivortex pair is strongly attractive, meaning that the vortices will always try to move towards each other and annihilate. If the initial tangential velocity is small, then the vortices will be drawn to the centre and annihilate, see \cref{fig:V-AV-orbit1}. For a fine-tuned velocity, we can show that the vortices can orbit the origin in a stable manner, see \cref{fig:V-AV-orbit2}.

We see in \cref{fig:V-AV-orbit1} that the vortices only manage to orbit the origin for half an orbit before the static force dominates and the vortices annihilate at the origin.

\begin{figure}
\centering
\begin{minipage}{0.49\linewidth}
\includegraphics[width=\linewidth]{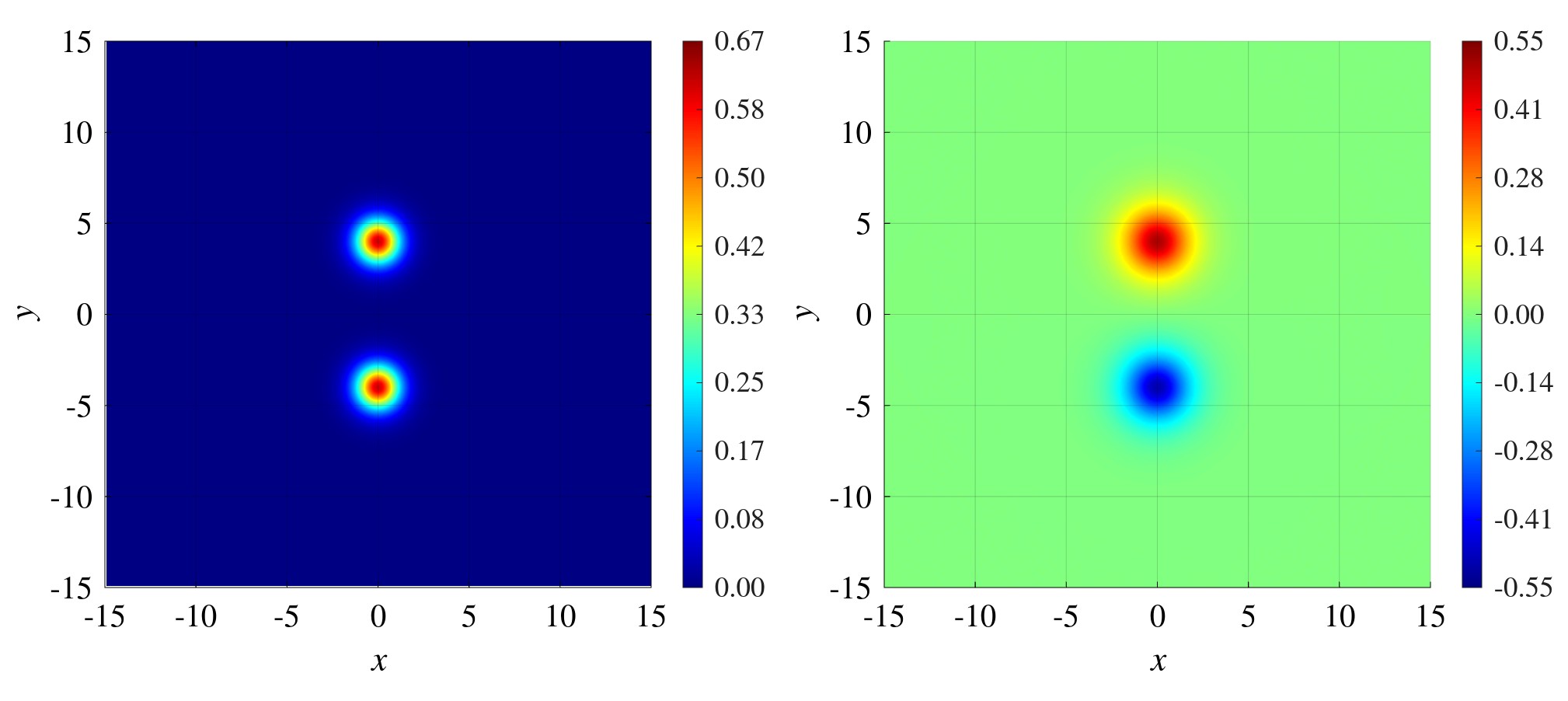}
\caption*{$t = 0$}
\end{minipage}
\hfill
\begin{minipage}{0.49\linewidth}
\includegraphics[width=\linewidth]{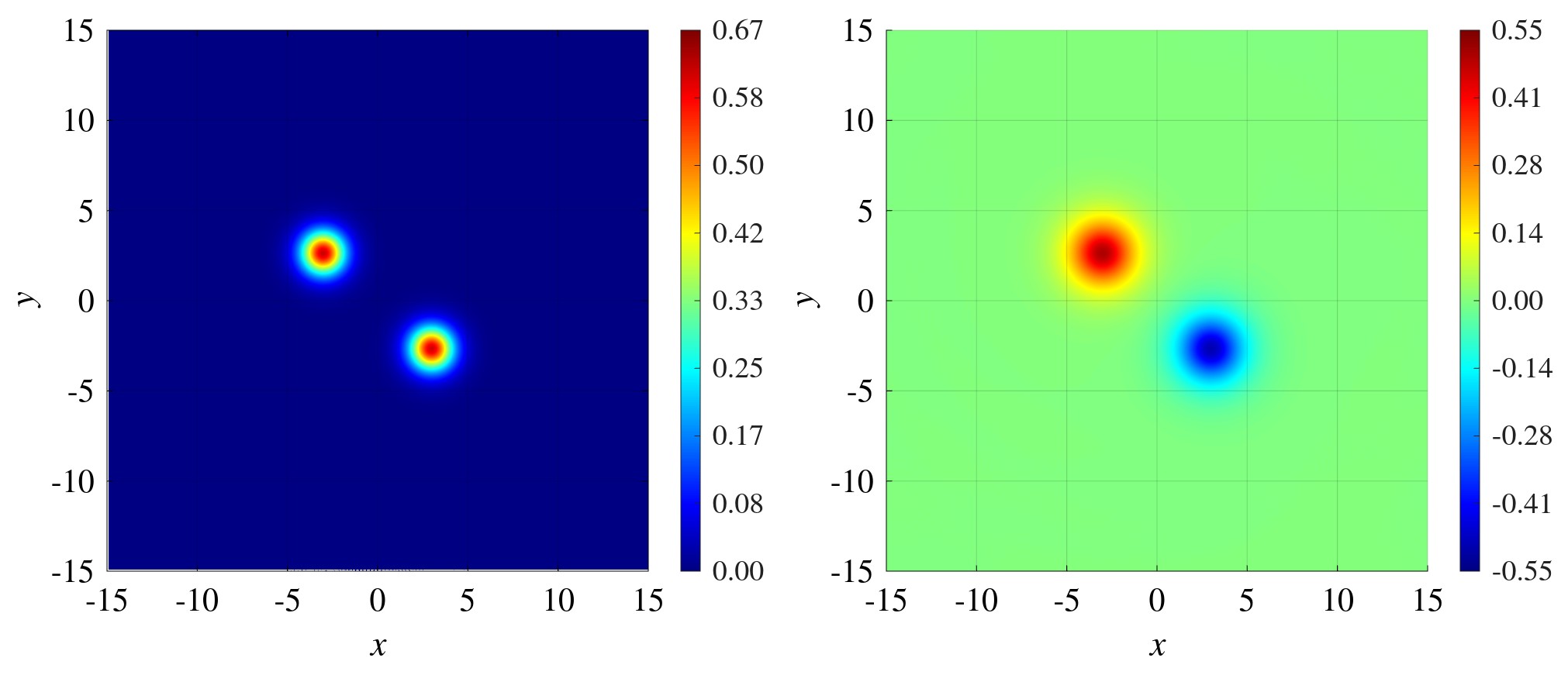}
\caption*{$t = 40$}
\end{minipage}
\begin{minipage}{0.49\linewidth}
\includegraphics[width=\linewidth]{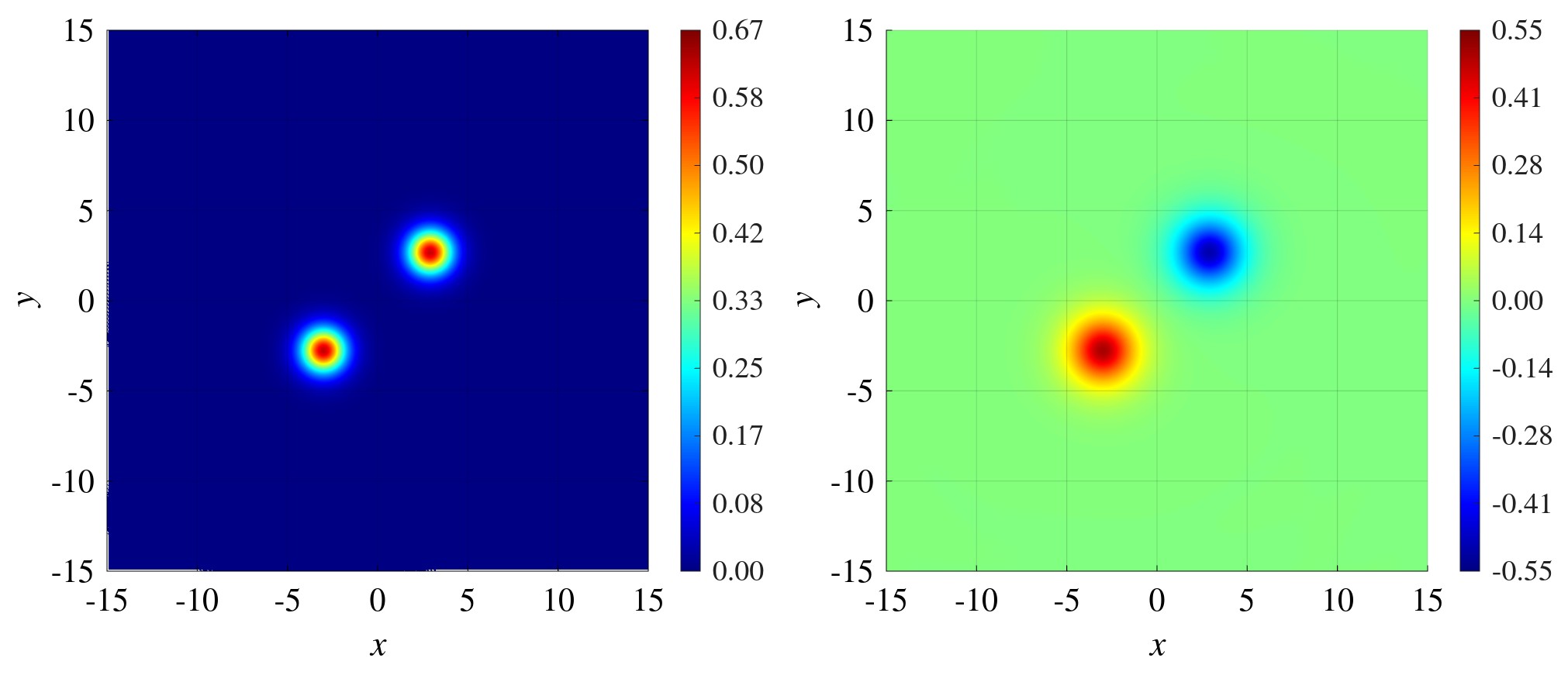}
\caption*{$t = 110$}
\end{minipage}
\hfill
\begin{minipage}{0.49\linewidth}
\includegraphics[width=\linewidth]{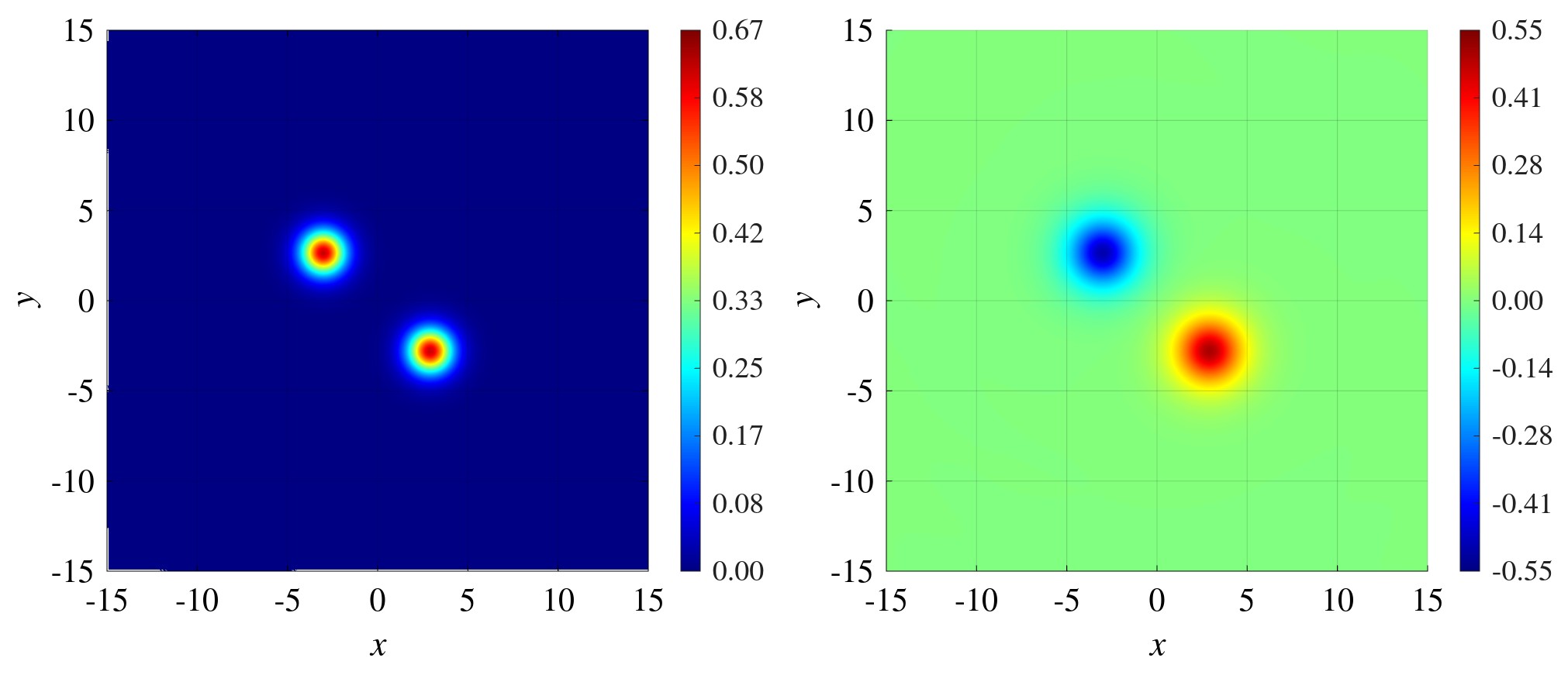}
\caption*{$t = 190$}
\end{minipage}
\begin{minipage}{0.49\linewidth}
\includegraphics[width=\linewidth]{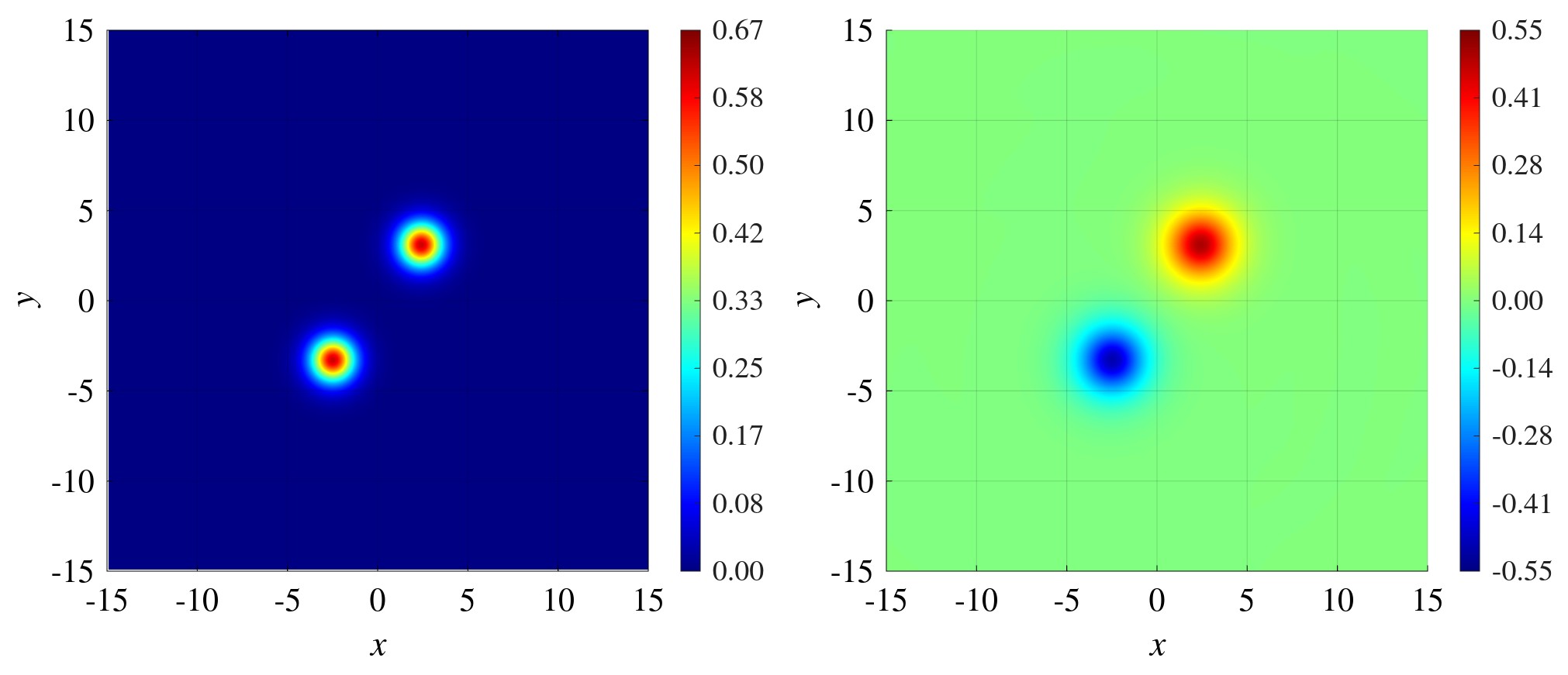}
\caption*{$t = 270$}
\end{minipage}
\hfill
\begin{minipage}{0.49\linewidth}
\includegraphics[width=\linewidth]{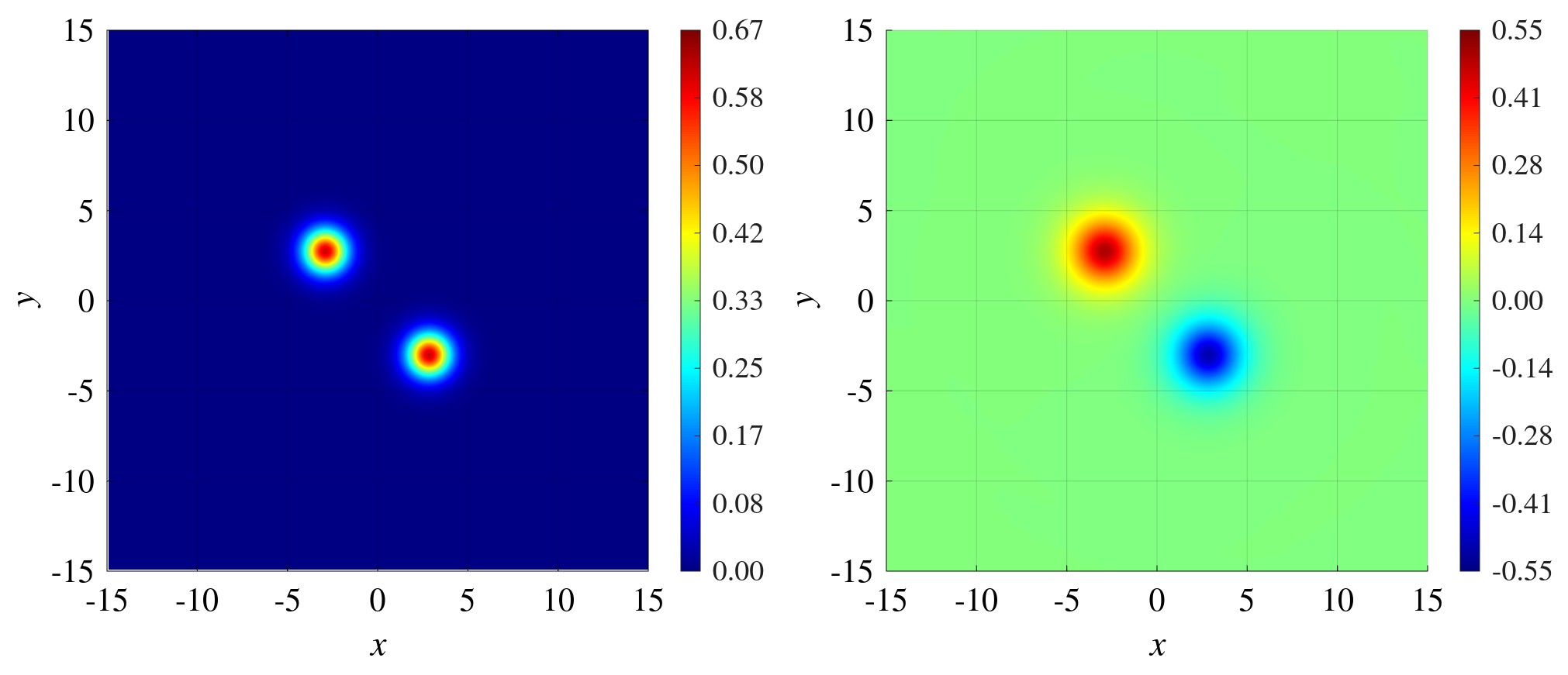}
\caption*{$t = 340$}
\end{minipage}
\begin{minipage}{0.49\linewidth}
\includegraphics[width=\linewidth]{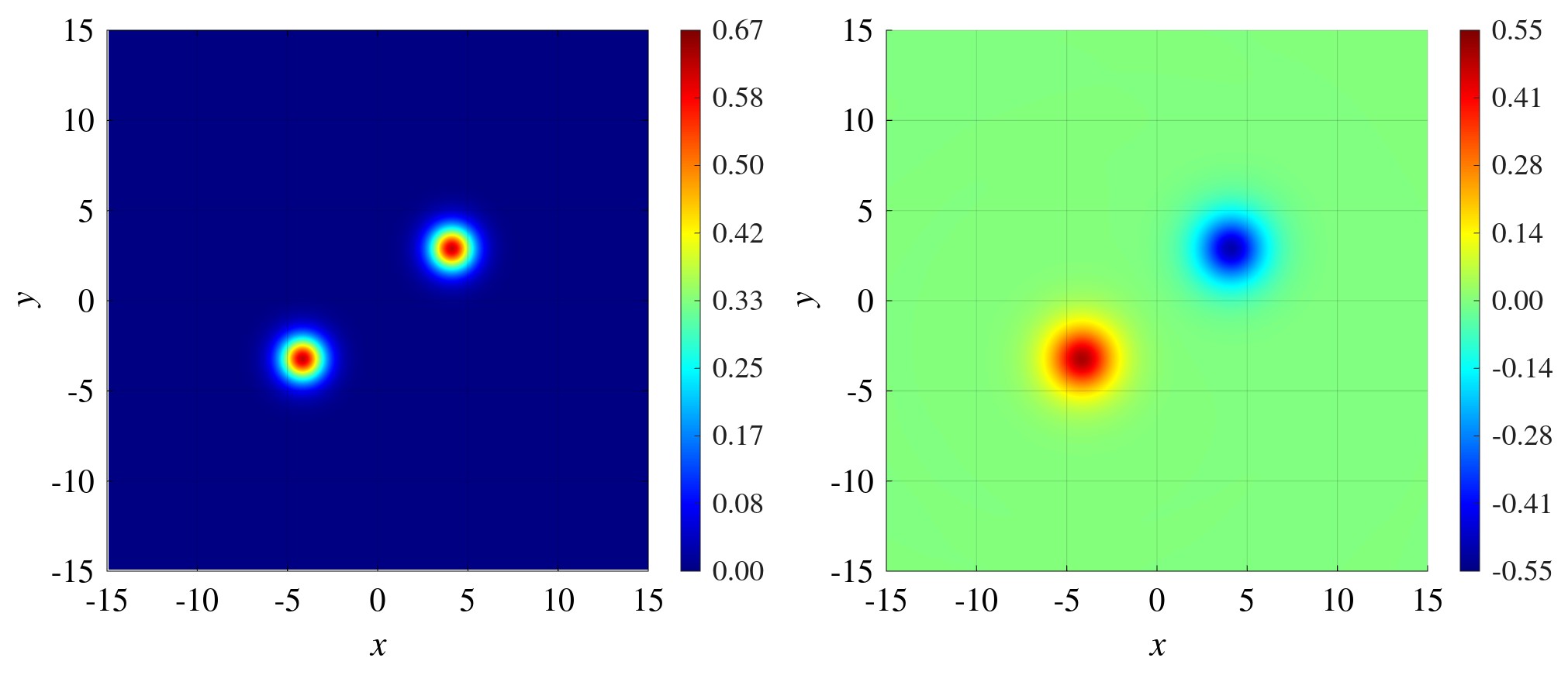}
\caption*{$t = 420$}
\end{minipage}
\hfill
\begin{minipage}{0.49\linewidth}
\includegraphics[width=\linewidth]{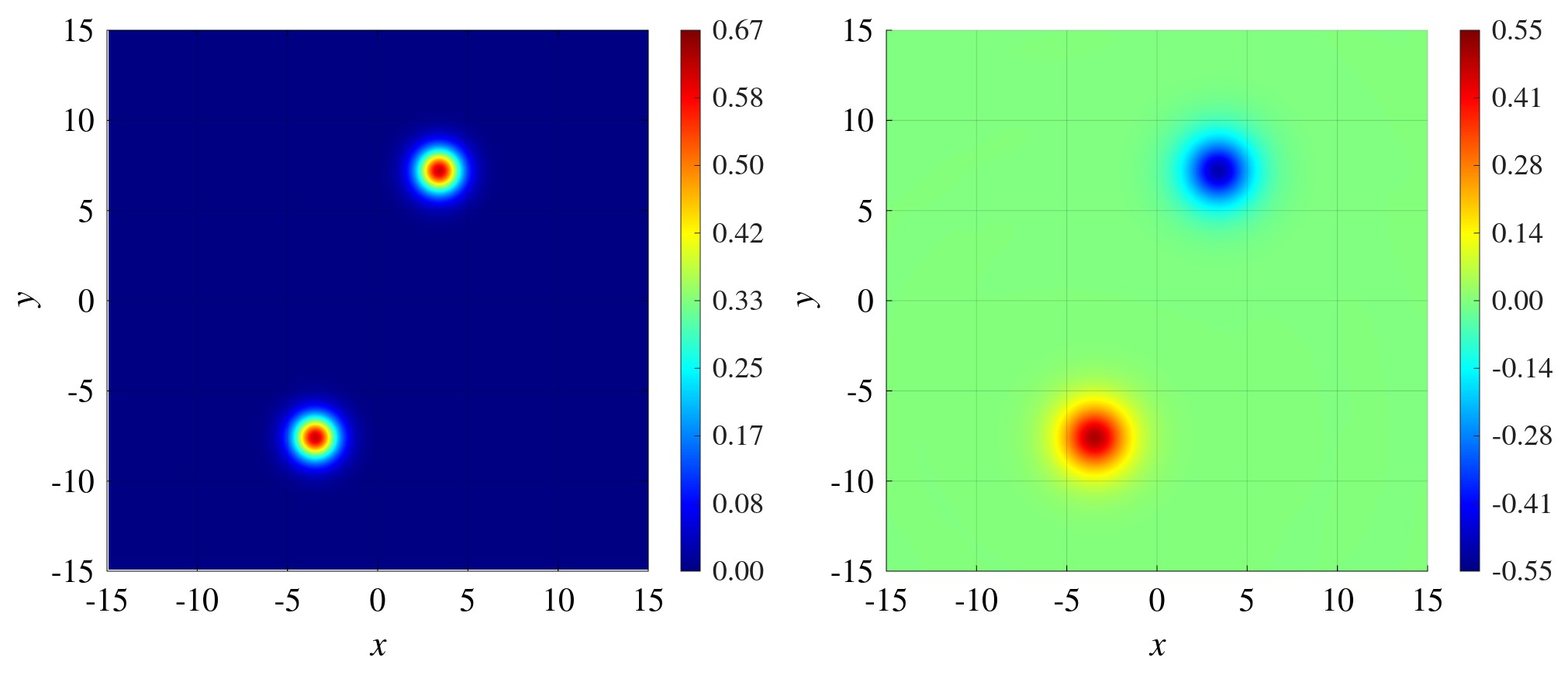}
\caption*{$t = 480$}
\end{minipage}
\caption{Heat plots of the energy density and magnetic field $B$ for a vortex-antivortex scattering ($v_{\rm in} = 0.08327$) with non-zero impact parameter, as seen in \cref{fig:V-AV-orbit2}. Red (positive) and blue (negative) indicate the magnetic field’s sign. We see that the vortex-antivortex form a long-lived orbit until  $t=420$, at which point the pair escape the orbit.}
\label{fig:VA-V_Orbit}
\end{figure}

We see in \cref{fig:V-AV-orbit2} that for a fine-tuned velocity, the vortices orbit the origin in a stable manner for one full orbit before escaping to infinity. It should be noted here that it is impossible for the vortices to escape due to the strong attractive force between a vortex anti-vortex pair, however they escape during the simulation due to damping boundary conditions which halt the velocity of the vortices. We can plot some dynamical snapshots of \cref{fig:V-AV-orbit2} displaying a heat plot of the Higgs field $|\phi|^2$ and magnetic field $B$ for a vortex-antivortex pair with non-zero impact parameter, see \cref{fig:VA-V_Orbit}.

This is not a comprehensive study of vortex-antivortex orbits, as this is beyond the scope of this paper. A systematic map of vortex-antivortex scattering scenarios in the unexcited case, including the resonant annihilation structure in the deep type \rom{2} regime, has recently been provided in \cite{bachmaier2026resonance}. Earlier numerical work on vortex-antivortex reconnection in the cosmic string context can be found in \cite{Matzner,VA}.

\section{Conclusions}
\label{sec:Conclusions}
We have investigated vortex dynamics in the Abelian-Higgs model away from critical coupling (\(\lambda \neq 1\)), extending our analysis of excited vortices at critical coupling \cite{SKMRTW}. Our numerical simulations focus on Type \rom{1} (\(\lambda < 1\)) and Type \rom{2} (\(\lambda > 1\)) regimes, as well as vortex-antivortex interactions, examining how static forces, mode-induced interactions, and centrifugal effects govern scattering and orbital behaviour.

In Type \rom{1} systems, where static attraction dominates, exciting out-of-phase shape modes introduces repulsive forces. We observed spectral walls, where mode frequencies enter the continuous spectrum, altering vortex trajectories. For \(\lambda = 0.9\), vortices bounce off these walls, with trajectories showing distinct velocity changes (\cref{fig:l0.9_trajectories1}). Quasi-stationary states emerge when static attraction balances mode-induced repulsion, causing vortices to stabilise at a fixed separation (\cref{fig:trajectories_localMax_l09}). For \(\lambda = 0.5\), spectral walls appear before collision, though their effect is subdued by strong static attraction (\cref{fig:l0.5_trajectories}).

In Type \rom{2} systems, characterised by static repulsion, in-phase shape mode excitations induce attractive forces. Our simulations reveal quasi-bound states where vortices scatter multiple times before separating or oscillating at a fixed distance (\cref{fig:l1.1_trajectories1,fig:trajectories_l11_localMin}). The interaction energy shows local extrema, trapping vortices in potential wells (\cref{fig:Int_11}). For \(\lambda = 2\), spectral walls influence dynamics as frequencies reach the continuous spectrum (\cref{fig:l2_trajectories,fig:l2_sepFreq}).

Orbital dynamics further highlight the interplay of forces. In Type \rom{1} systems, stable circular orbits form when tangential velocities counterbalance static attraction, with fine-tuned initial conditions yielding sustained orbits (\cref{fig:l0.9_orbit3}). In Type \rom{2} systems, mode-induced attraction supports bounded orbits within a potential well, though excitation decay limits stability (\cref{fig:l1.1_orbit2}). Vortex-antivortex pairs achieve brief orbits with carefully tuned velocities before annihilation or numerical escape (\cref{fig:V-AV-orbit2}).

These results demonstrate that the interplay of static, mode-induced, and centrifugal forces produces complex scattering and orbital behaviours, with spectral walls and quasi-stationary states enriching vortex dynamics away from critical coupling. Our findings provide a framework for understanding topological soliton interactions, bridging theoretical predictions and numerical observations.

Future work could explore vortex dynamics in the presence of impurities, which may act as trapping sites and lead to novel bound states or chaotic scattering, as suggested by \cite{Steffen2,Ashcroft:2018}. The Schrödinger-Chern-Simons model, with its conservative first-order dynamics, offers another avenue, where mode excitations could reveal new dynamical phenomena, building on numerical studies like \cite{Krusch:2005wr}. Additionally, investigating vortex lattices in the presence of mode excitations or impurities, as inspired by \cite{WinyardSpeight2024,SpeightWinyard2023}, could connect these dynamical phenomena to ordered macroscopic states in superconducting systems.
It would also be of interest to extend our analysis to vortices in non-trivial backgrounds, such as the oscillating axion background considered in \cite{kitajima2025abelian}, or to BPS impurity models, where the static intervortex force can be tuned independently of the mode structure \cite{BPS-imp-vortex-1,BPS-imp-vortex-3}. Similar mode-driven phenomena are anticipated for higher-dimensional BPS solitons. Recent simulations of magnetic monopole collisions \cite{bachmaier2025simulations} provide a natural setting for such an extension, with possible cosmological implications for the gravitational wave signatures of cosmic string networks \cite{LISA,Hind1}.

\section*{Acknowledgements}
All numerical calculations were run using the high performance computing systems provided by the University of Kent. Morgan Rees acknowledges the UK Engineering and Physical Sciences Research Council (EPSRC) for a PhD studentship. Thomas Winyard would like to thank the School of Mathematics at the University of Edinburgh for funding his postdoctoral position for the majority if this work.

\appendix
\section{Spectrum ODE Discretisation}\label{appendix:dis}
The discretised ODEs are
\begin{align}
    \omega^2 v_i &= -\frac{v_{i+1} - 2 v_i + v_{i-1}}{h^2} - \frac{1}{\rho_i} \frac{v_{i+1} - v_{i-1}}{2 h} + \left( \frac{k^2}{\rho_i^2} + f^2(\rho_i) \right) v_i + 2 \left( f(\rho_i) + \rho_i f'(\rho_i) \right) w_i, \label{eq:D_v} \\
    \omega^2 u_i &= -\frac{u_{i+1} - 2 u_i + u_{i-1}}{h^2} - \frac{1}{\rho_i} \frac{u_{i+1} - u_{i-1}}{2 h} + \left( \frac{k^2 + (N - a_\theta(\rho_i))^2}{\rho_i^2} + \frac{3\lambda}{2} f^2(\rho_i) - \frac{\lambda}{2} \right) u_i \notag \\
    &\quad - \frac{2 (N - a_\theta(\rho_i)) (k^2 + \rho_i^2 f^2(\rho_i))}{\rho_i^2} w_i + \frac{2 (N - a_\theta(\rho_i)) f(\rho_i)}{\rho_i} \frac{v_{i+1} - v_{i-1}}{2 h}, \label{eq:D_u} \\
    \omega^2 w_i &= -\frac{w_{i+1} - 2 w_i + w_{i-1}}{h^2} - \frac{1}{\rho_i} \frac{w_{i+1} - w_{i-1}}{2 h} + \left( \frac{k^2 + (N - a_\theta(\rho_i))^2}{\rho_i^2} + \left( 1 + \frac{\lambda}{2} \right) f^2(\rho_i) - \frac{\lambda}{2} \right) w_i \notag \\
    &\quad - \frac{2 (N - a_\theta(\rho_i))}{\rho_i^2} u_i + \frac{2 f'(\rho_i)}{\rho_i} v_i, \label{eq:D_w}
\end{align}
where $h$ is the step size in $\rho$, $i = 1, 2, \ldots, M+1$, and for $k=0$, the $w_i$ terms are omitted. We solve the discretised eigenvalue problem on the interval $\rho \in [0, L]$ with domain size $L = 30$ and grid $h = 0.002$, giving $M = 15001$ interior grid points. The eigenvalues are computed using MATLAB.
\section{Spectrum Boundary Conditions}\label{appendix:bc}
To ensure regularity at $\rho = 0$, we apply the Frobenius method to the ODEs \cref{eq:k0_v,eq:k0_u} for $k=0$ and \cref{eq:kn0_v,eq:kn0_w} for $k \neq 0$, assuming solutions of the form
\begin{equation}
    g(\rho) = \rho^s \sum_{j=0}^\infty u_j^{(i)} \rho^j, \label{eq:Frobenius}
\end{equation}
where $i \in \{v, u\}$ for $k=0$, $i \in \{v, u, w\}$ for $k \neq 0$, and $u_0^{(i)} \neq 0$. Using the asymptotic behaviour $f(\rho) \approx c \rho^N$, $a_\theta(\rho) \approx d \rho^2$ near $\rho = 0$ from \cref{eq:polarAnsatz}, we derive the indicial equations.

For $k=0$, substituting \cref{eq:Frobenius} into \cref{eq:k0_v,eq:k0_u}, the leading terms (at order $\rho^{s-2}$) give
\begin{align}
    v: & \quad -s^2 + 1 = 0 \implies s = \pm 1, \label{eq:indicial_v_k0} \\
    u: & \quad -s^2 + (N - a_\theta)^2 \approx -s^2 + N^2 = 0 \implies s = \pm N. \label{eq:indicial_u_k0}
\end{align}
Choosing $s = 1$ for $v$ and $s = N$ for $u$ to ensure regularity, the solutions are
\begin{align}
    v(\rho) &\approx u_0^{(v)} \rho + u_2^{(v)} \rho^3 + \cdots, \label{eq:frob_v_k0} \\
    u(\rho) &\approx u_0^{(u)} \rho^N + u_2^{(u)} \rho^{N+2} + \cdots. \label{eq:frob_u_k0}
\end{align}
The recurrence relation for $v$ from \cref{eq:k0_v}, considering the $\rho^j$ term, is
\begin{equation}
    \left[ -(s+j)^2 + 1 \right] u_j^{(v)} + \text{higher-order terms} = -\omega^2 u_{j-2}^{(v)}, \label{eq:recurrence_v_k0}
\end{equation}
where for $j=0$, the indical equation confirms $s = \pm 1$, and for $j=2m$, even coefficients satisfy
\begin{equation}
    u_{2m}^{(v)} = -\frac{\omega^2}{-4 (m+1)} u_{2m-2}^{(v)}, \quad m = 1, 2, \ldots. \label{eq:recurrence_even_v_k0}
\end{equation}

For $k \neq 0$, substituting \cref{eq:Frobenius} into \cref{eq:kn0_v,eq:kn0_w}, the leading terms give
\begin{align}
    v: & \quad -s^2 + k^2 = 0 \implies s = \pm k, \label{eq:indicial_v_kn0} \\
    u: & \quad -s^2 + k^2 + (N - a_\theta)^2 \approx -s^2 + k^2 + N^2 = 0 \implies s = \pm \sqrt{k^2 + N^2}, \label{eq:indicial_u_kn0} \\
    w: & \quad -s^2 + k^2 + (N - a_\theta)^2 \approx -s^2 + k^2 + N^2 = 0 \implies s = \pm \sqrt{k^2 + N^2}. \label{eq:indicial_w_kn0}
\end{align}
Choosing $s = |k|$ for $v$ (since $k \in \mathbb{Z}$), and $s = \sqrt{k^2 + N^2}$ for $u$ and $w$ to ensure regularity, the solutions are
\begin{align}
    v(\rho) &\approx u_0^{(v)} \rho^{|k|} + u_2^{(v)} \rho^{|k|+2} + \cdots, \label{eq:frob_v_kn0} \\
    u(\rho) &\approx u_0^{(u)} \rho^{\sqrt{k^2 + N^2}} + u_2^{(u)} \rho^{\sqrt{k^2 + N^2} + 2} + \cdots, \label{eq:frob_u_kn0} \\
    w(\rho) &\approx u_0^{(w)} \rho^{\sqrt{k^2 + N^2}} + u_2^{(w)} \rho^{\sqrt{k^2 + N^2} + 2} + \cdots. \label{eq:frob_w_kn0}
\end{align}
The recurrence relation for $v$ from \cref{eq:kn0_v} is
\begin{equation}
    \left[ -(s+j)^2 + k^2 \right] u_j^{(v)} + \text{higher-order terms} = -\omega^2 u_{j-2}^{(v)}, \label{eq:recurrence_v_kn0}
\end{equation}
where for $j=0$, $s = \pm k$, and for $j=2m$, even coefficients satisfy
\begin{equation}
    u_{2m}^{(v)} = -\frac{\omega^2}{-4 (m + |k|)} u_{2m-2}^{(v)}, \quad m = 1, 2, \ldots. \label{eq:recurrence_even_v_kn0}
\end{equation}

For boundary conditions at $\rho = 0$ ($i=1$), if $g(0) = 0$, we set $g_1 = 0$ for $g \in \{v, u\}$ ($k=0$) or $g \in \{v, u, w\}$ ($k \neq 0$). Otherwise, if $g(0)$ is constant, regularity requires $g'(0) = 0$, so using a second-order forward difference formula \cite{Alonso_Izquierdo_2016}
\begin{equation}
    g'(0) = \frac{-g^{(2)} + 4 g^{(1)} - 3 g^{(0)}}{2 h} = 0 \implies g^{(0)} = \frac{4 g^{(1)} - g^{(2)}}{3}. \label{eq:forward_diff}
\end{equation}
Thus, for $k=0$, the boundary conditions at $i=1$ are
\begin{align}
    \omega^2 v_1 &= -\frac{4}{3} \frac{v_2 - v_1}{h^2} + \left( \frac{1}{h^2} + f^2(h) \right) v_1 + \frac{2 N}{h} \left( 1 - \frac{a_\theta(h)}{N} \right) f(h) u_1, \label{eq:bc_v_k0} \\
    \omega^2 u_1 &= -\frac{4}{3} \frac{u_2 - u_1}{h^2} + \left( \frac{(N - a_\theta(h))^2}{h^2} + \frac{3\lambda}{2} f^2(h) - \frac{\lambda}{2} \right) u_1 + \frac{2 N}{h} \left( 1 - \frac{a_\theta(h)}{N} \right) f(h) v_1, \label{eq:bc_u_k0}
\end{align}
where $h$ is sufficiently small. For $k \neq 0$, the boundary conditions at $i=1$ are
\begin{align}
    \omega^2 v_1 &= -\frac{4}{3} \frac{v_2 - v_1}{h^2} + \left( \frac{k^2}{h^2} + f^2(h) \right) v_1 + 2 \left( f(h) + h f'(h) \right) w_1, \label{eq:bc_v_kn0} \\
    \omega^2 u_1 &= -\frac{4}{3} \frac{u_2 - u_1}{h^2} + \left( \frac{k^2 + (N - a_\theta(h))^2}{h^2} + \frac{3\lambda}{2} f^2(h) - \frac{\lambda}{2} \right) u_1 \notag \\
    &\quad - \frac{2 (N - a_\theta(h)) (k^2 + h^2 f^2(h))}{h^2} w_1 + \frac{2 (N - a_\theta(h)) f(h)}{h} \frac{v_2 - v_0}{2 h}, \label{eq:bc_u_kn0} \\
    \omega^2 w_1 &= -\frac{4}{3} \frac{w_2 - w_1}{h^2} + \left( \frac{k^2 + (N - a_\theta(h))^2}{h^2} + \left( 1 + \frac{\lambda}{2} \right) f^2(h) - \frac{\lambda}{2} \right) w_1 \notag \\
    &\quad - \frac{2 (N - a_\theta(h))}{h^2} u_1 + \frac{2 f'(h)}{h} v_1, \label{eq:bc_w_kn0}
\end{align}
where $v_0 = u_0 = w_0 = 0$ if the series starts at a positive power. At $i = M+1$, we impose
\begin{equation}
    v_{M+1} = u_{M+1} = w_{M+1} = 0. \label{eq:bc_infinity}
\end{equation}

\label{Bibliography}
    \bibliographystyle{abbrv}  
    \bibliography{Bibliography}  

\end{document}